\documentclass[a4paper,fleqn,usenatbib]{mnras}

\usepackage{newtxtext,newtxmath}

\usepackage[T1]{fontenc}
\usepackage{ae,aecompl}

\usepackage{graphicx}	
\usepackage{amsmath}	
\usepackage{amssymb}	
\usepackage{siunitx}
\usepackage{tabularx}
\usepackage{subfigure}
\usepackage{microtype}
\usepackage{multirow}
\usepackage[percent]{overpic}
\usepackage{color}
\usepackage[outdir=./]{epstopdf}
\usepackage{nicefrac}
\DeclareSIUnit \parsec {pc}
\DeclareSIUnit \solarmass {\mbox{M$_{\sun}$}}
\DeclareSIUnit \au {AU}
\DeclareSIUnit \year {yr}
\DeclareSIUnit \jansky {Jy}
\DeclareSIUnit \degree {$^{\circ}$}
\newcommand{\referee}{}

\title[Synthetic molecular line observations of the first hydrostatic core]{Synthetic molecular line observations of the first hydrostatic core from chemical calculations}

\author[A. K. Young et al.]{
Alison K. Young,\thanks{E-mail: ayoung@astro.ex.ac.uk (AKY)}
Matthew R. Bate,
Tim J. Harries
and David M. Acreman
\\
Department of Physics and Astronomy, University of Exeter, Stocker Road, Exeter EX4 4QL, UK\\
}

\date{Accepted XXX. Received YYY; in original form ZZZ}

\pubyear{2019}

\begin{document}
\label{firstpage}
\pagerange{\pageref{firstpage}--\pageref{lastpage}}
\maketitle

\begin{abstract}
The first stable object to develop in the low--mass star formation process has long been predicted to be the first hydrostatic core (FHSC). Despite much effort, it has still yet to be definitively observed in nature. More specific observational signatures are required to enable observers to distinguish the FHSC from young, faint, but more evolved protostars. Here we present synthetic spectral line observations for CO, SO, CS and HCO$^+$ that were calculated from radiation (magneto)hydrodynamical models, chemical modelling and Monte Carlo radiative transfer. HCO$^+$~$(1-0)$ and SO~$(8_7 - 7_6)$ spectra of the FHSC show variations for observations at a low inclination which may allow a candidate FHSC to be distinguished from a more evolved object. We find that the FHSC outflow is unlikely to be detectable with ALMA, which would discount the observed sources with slow outflows that are currently identified as candidate FHSCs. We compare the results of simulated ALMA observations with observed candidate FHSCs and recommend Oph A SM1N and N6-mm as the most promising candidates to follow up.
\end{abstract}




\section{Introduction}

The first hydrostatic core (FHSC) is the first stable object predicted to form during the gravitational collapse of a pre-stellar core \citep{larson1969}. The FHSC has a typical radius of $\approx$~5~AU but larger radii are possible with rotation (e.g. \citealt{bate1998,st2006,bate2011,tomida2015}). Magnetohydrodynamic (MHD) models have shown that the FHSC can launch a slow outflow (e.g. \citealt{tomisaka2002aa,banerjee2006aa,machida2008aa,bate2014,lewis2017,wurster2018aa}) and that a magnetically supported pseudo--disc can form (e.g. \citealt{hennebelle2008aa,commercon2012b}). The FHSC lasts for a few hundred to a few thousand years, increasing in central density from $\sim$~\SI{e-12}{\gram\per\centi\meter\cubed} to $\sim$~\SI{e-8}{\gram\per\centi\meter\cubed}. When the central temperature reaches $\sim$\SI{2000}{\kelvin}, the molecular hydrogen dissociates and the second collapse begins. The central temperature and density increase rapidly and the second hydrostatic core, or stellar core, forms with a density of $\sim$~\SI{e-2}{\gram\per\centi\meter\cubed}.

This key stage of star formation is still yet to be observed definitively. The observational search for the FHSC has, however, produced several ``candidate FHSCs'', for example B1-bN/bS \citep{pezzuto2012,hirano2014,gerin2015,fuente2017aa}, CB17-MMS \citep{chen2012} and Aqu-MM1 \citep{Maury:2011aa,young2018aa}. A low luminosity source ($\lesssim 0.1$~L$_{\odot}$) is usually identified as a candidate FHSC if it is faint or undetected at wavelengths $\lambda <$ \SI{70}{\micro\metre} and no outflow faster than 10~km~s$^{-1}$ is observed.

In recent years, attempts have been made to solve the problem of how to distinguish an FHSC from very faint, young protostars by simulating observations. \citet{tomisaka2011} simulated CS line emission before and after FHSC formation using a fixed CS abundance and found the blue asymmetry characteristic of infall, signatures of a rotating outflow and that the linewidth increases after FHSC formation. \citet{commercon2012a} examined the evolution of the spectral energy distribution of the FHSC and were not able to distinguish the FHSC from the newly formed stellar core. \citet{commercon2012b} subsequently calculated synthetic \textit{Atacama Large Millimeter/Submillimeter Array} (ALMA) dust emission maps which showed that ALMA would be able to resolve a fragmentating FHSC but could not reveal any difference between an FHSC and a stellar core. \citet{young2018aa} produced synthetic SEDs for FHSCS with a range of physical properties and evolutionary stages and showed that SEDs are useful for selecting sources most likely to be FHSCs and placing constraints on their rotation rate, for example. Like \citet{commercon2012a}, they did not find any features to distinguish between SEDs of FHSCs and the stellar core. The above results indicate that high resolution observations of molecular line emission are needed to identify the FHSC.


The physical conditions within the FHSC are very different to those of the collapsing core and so different chemical species are expected to form in the FHSC. \citet{furuya2012aa} modelled chemical evolution from a molecular cloud core to the FHSC and found that large organic molecules could be good tracers of the FHSC. \citet{hincelin2016} also calculated chemical abundances as a post--process to a magnetohydrodynamical model of a core collapse up to FHSC formation and examined the differences in chemical abundances between the different components, such as the core and disc. This work illustrates that different species may indeed trace different structures. The next step is to use calculated chemical abundances to simulate observations which could provide detailed diagnostics of the age of FHSCs and of the kinematic structures associated with them.

In this paper we calculate chemical abundances by post--processing high resolution smoothed particle hydrodynamics simulations of the collapse of a pre--stellar core up to the formation of the stellar core. We then use these chemical abundances to simulate observations of selected molecular species.


\section{Method}

\subsection{Hydrodynamical modelling}

We model the collapse of a dense core until just after the formation of the stellar core using the smoothed particle hydrodynamics (SPH) code {\sc sphNG} which originated from the code of \citet{benz1990aa} and was subsequently developed as described in \citet{bate1995aa}, \citet{whitehouse2005}, \citet{Whitehouse2006} and \citet{Price2007aa}. The models employ the radiative transfer method presented in \mbox{\citet{bateketo}} which combines the flux--limited diffusion method with additional interstellar medium (ISM) physics. This treats gas and dust temperatures separately and includes heating by the interstellar radiation field (ISRF) and cooling via line emission. Ideal magnetohydrodynamics (MHD) is included using the formalism of \citet{price2005aa}, divergence cleaning \citep{tricco2012aa} and artificial resistivity \citep{tricco2013aa}.

Two key types of structure associated with the FHSC are nonaxisymmetric disc structures (e.g. spiral arms) and outflows. Magnetic fields are required to drive outflows (e.g. \citealt{bate2014,lewis2015}) but the magnetic braking effect of ideal MHD suppresses the formation of rotational structures (e.g. \citealt{bate2014}). For this reason we performed two simulations: firstly a radiation hydrodynamical model, which we will refer to as the RHD model, to examine rotational structures and secondly a radiation magnetohydrodynamical model, which we will refer to as the MHD model, to simulate observations of the outflow.

{\referee Both simulations start with a dense core that is} modelled as a 1~M$_{\odot}$ Bonnor-Ebert sphere of radius \SI{7e16}{\centi\metre} ($\sim$~\SI{4700}{\au}), central density \SI{8.34e-18}{\gram\per\centi\metre\cubed} using \num{3.5e6} SPH particles. The density contrast between the central and outer regions of the core was 15.1.

For the RHD model an initial uniform rotation of $\Omega =$~\num{2.02e-13}~rad~s$^{-1}$ was used, which gives a rotational to gravitational energy ratio $\beta = 0.02$, to produce an FHSC with non-axisymmetric features.

{\referee For the MHD model, we place the Bonnor-Ebert sphere in a warm, low density, cubic ambient medium of side length \SI{2.8e17}{\centi\metre}. The ratio of the density of the outer regions of the core and the ambient medium (the box) was 30:1, similar to the models of \citet{bate2014}}. The initial mass--to--flux ratio is $\mu= 5$, $\Omega =$~\num{3.44e-13}~rad~s$^{-1}$ and $\beta = 0.05$. An outflow is launched from the FHSC and this extends to $\sim$~\SI{150}{\au} by the time the stellar core forms.

\begin{figure}
\centering
\includegraphics[width=8cm,trim= 0cm 0.2cm 0cm 0cm,clip]{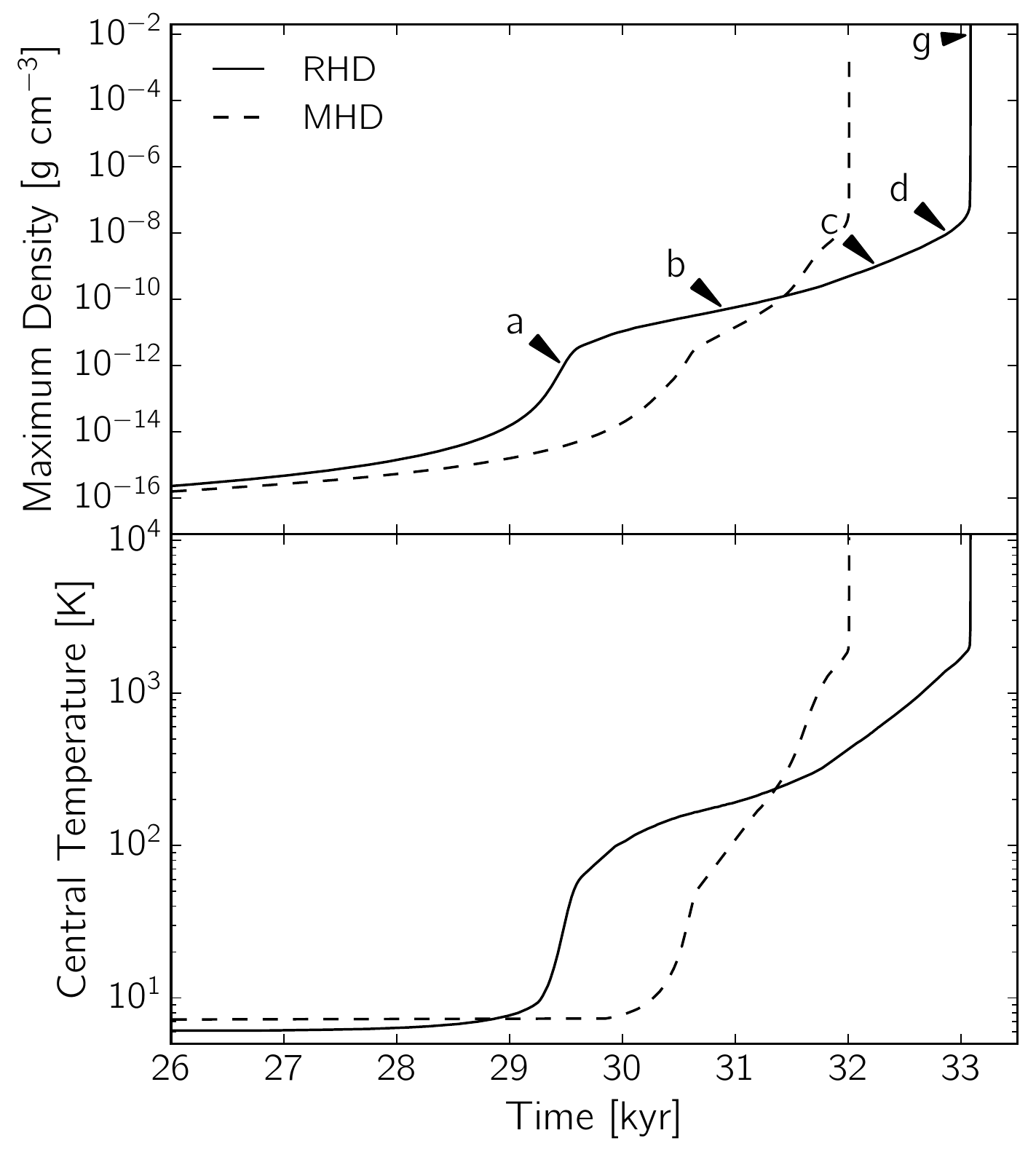}
\caption{The time evolution of the maximum density (top) and central temperature (bottom) for the RHD model (solid lines) and MHD model (dashed lines). Snapshots were taken from these models at certain values of maximum density for comparing evolutionary stages and simulating observations. These are indicated for the RHD model. Snapshot (a) is taken just before the FHSC forms. The density and temperature increase more slowly after this point before rapidly increasing again during the second collapse.}
\label{fig:rhoTevo}
\end{figure} 

\begin{figure}
\centering
\includegraphics[width=8.5cm]{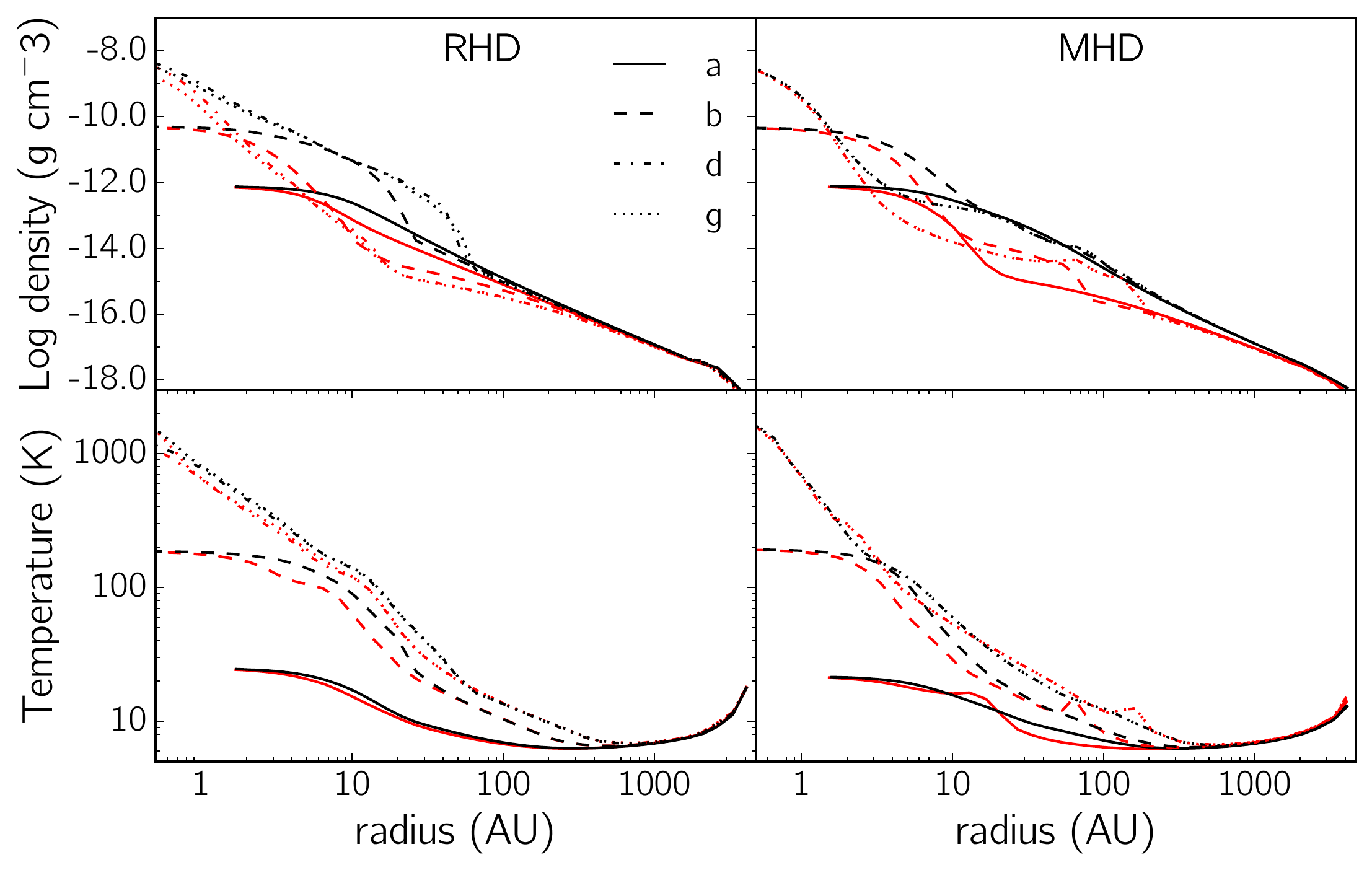}
\caption{Density and temperature radial profiles for the RHD and MHD models at snapshots before FHSC formation (a), during the FHSC stage (b, d) and after the formation of the stellar core (g). Profiles are shown parallel to the rotation axis (black) and perpendicular to the rotation axis (red).}
\label{fig:rhoTprofiles}
\end{figure}

\subsection{Chemical modelling}
Chemistry calculations were performed using {\sc krome} \citep{grassi2014aa}, which is a publicly available code for solving the kinetic equations of non--equilibrium gas--phase chemistry and is based on the {\sc dlsodes} solver \citep{hindmarsh1983aa}. Adsorption and desorption reactions are included by considering the frozen--out species as separate species with their own abundances. 

We calculate the chemical abundances by post--processing the SPH simulations. The total mass density, gas temperature, dust temperature and extinction for each particle are provided by the SPH calculation for each timestep. {\sc krome} is called for each SPH particle and the chemistry is evolved for a time equal to the difference between the current and previous hydrodynamical outputs.

\subsubsection{Chemical network}
\begin{table}
\centering
\caption{Elemental abundances used for the calculation of initial ISM abundances. {\referee These were mostly taken from \citet{reboussin2014aa} and references therein and are very similar to the initial abundances of \citet{hincelin2016} except for S$^+$. We use the higher $S^+$ abundance from the latter work. a) \citet{wakelam2008aa}, b) \citet{jenkins2009aa}, c) \citet{hincelin2011aa} d) \citet{graedel1982aa}. }}
\label{tab:abunds}
\begin{tabular}{ll}
Element & Abundance (n$_{\mathrm j}/n_{\mathrm{H}}$) \\
\hline
H$_2$  & 0.5 \\
He   & \num{9e-2} $^\mathrm{a}$\\
N    & \num{6.2e-5} $^\mathrm{b}$\\
O    & \num{1.4e-4}  $^\mathrm{c}$\\
C$^+$   & \num{1.7e-4}  $^\mathrm{b}$\\
S$^+$   & \num{8e-8}   $^\mathrm{d}$\\
Si$^+$  & \num{8e-9}   $^\mathrm{d}$\\
Fe$^+$  & \num{3e-9}   $^\mathrm{d}$\\
Na$^+$  & \num{2e-9}   $^\mathrm{d}$\\
Mg$^+$  & \num{7e-9}   $^\mathrm{d}$\\
P$^+$   & \num{2e-10}  $^\mathrm{d}$\\
Cl$^+$  & \num{1e-9}   $^\mathrm{d}$\\

\end{tabular}
\end{table}

We combine the gas--phase KIDA 2011 network \citep{wakelam2012aa}, including grain charge transfer reactions, with gas--grain reactions from the network of \citet{reboussin2014aa} which was based on that of \citet{garrod2007aa}. {\referee This includes both thermal desorption reactions and desorption induced by cosmic rays.} The KIDA 2011 network was chosen over the 2014 network because it better matches the observed abundances of a greater number of species that we wished to model (with the exception of {\referee SO}) \citep{wakelam2015aa}. This gives a gas--grain network of 7009 reactions and 651 chemical species, where species frozen out onto the grain surface are counted separately. We neglect grain surface reactions{\referee , and therefore also desorption via exothermic surface reactions,} to simplify the calculations and because they are much less significant for the simpler molecules considered here. The exception is the formation of H$_2$, for which we use the parameterization of the KIDA 2011 network.

The network contains reactions for temperatures up to \SI{800}{\kelvin} but \citet{wakelam2012aa} acknowledge that the gas phase network has only been tested up to \SI{300}{\kelvin}. This network is appropriate for the FHSC stage because gas temperatures only exceed \SI{300}{\kelvin} within the FHSC itself. At stellar core formation, the gas temperature only exceeds \SI{300}{\kelvin} within $\sim$~\SI{3}{\au} of the centre and exceeds \SI{800}{\kelvin} within $\sim$ \SI{1}{\au}. These scales are unobservable due to the high densities. Emission from the very centre is mostly reprocessed by the surrounding material. Chemical abundances did not converge at the highest temperatures and densities within the FHSC so the last converged values are retained along with the time for which they were calculated. This is explained further in Section~\ref{sec:methodCE}.

\subsubsection{Chemical evolution}
\label{sec:methodCE}

First we calculate the chemical evolution for dense, cold ISM conditions: $T=$~\SI{10}{\kelvin}, $\rho=$~\SI{4e-18}{\gram\per\centi\metre\cubed} $(n_{\mathrm{H2}}\approx$~\num{e6}~cm$^{-3})$, $A_v=20$, cosmic ray ionisation rate $\zeta=$~\SI{1.3e-17}{\per\second} and the elemental abundances given in Table~\ref{tab:abunds} which were taken from {\referee \citet{reboussin2014aa} and \citet{hincelin2016}}. The dust grain abundance is calculated assuming a dust-to-gas ratio of 0.01 and that all dust grains have a negative charge initially. The initial electron abundance is found from the difference between the number density of cations and that of negatively charged grains such that the gas and dust have no net charge.

The chemical abundances calculated under the above conditions most closely matched those observed in the dense cores TMC-1 and L134N \citep{agundez2013aa} at $t=$~\num{1.2e5}~years. These abundances are therefore taken as the initial values and assigned to every SPH particle in the simulated dense core. Next, the chemistry was evolved for each particle in the initial Bonnor-Ebert sphere under their individual density, temperature and extinction values for \num{60000} years, which is approximately the free--fall time of the core, to calculate the initial chemical abundances throughout the simulated dense core.

From this point, the new values of $\rho$, $T_{\mathrm{gas}}$, $T_{\mathrm{dust}}$ and $A_v$ are provided by the SPH model for the next timestep and the chemical abundances calculated for the time between the current and previous timesteps. For the first $\sim$~\num{30000}~years, during the first collapse, the physical conditions change little so it is unnecessary to calculate chemical abundances after every hydrodynamical timestep.  After each chemistry timestep, the next minimum chemistry timestep is set relative to the local free-fall time: $\delta t = \frac{1}{3} t_{\rm{ff}}$, where $t_{\rm{ff}}=\sqrt{3\pi/32G\rho}$. The chemistry timesteps therefore decrease as the maximum density increases. During the first collapse stage the chemistry timestep decreases from $\delta t \approx$~\num{8000}~years to a few hundred years. During FHSC phase, the chemistry timestep is $\sim$20-80 years.

We note that the very centre of the core is $\sim$~\SI{7}{\kelvin} initially which means that the network was extrapolated below its minimum temperature of \SI{10}{\kelvin}. In the central regions of the core the visual extinction of the ISRF becomes very large so a cap of $A_v=86$ is implemented. This leads to effectively zero rates for reactions driven by UV photons for these regions.

As mentioned in the previous section, the chemical solver failed to converge at high densities. If the chemistry fails to converge for a particle the chemistry timestep is halved. If the chemistry does converge with the smaller timestep, {\sc krome} is called a second time for the particle with the same half--timestep such that the chemistry is evolved for the same net timestep. If the chemistry does not converge with the smaller timestep, the previous abundances are retained and the number of times the particle has failed to converge is recorded. 
{\referee The chemical model fails to converge within the centre of the FHSC, where $\rho \gtrsim $~\SI{e-10}{\gram\per\centi\meter\cubed}. This value is exceeded midway through the FHSC phase and within 2-3~AU of the centre, or within 8~AU near midplane of the disc in the RHD model. If the chemistry for a given SPH particle has failed to converge 5 times, or if $\rho > $~\SI{e-10}{\gram\per\centi\meter\cubed}, the chemistry calculation is skipped and the current abundances are retained. All unconverged particles lie inside the FHSC and, for the RHD model, comprise only 1.7 per cent of the total mass during the FHSC stage and 2.8 per cent of the total mass after stellar core formation. These values are very similar for the MHD model. Since we are interested in modelling observational properties, the chemistry of the centre of the FHSC is inconsequential because the FHSC is extremely optically thick and submillimetre emission from this region would be reprocessed before reaching the observer.}

\begin{figure*}
\centering
\includegraphics[width=17cm,trim= 0cm 13.2cm 4.8cm 0cm,clip]{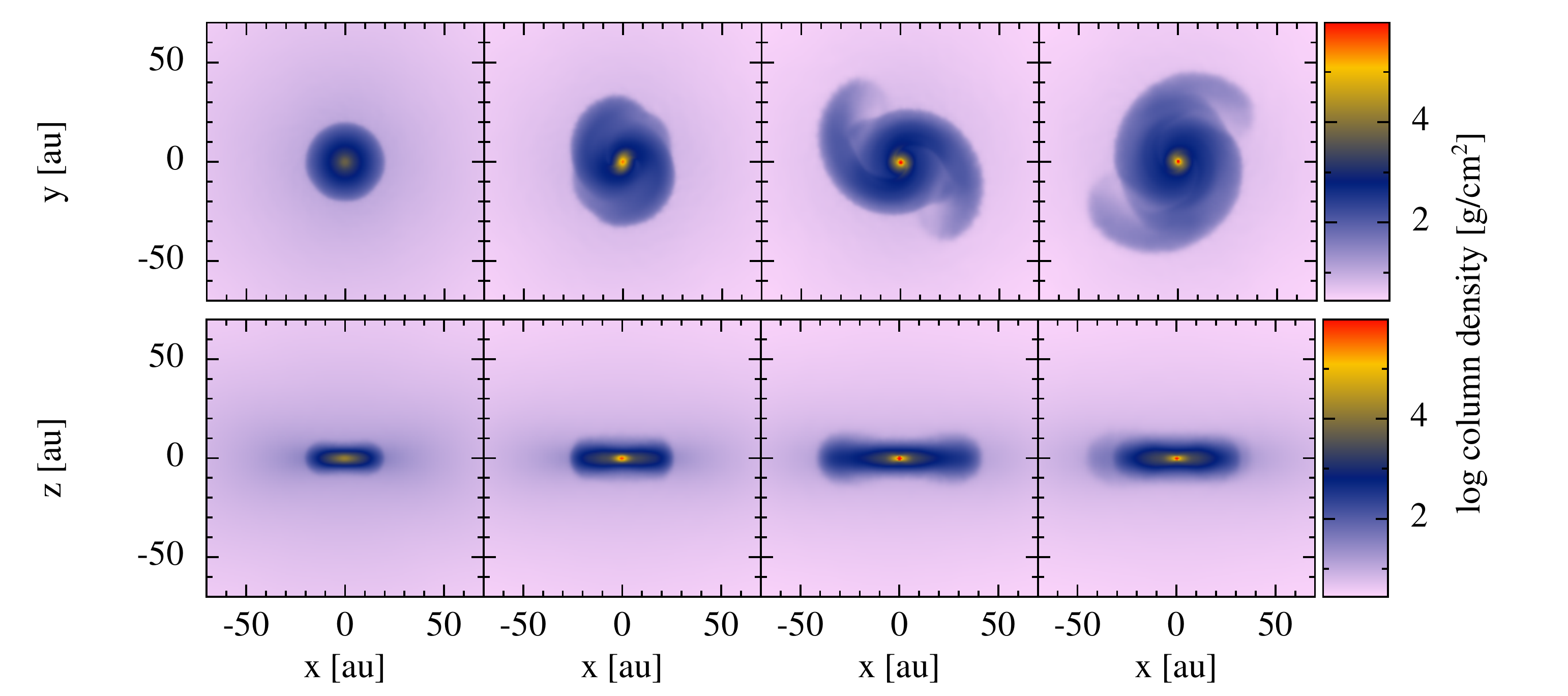}
\includegraphics[width=17cm,trim= 0cm 13.2cm 5cm 0cm,clip]{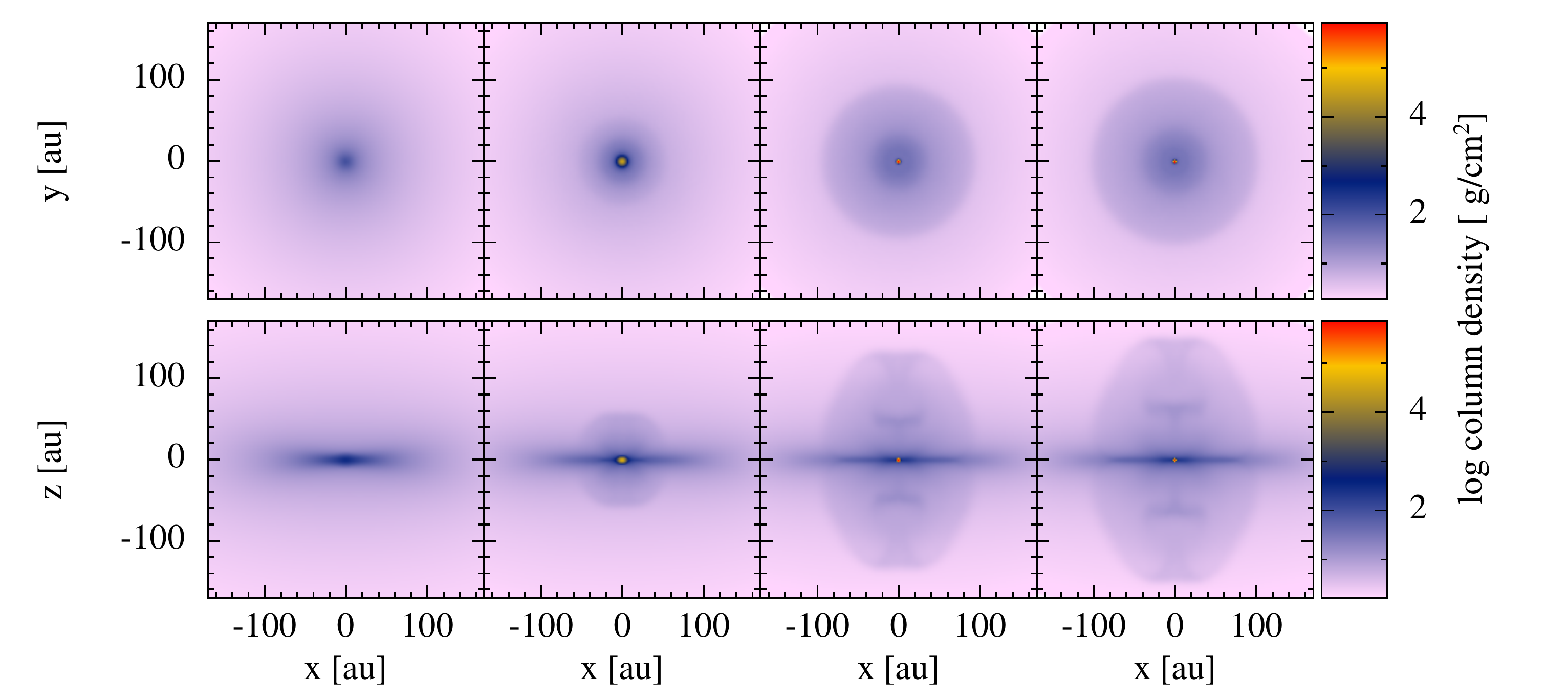}
\caption{Evolution of the RHD model viewed at $i=0^{\circ}$ and $i=90^{\circ}$ (upper two rows) and the MHD model also viewed at the same inclination angles (lower two rows). From left to right, panels show FHSC snapshots (b), (c) and (d) and the stellar (second) core snapshot (g). The development of spiral structure is apparent in the RHD model and the outflow is seen in the MHD model at $i=90^{\circ}$.}
\label{fig:columndens}
\end{figure*}
\begin{figure}
\centering
\includegraphics[width=7.3cm,trim= 0cm 0cm 3cm 0cm,clip]{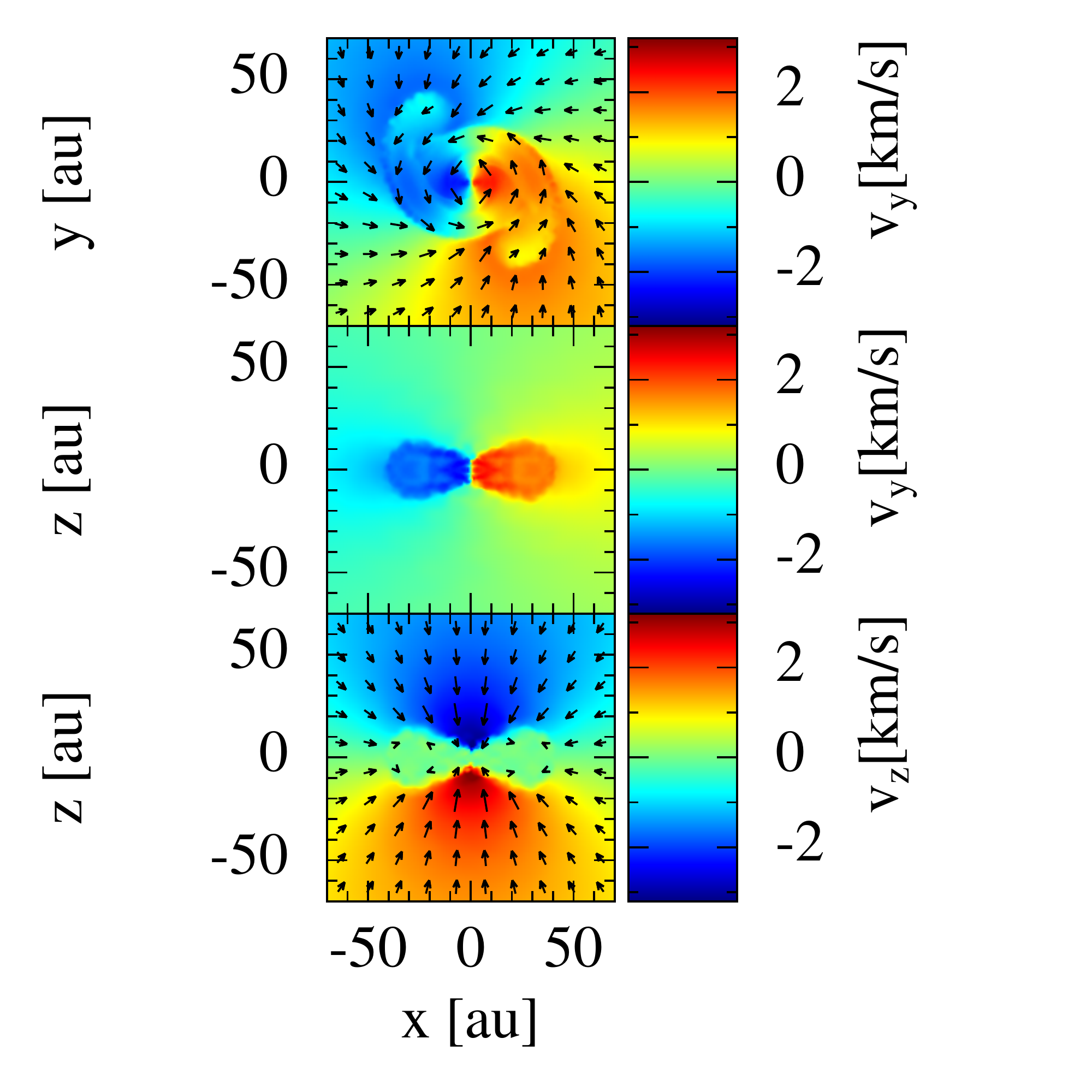}\\
\includegraphics[width=7.3cm,trim= 0cm 0cm 3cm 0cm,clip]{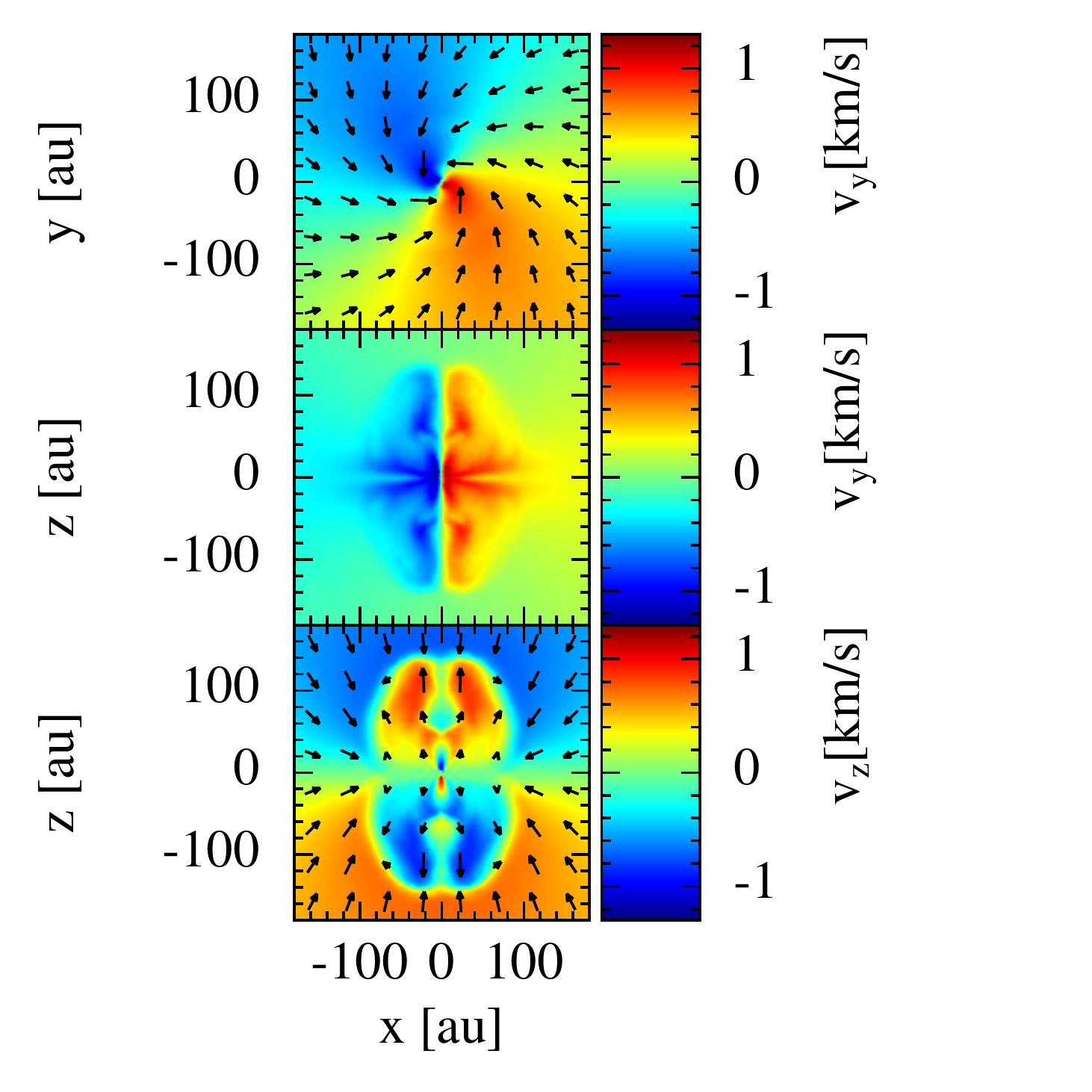}
\caption{Velocity structure at snapshot (d) (late FHSC) Upper panels: RHD model. Cross-section of the velocity component approaching an observer viewing the system edge--on, $v_y$, in the midplane (top) showing a rotationally supported disc; $v_y$, as viewed edge--on (centre); the vertical component of the velocity, $v_z$, (bottom). Arrows show the direction of the net velocity. Lower panels: As above but for the MHD model. The key features are the rotating pseudo-disc and outflow. The highest velocities are due to infall within $\pm 20$~AU. }
\label{fig:velcuts}
\end{figure}

\subsection{Radiative transfer}

We took snapshots from the chemical post--processing of the hydrodynamical models to synthesize the line emission of selected molecular species with {\sc torus}. {\sc torus} is a Monte Carlo radiative transfer code that has been frequently applied to {\referee dust continuum} and line--transfer simulations of star--formation and protostellar discs (e.g. \citealt{harries2000,harries2011aa}). First the adaptive mesh refinement (AMR) grid is populated from the SPH particles following the method described by \citet{rundle2010}. The SPH particles were mapped onto the cubical AMR grid of side length \SI{6e15}{\centi\meter} (400~AU) and mass resolution of \SI{e27}{\gram} per grid cell for the radiative transfer modelling, with the exception of the HCO$^+$ models. The density and temperature are low enough outside this region that the contribution to the spectrum is negligible relative to the contribution from the inner few 10s of AU. There is, however, a significant contribution from the envelope to the HCO$^+ (1-0)$ spectrum, therefore a larger grid of \SI{8e16}{\centi\meter} (5350~AU) with a coarser mass resolution (\SI{e28}{\gram} per cell) was used for this transition.

The level populations were calculated assuming local thermodynamic equilibrium (LTE) and are calculated iteratively from the radiation incident on each cell. This is a valid assumption here because the density in the simulated collapsing cloud is higher than the critical density for the species and transitions considered. A Monte Carlo method is used to follow the emission and absorption of continuum and line photon packets through the model to an observer. {\referee A turbulent line broadening of \SI{0.1}{\kilo\metre\per\second} is applied.} A position-position-velocity (PPV) FITS image cube is produced for an observer at 150~pc. The PPV cubes have a side length of 370~AU and 81 channels from \SIrange{-4}{+4}{\kilo\meter\per\second}, giving a channel width of \SI{0.1}{\kilo\metre\per\second}.

{\referee The silicate dust grain type of \citet{drainelee1984} with a power law distribution of grain sizes between \SI{0.1}{\micro\metre} and \SI{1}{\micro\metre} of $n(a) \propto a^{-q}$ with $q=3.5$ \citep{mathis1977} was used to calculate the dust extinction, continuum emission for the continuum images, and the continuum contribution to the spectra.}

\subsection{Processing image files}
\label{sec:processing}
The PPV cubes generated straight from {\sc torus} are noise--free so it is sufficient to perform the continuum subtraction with just one line--free channel. Integrated intensity maps were constructed from the continuum--subtracted PPV cubes by summing the channels as follows, where $\Delta v$ is the channel width:
\begin{equation}
I = \sum_{v=1}^{n} I_v \Delta v .
\label{eq:mom0}
\end{equation}

Spectra were calculated by averaging pixels within a 0.35~arcsec diameter aperture for CS and SO, and within a 0.67~arcsec aperture for CO and HCO$^+$ for which the emission is more extended.

ALMA observations were simulated from the PPV cubes using the {\tt simalma} routine in CASA \citep{mcmullin2007aa}. Channels corresponding to \SIrange{-4.0}{-3.5}{\kilo\meter\per\second} and \SIrange{+3.5}{+4.0}{\kilo\metre\per\second} for which there was no line emission were averaged and subtracted from all channels to remove the continuum emission.

Position--velocity diagrams were constructed for CO PPV cubes to examine rotation. The PV cuts were taken perpendicular to the rotation axis, through the midplane of the disc or pseudo--disc. The brightness of each velocity channel at each position was calculated by adding the intensity of pixels within $\pm$~0.15~arcsec above and below this centre line.

\section{Results}
\subsection{Morphology and velocity structure of the hydrodynamical models.}
\label{sec:morphology}

The evolution of the maximum density and central temperature are presented in Fig.~\ref{fig:rhoTevo}. Initially the temperature is lowest at the centre of the core and the temperature remains $\approx$~15~K at the edge throughout the simulation due to heating by the ISRF. It is therefore more useful to compare the central temperature rather than maximum temperature. The FHSC forms after \num{29500} years in the RHD model and after \num{30700} years in the MHD model. The FHSC is deemed to have formed when the rapid increase in central density and temperature that occurs during first collapse ceases and the density and temperature continue increasing at a slower rate (just after point (a) in Fig.~\ref{fig:rhoTevo}). The FHSC lasts for \num{3500} years in the RHD model and for \num{1300} years in the MHD model. The lifetime is shorter in the MHD model because magnetic braking slows the rotation such that the rotational support is reduced.

We take snapshots from the hydrodynamical models at selected times to allow comparison of the synthetic observations for different evolutionary stages. These are taken at the following values of the central density: (a)\SI{e-12}{\gram\per\centi\metre\cubed}, (b) \SI{5e-11}{\gram\per\centi\metre\cubed}, (c) \SI{e-9}{\gram\per\centi\metre\cubed}, (d) \SI{e-8}{\gram\per\centi\metre\cubed}, (g) \SI{e-2}{\gram\per\centi\metre\cubed} (these are the same values as used in \citealt{young2018aa}). Snapshot (a) is taken just before FHSC formation, (b), (c) and (d) are during the FHSC stage and (g) is taken just after the formation of the stellar core. Density and temperature profiles for four key evolutionary stages are presented in Fig.~\ref{fig:rhoTprofiles}. It is important to note that the temperature remains $<100$~K at $r>10$~AU even at snapshot (g), i.e. shortly after stellar core formation, and $T<300$~K at $r>4$~AU (see Fig.~\ref{fig:rhoTprofiles}) which means the chemical network is valid for observable size scales.

The morphology of the RHD and MHD models is shown in Fig.~\ref{fig:columndens}. In the RHD model (upper two panels), the FHSC has the form of a rotationally supported disc which increases in radius from $\sim$~20~AU to $\sim$~35~AU (c.f. \citealt{bate1998,bate2011}). Late in the FHSC stage, after \num{32000} years, a spiral instability develops. The disc is also apparent in the diagrams of the velocity structure shown in Fig.~\ref{fig:velcuts}, upper panels, as well as the infall perpendicular to the plane of the disc. After second collapse and the formation of the second (stellar) core, the disc and spiral structures persist.

In the MHD model, the FHSC forms after \num{30700} years. The cloud core is initially rotating quickly in this model but the rotation is significantly reduced by magnetic braking and the FHSC is not rotationally supported. A pseudo-disc forms around the FHSC, comprised of gas with infalling, rotating motions (see Fig.~\ref{fig:velcuts}). An outflow is launched around 400 years after FHSC formation, the structure of which is shown in Fig.~\ref{fig:columndens}, lower panels, and the lower panels of Fig.~\ref{fig:velcuts}. Outflow velocities reach \SIrange{0.5}{1.0}{\kilo\metre\per\second} at distances of 20 to 150~AU from the midplane. Infall continues through the pseudo--disc at $|v|\leq$~\SI{0.5}{\kilo\metre\per\second} and also vertically within $r<10$~AU with velocities exceeding \SI{1}{\kilo\metre\per\second}. The line--of--sight rotational velocities in the outflow and pseudo--disc are similar to the outflow and infall velocities which means that all of these regions must be considered to understand the nature of the spectra.

Infall velocities reach $\sim 3$~\si{\kilo\metre\per\second} in the RHD model but are lower in the MHD model due to additional magnetic pressure and remain $< 1.5$~\si{\kilo\metre\per\second} after stellar core formation.
\subsection{Chemical evolution}
\label{sec:chem_evo}

\begin{figure*}
\centering
\includegraphics[width=17.2cm]{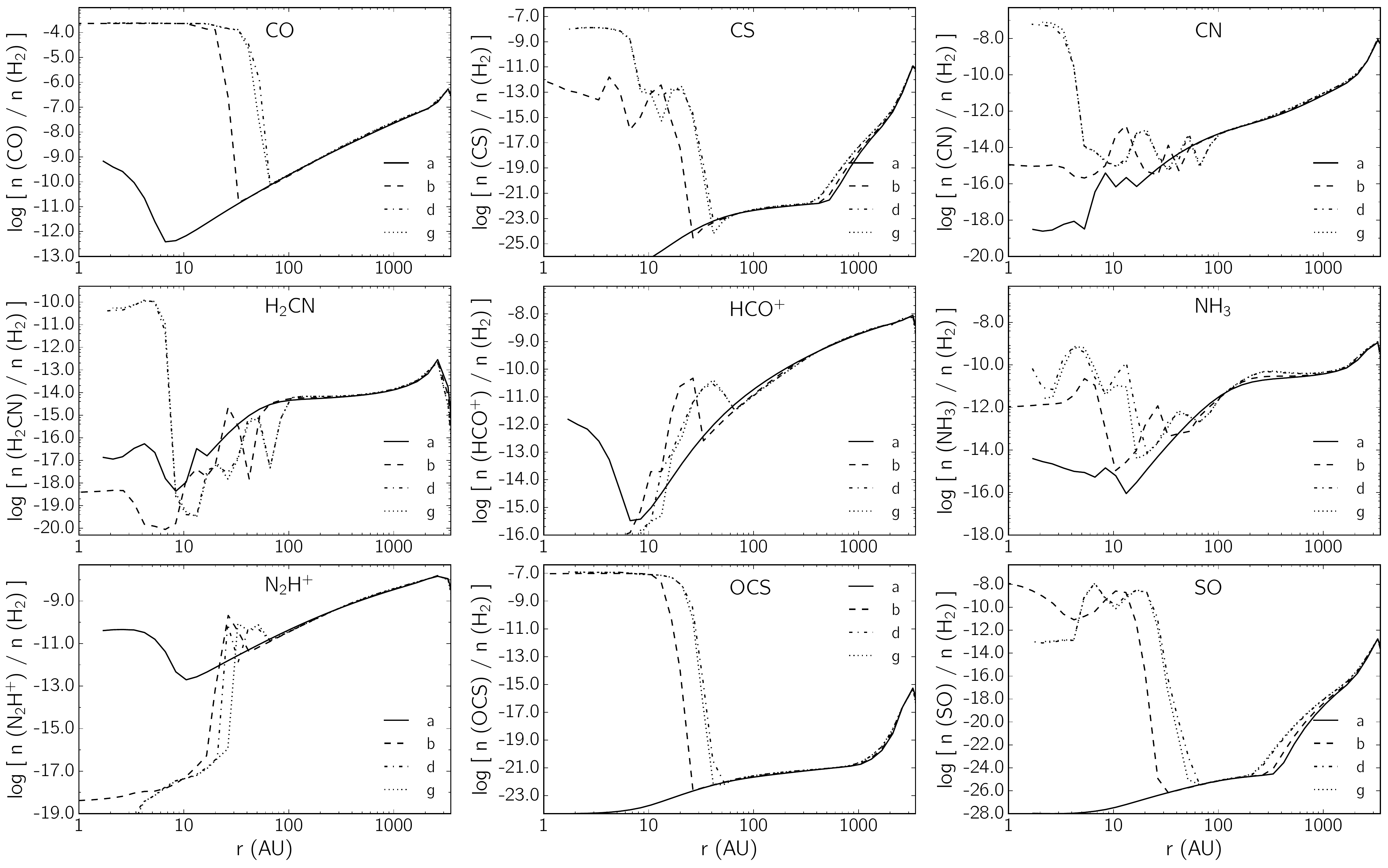}
\caption{Average abundances of selected species calculated for the RHD model perpendicular to the rotation axis.}
\label{fig:B02abundh}
\end{figure*}

\begin{figure*}
\centering
\includegraphics[width=17.2cm]{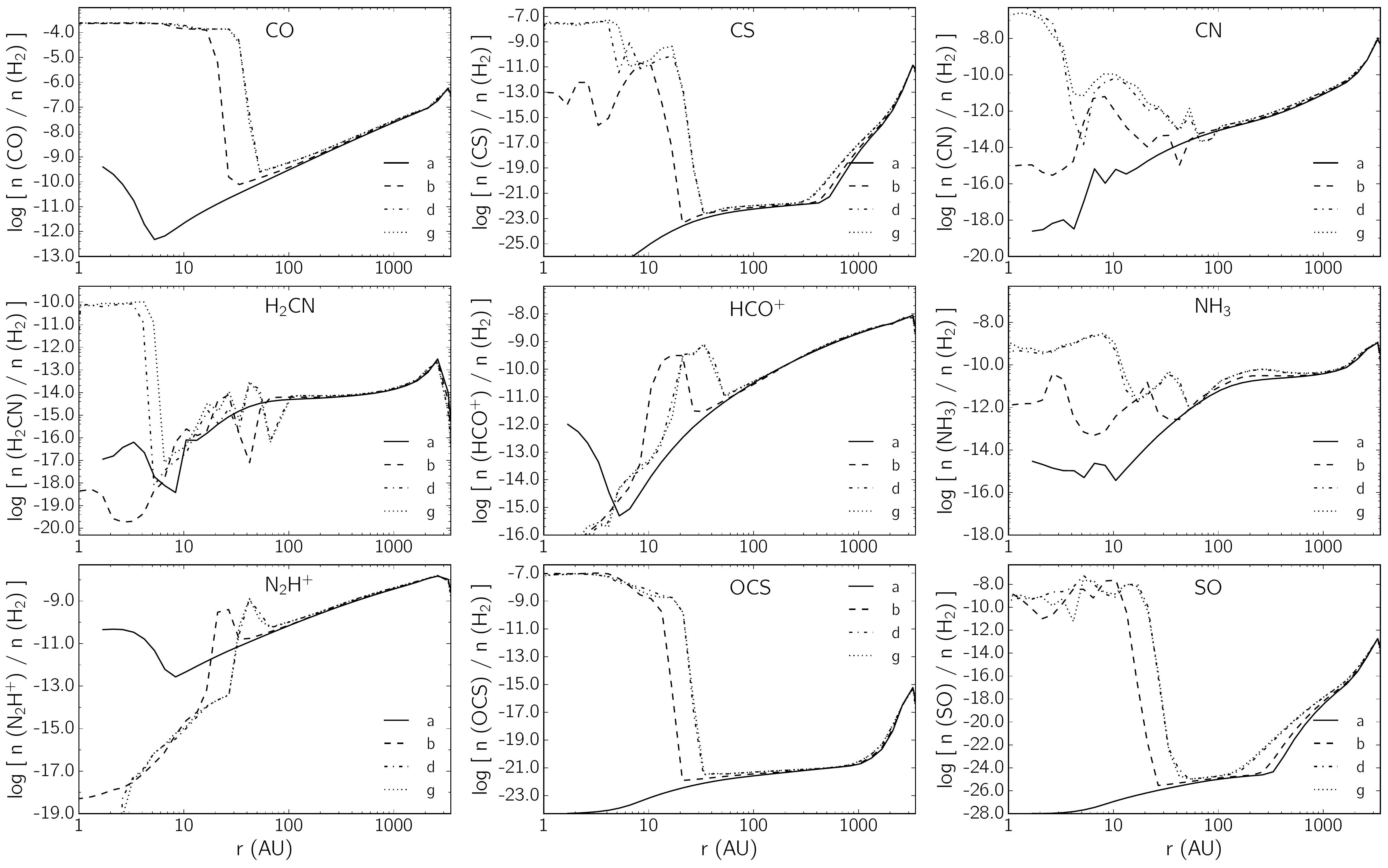}
\caption{Same as for Fig.~\ref{fig:B02abundh} but in the vertical direction (parallel to the rotation axis).}
\label{fig:B02abundv}
\end{figure*}

\begin{figure*}
\centering
\includegraphics[width=17.2cm]{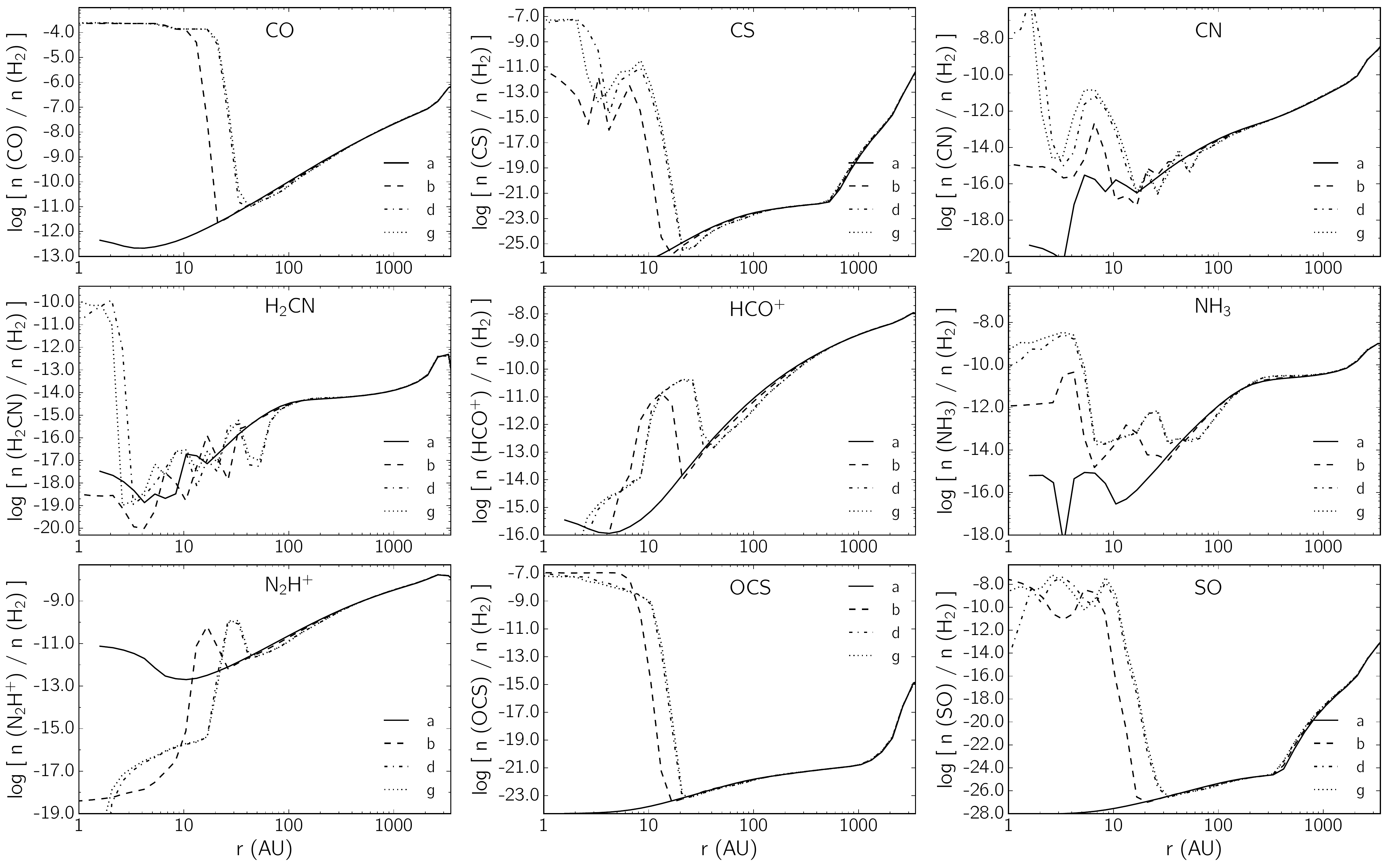}
\caption{Average abundances of selected species from the MHD model perpendicular to the rotation axis.}
\label{fig:Mu5abundh}
\end{figure*}

\begin{figure*}
\centering
\includegraphics[width=17.2cm]{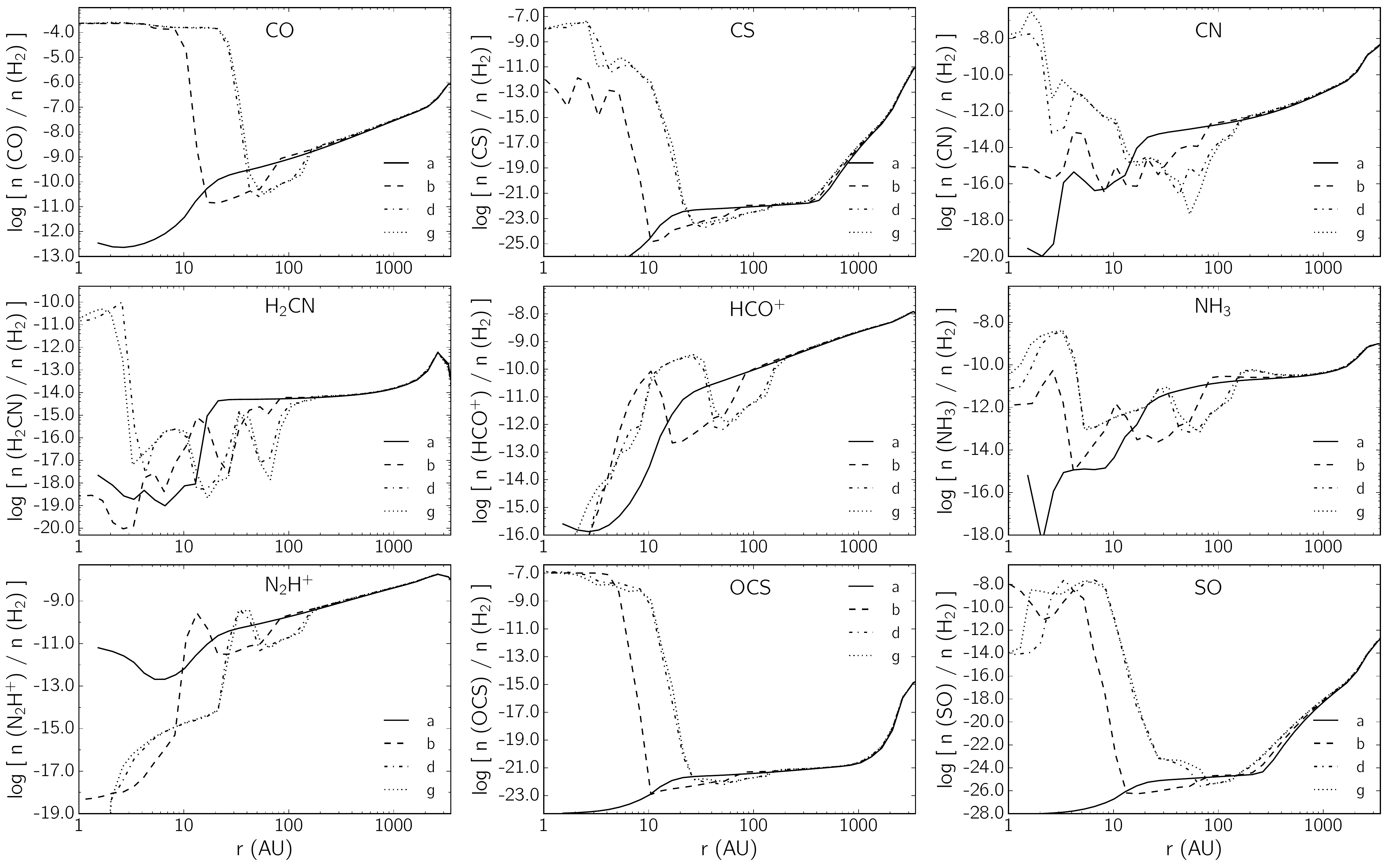}
\caption{Same as for Fig.~\ref{fig:Mu5abundh} in the vertical direction.}
\label{fig:Mu5abundv}
\end{figure*}
\begin{figure*}
 \centering
  \includegraphics[width=17cm,trim= 0cm 9.2cm 0cm 0.1cm,clip]{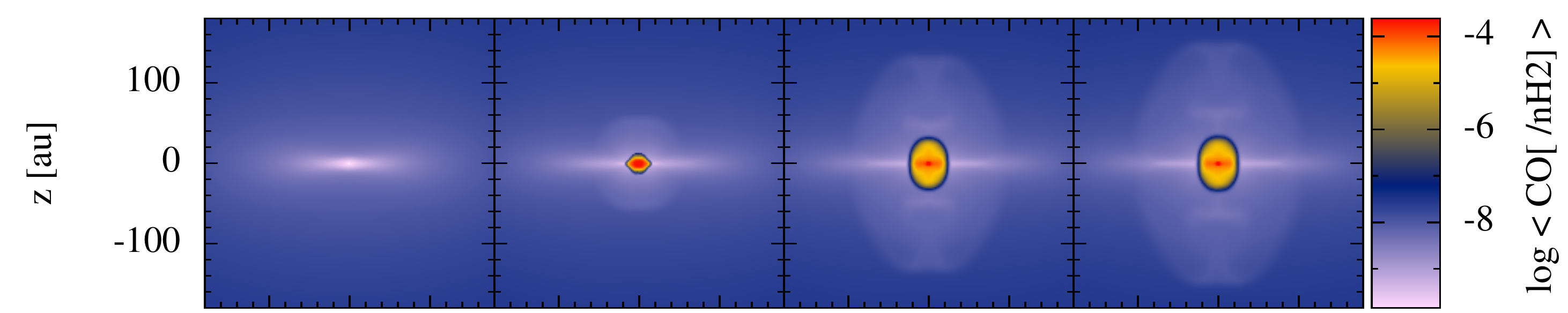}
  \includegraphics[width=17cm,trim= 0cm 9.2cm 0cm 0.1cm,clip]{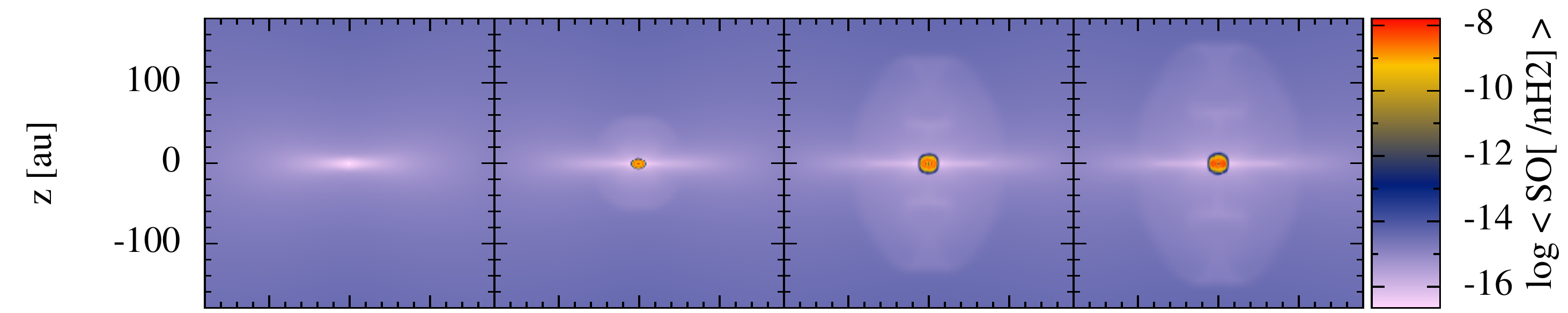}
 \includegraphics[width=17cm,trim= 0cm 9.2cm 0cm 0.1cm,clip]{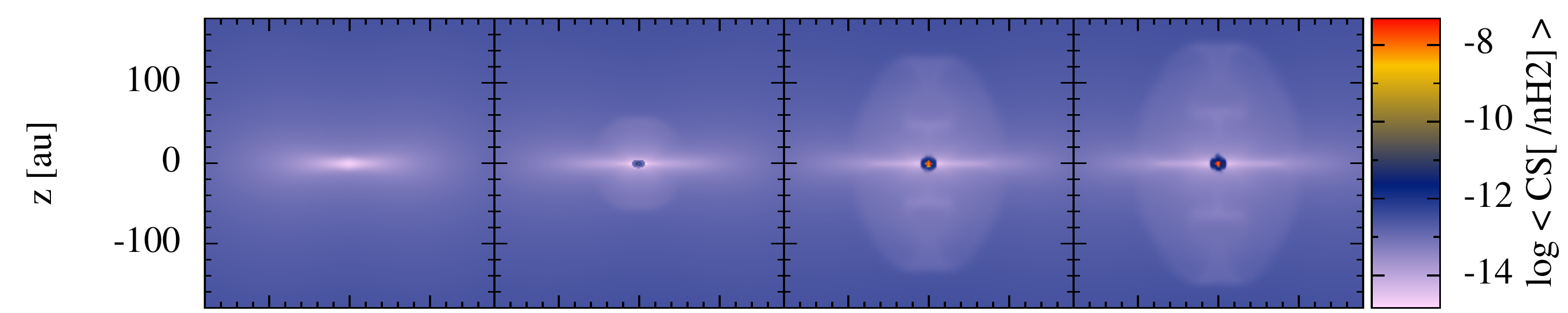}
  \includegraphics[width=17cm,trim= 0cm 7cm 0cm 0.1cm,clip]{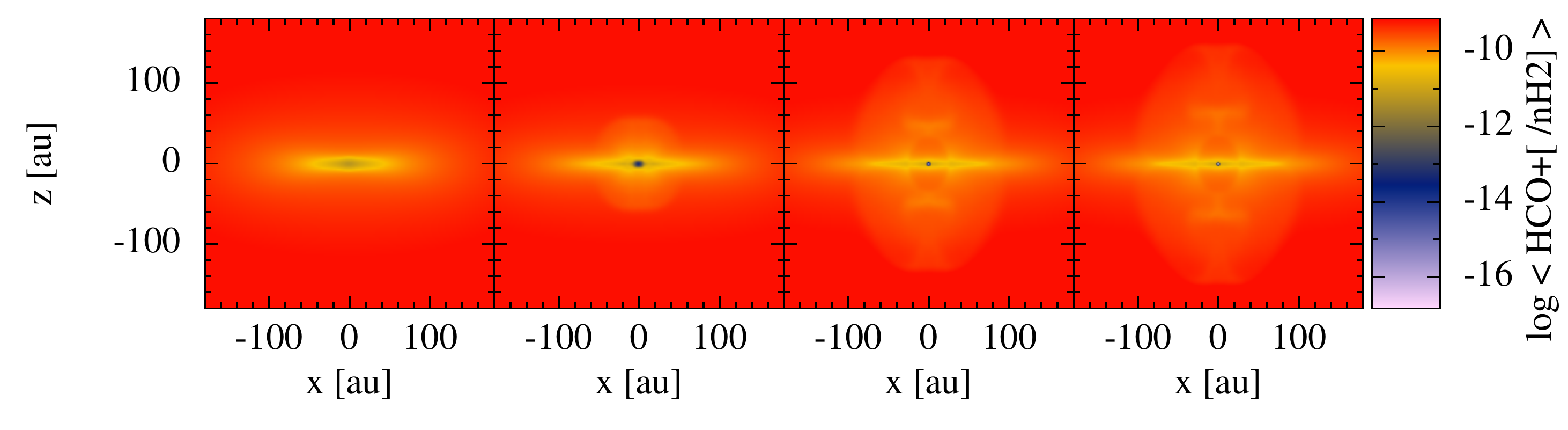}
\caption{Mass weighted average abundances (relative to n$_{\mathrm{H}_2}$) of selected species from the MHD model, viewed at $i=90^{\circ}$. From left to right, the snapshots are of the first collapse (a), early FHSC (b), late FHSC (d), and just after stellar core formation (g).}
 \label{fig:MuChem}
\end{figure*} 

Chemical abundances as a function of radius for the RHD case are shown in Figs.~\ref{fig:B02abundh} and \mbox{\ref{fig:B02abundv}}. From here on we refer to chemical abundances relative to the abundance of molecular hydrogen. Other species with high abundances not shown here include CCH, HCN, HNC and H$_2$O. Some interesting species such as methanol and formaldehyde form primarily through grain surface reactions and so we cannot calculate realistic abundances for these. The abundances beyond \SI{100}{\au} remain mostly unchanged throughout FHSC phase and stellar core formation because the temperature and density change very little. However, for nearly all the species shown, there is a dramatic change at $r\lesssim$~\SI{30}{\au} when the FHSC forms and the gas--phase abundances increase quickly. We note that the abundances of the species discussed here are also high at large radii in the envelope and surrounding molecular cloud. However, if we select transitions that are excited at the higher temperatures of the warm core then it is only the abundances towards the centre of the cloud core that matter for observations. 

The abundance of CS peaks sharply within a radius of a few AU, indicating that it traces the warmest region at the centre of the FHSC. The peak abundance increases significantly during the FHSC stage so we may expect to see CS lines brighten as the core evolves. CN traces a similar region but its central abundance increases later than CS. The abundance of H$_2$CN increases rapidly within the FHSC and becomes 4--5 orders of magnitude higher than in the surrounding envelope, which may provide an indication of the FHSC.

\begin{figure*}
\centering
\includegraphics[height=5.8cm,trim= 1cm -0.1cm 1.3cm 0cm,clip]{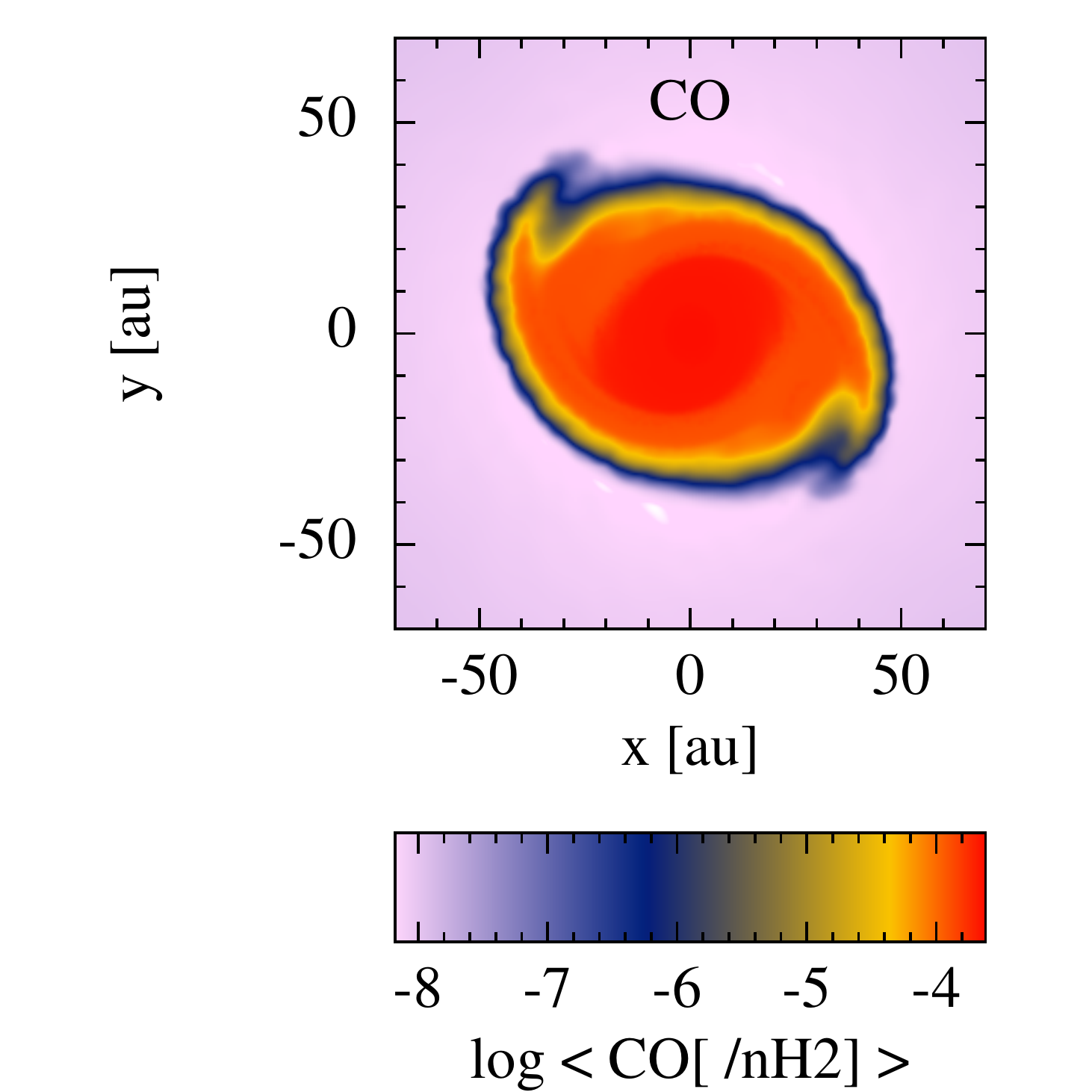}
\includegraphics[height=5.8cm,trim= 3.1cm -0.1cm 1.3cm 0cm,clip]{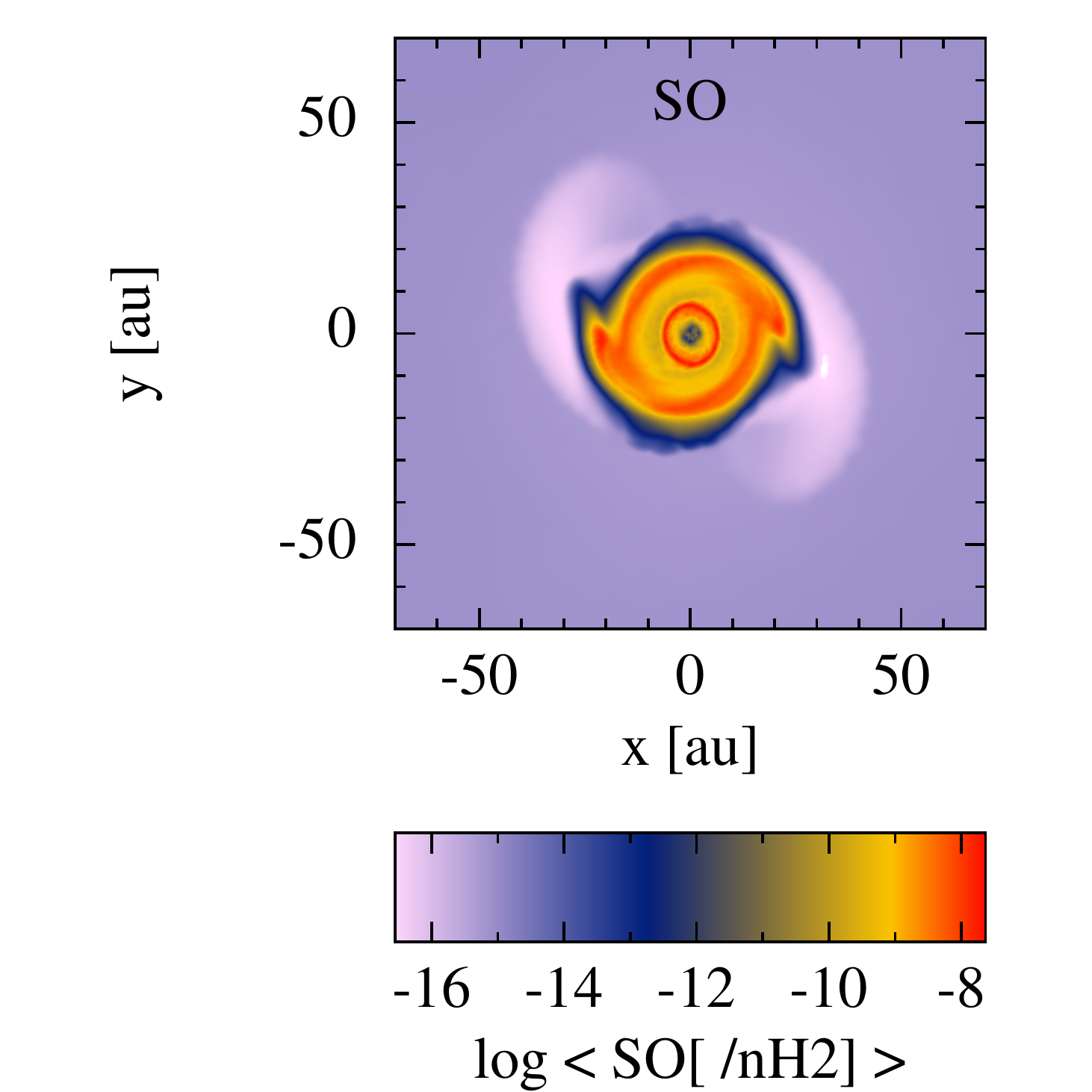}
\includegraphics[height=5.8cm,trim= 3.1cm -0.1cm 1.3cm 0cm,clip]{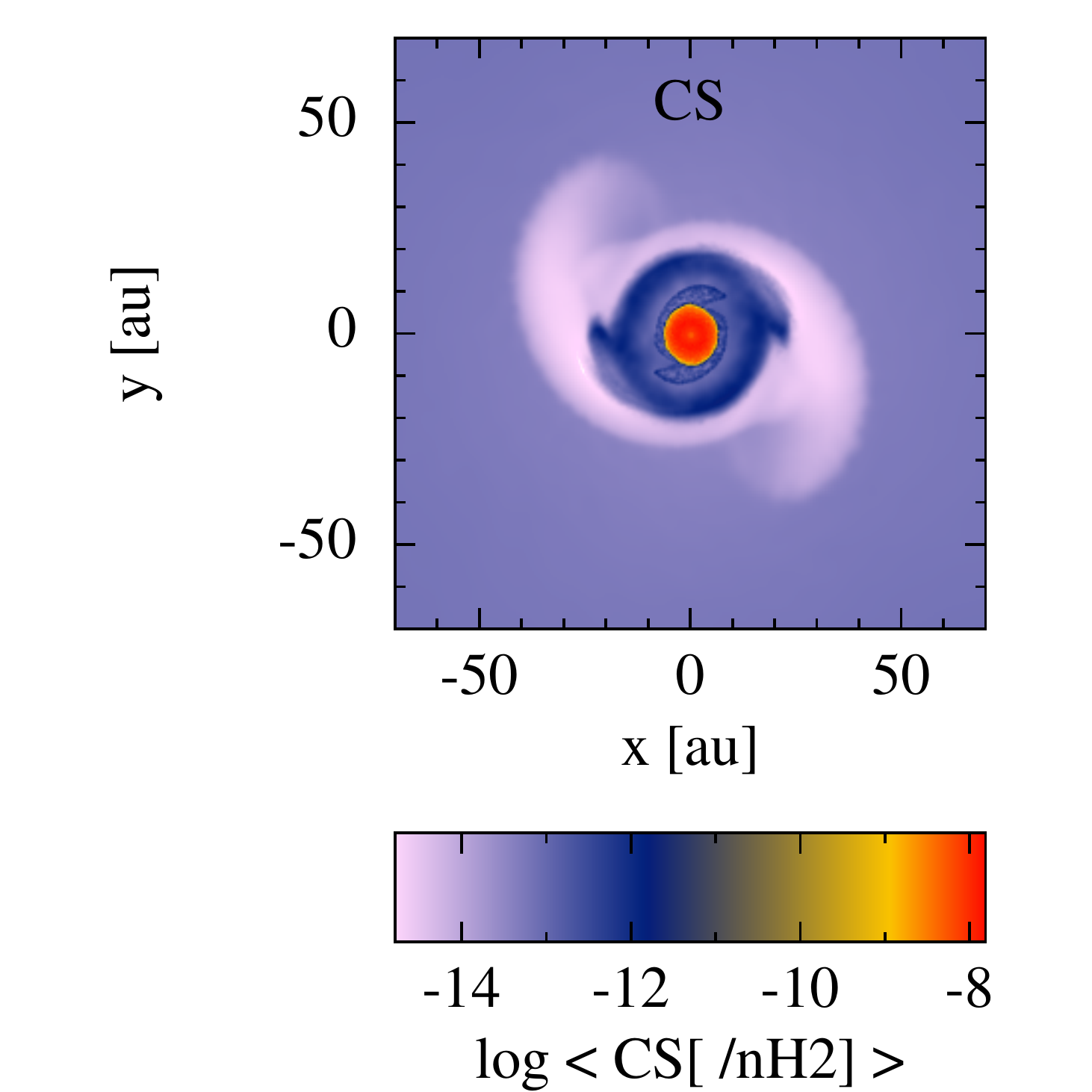}
\includegraphics[height=5.8cm,trim=3.1cm -0.1cm 1.3cm 0cm,clip]{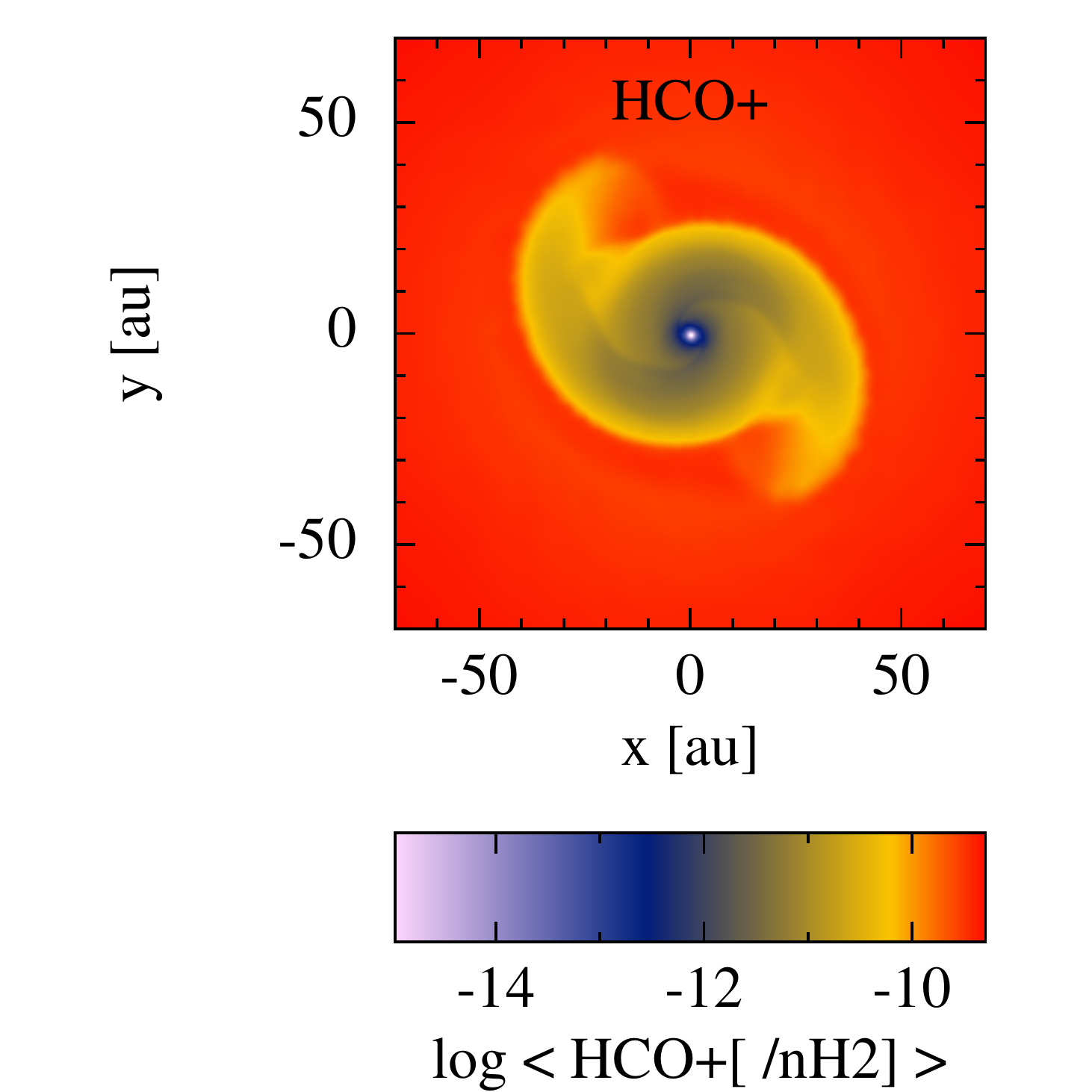}
\caption{Mass weighted average abundance relative to n$_{\mathrm{H}_2}$ for snapshot (d), late FHSC for CO, SO, CS and HCO$^+$ respectively. CO and CS trace the spiral structure of the FHSC and CS traces just the central few AU. The HCO$^+$ abundance is lower in the spiral arms than in the envelope.}
\label{fig:spiralabunds}
\end{figure*}

OCS, SO and CO trace the FHSC but to different radii. In the RHD model, these species trace regions within $\sim$ 15, 20 and 30~AU respectively in the disc plane. For these species, the peak abundance does not change after the FHSC forms but the radius within which the abundance is high increases as the disc warms up. For example, CO desorbs from dust grains as soon as the FHSC forms and the size of the disc traced by CO increases from $\sim$~\SI{15}{\au} to $\sim$~\SI{30}{\au} in the first half of the FHSC phase. OCS follows the same pattern but traces the disc to a smaller ($<$~\SI{15}{\au}) radius.

Abundances in the MHD model (Figures \mbox{\ref{fig:Mu5abundh}}, \mbox{\ref{fig:Mu5abundv}} and \mbox{\ref{fig:MuChem}}) are similar except in the vertical direction after the outflow is launched. HCO$^+$ and CO are abundant out to larger radii in the vertical direction than the horizontal due to the outflow. SO still appears to trace the central core well since the abundance decreases by several orders of magnitude at 10~\si{\au} even in the direction of the outflow. In the MHD model, the peak abundance of SO increases by a factor of 10 during FHSC stage.

The abundance of HCO$^+$ is lowest in the centre of the core after FHSC formation and increases with radius. There is a region of enhanced HCO$^+$ abundance within a shell at $\sim$10~-~50~\si{\au}. The abundance of HCO$^+$ is lower in the outflow than surrounding envelope and lower still in the pseudo--disc. HCO$^+$ is therefore likely to trace the inner envelope structure, but not the disc or pseudo--disc.

The abundance of NH$_3$ peaks within the FHSC in the MHD model and to a lesser extent also in the RHD model. NH$_3$ is depleted to a radius of $\sim$100~AU within the pseudo--disc and within the outflow. In the RHD model, NH$_3$ is depleted within the spiral arms but the abundance is high within a radius of $r \lesssim 30$~AU and is high in the envelope.

Abundances of the other species here peak sharply in the centre, except for SO which traces out to 8~AU above the midplane and CO which traces out to 20~AU. At this distance, CO should trace the base of the outflow but SO is more likely to trace the infall close to the centre of the core.

The chemical abundances within and close to the FHSC are very different to the abundances in the centre of the pre-stellar core before FHSC formation because many species desorp from the dust grains with the rapid increase in temperature. For both models, the species that show significant evolution in abundance during the FHSC stage are CS, NH$_3$ and SO. The abundance of OCS in the centre of the core decreases during FHSC stage in the MHD model.

For all species considered here, clues to the evolutionary stage of the core are likely to be given by changes in envelope structures revealed by certain species rather than by changes in chemical abundances alone. \citet{murillo2018aa} conclude from observations of two Class 0 protostars that temperature is a key factor in driving the chemical composition of a protostellar envelope. Since the temperature structures are similar, it is likely that the kinematics rather than the chemical abundances will prove a better diagnostic of the FHSC.

\subsection{Synthetic observations}
In this section we present synthetic observations of {\referee CO $(4-3)$ (461.041~GHz), SO $(8_7 - 7_6)$ (340.714~GHz), CS $(8-7)$ (391.847~GHz) and HCO$^+$ $(1-0)$ (89.198~GHz)}. These species are expected to have high enough abundances to be detectable and to trace different structures. CS also shows a significant change in abundance as the FHSC grows. CN, NH$_3$ and N$_2$H$^+$ are also likely to trace FHSC structures. However, at the moment we are unable to compute the hyperfine structure lines necessary to model the emission from these molecules.

\subsubsection{Continuum}
\begin{figure}
\centering
\includegraphics[height=4.5cm,trim= 0cm 0cm 0cm 0cm,clip]{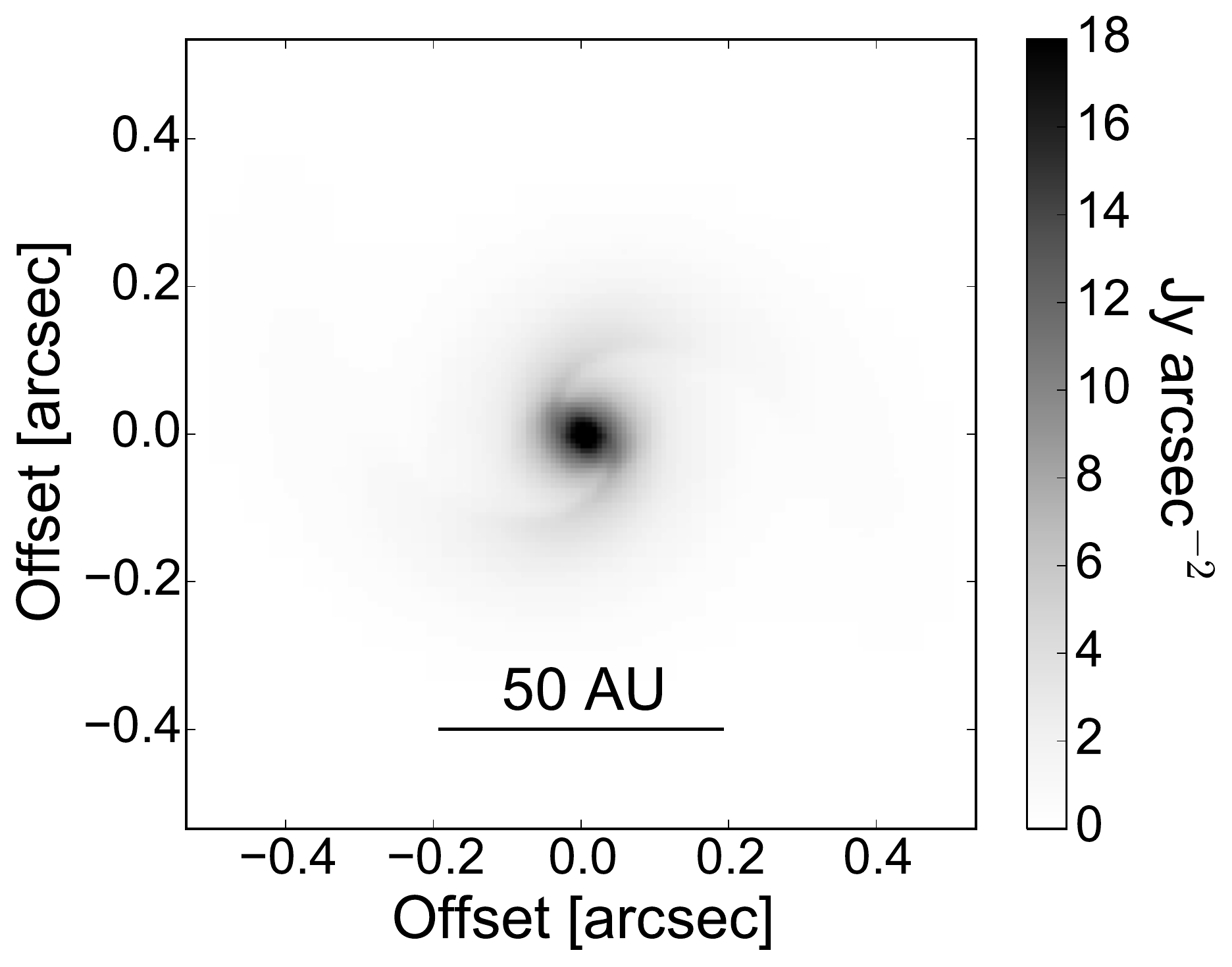}
\caption{230~GHz (1.3mm) continuum image for snapshot (d) from the RHD model. At millimetre wavelengths the continuum emission is brightest in the centre of the FHSC only and the spiral features are faint. The brightness scale has been capped at \num{18.0}~Jy~arcsec$^{-2}$ but the maximum intensity is \num{20.2}~Jy~arcsec$^{-2}$.}
\label{fig:rhocont}
\end{figure}

The \SI{230}{\giga\hertz} continuum image of the late FHSC from the RHD model is presented in \mbox{Fig.~\ref{fig:rhocont}} to allow comparison with line observations. The continuum emission here is tracing the denser regions such that the centre of the FHSC appears bright and the spiral structures are only visible within 20~AU. 

%
\subsubsection{CO}
\label{sec:COresults}
\begin{figure}
\centering
\includegraphics[width=\columnwidth,trim= 1cm 1.5cm 1cm 0cm,clip]{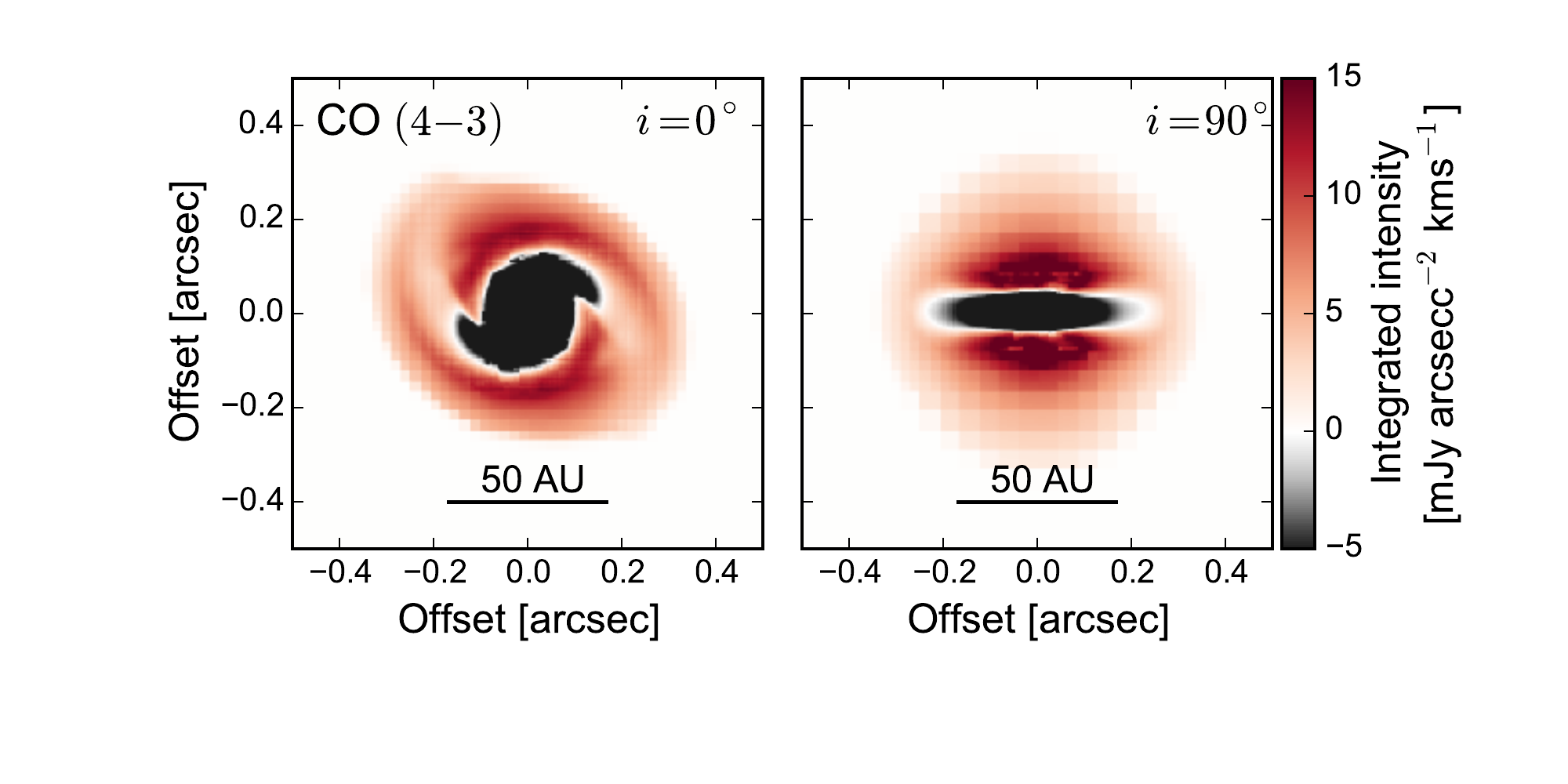}\\
\includegraphics[width=\columnwidth,trim= 1cm 1.5cm 1cm 0cm,clip]{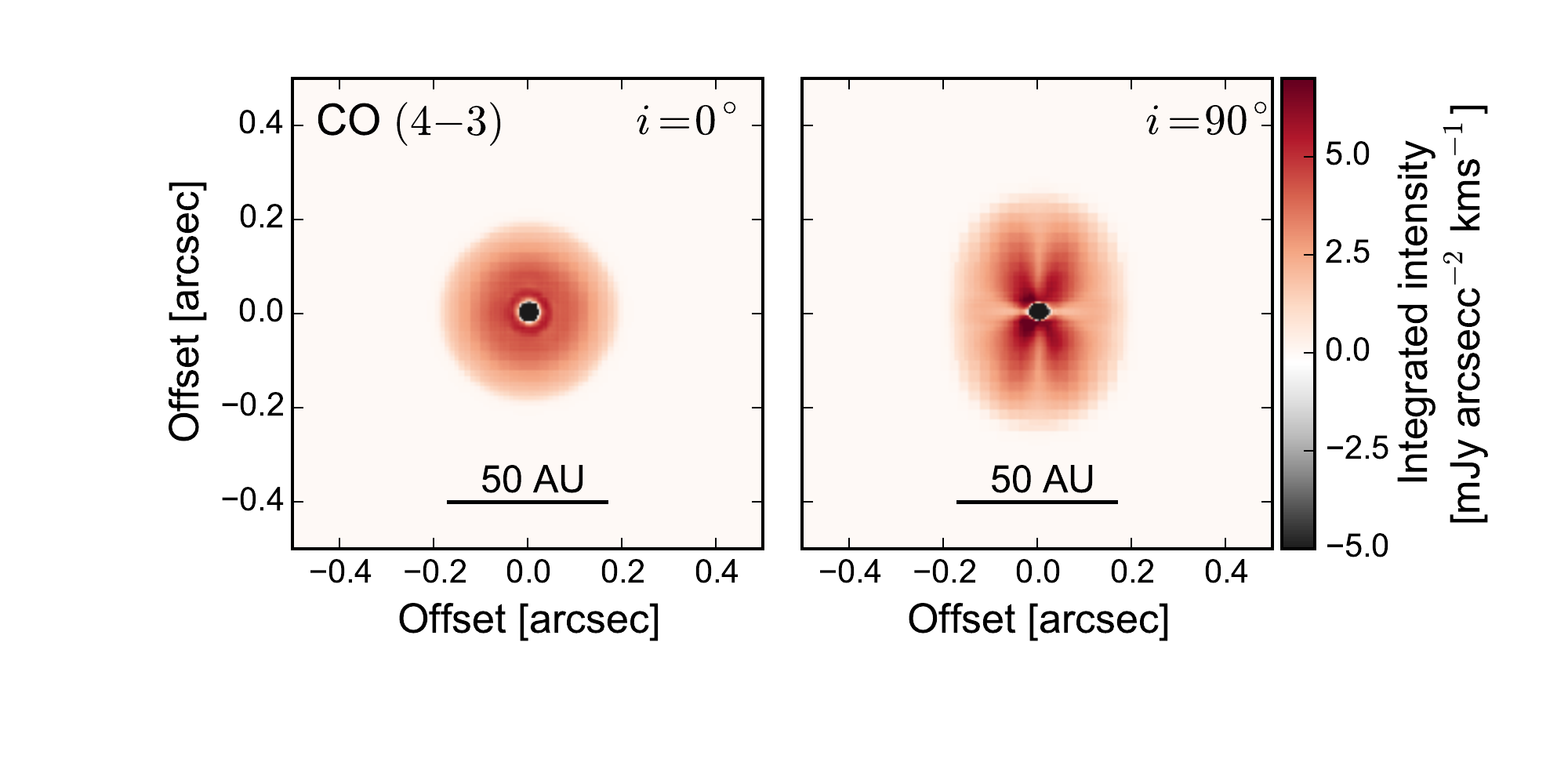}
\caption{ Top: Integrated intensity map for late FHSC in the RHD model, after subtracting continuum emission. Bottom: As above but for the MHD model.}
\label{fig:CO_mom0}
\end{figure}

\begin{figure}
\centering
\includegraphics[width=\columnwidth,trim= 0.5cm 3.2cm 1cm 1cm,clip]{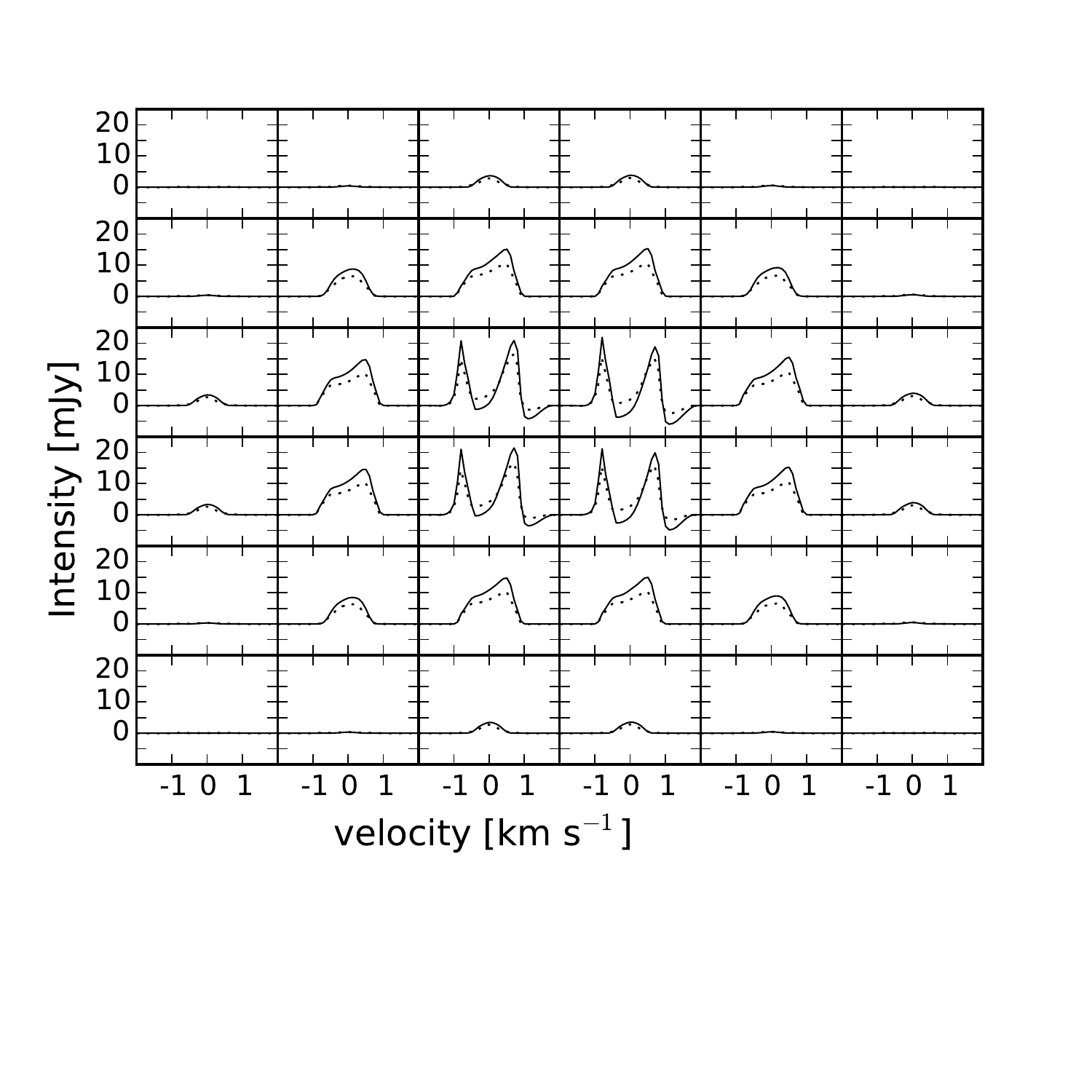}
\caption{Spectra for CO~$(4-3)$ (solid lines) and CO~$(3-2)$ (dotted lines) at $i=0^{\circ}$ (face-on) during late FHSC stage (d) of the MHD model. Each panel covers an area of $0.08 \times 0.08$~arcsec$^2$. The outflow has recently been launched and is apparent from the double peak.}
\label{fig:CO_line0deg}
\end{figure}
%
%
\begin{figure*}
\centering
\includegraphics[width=15cm]{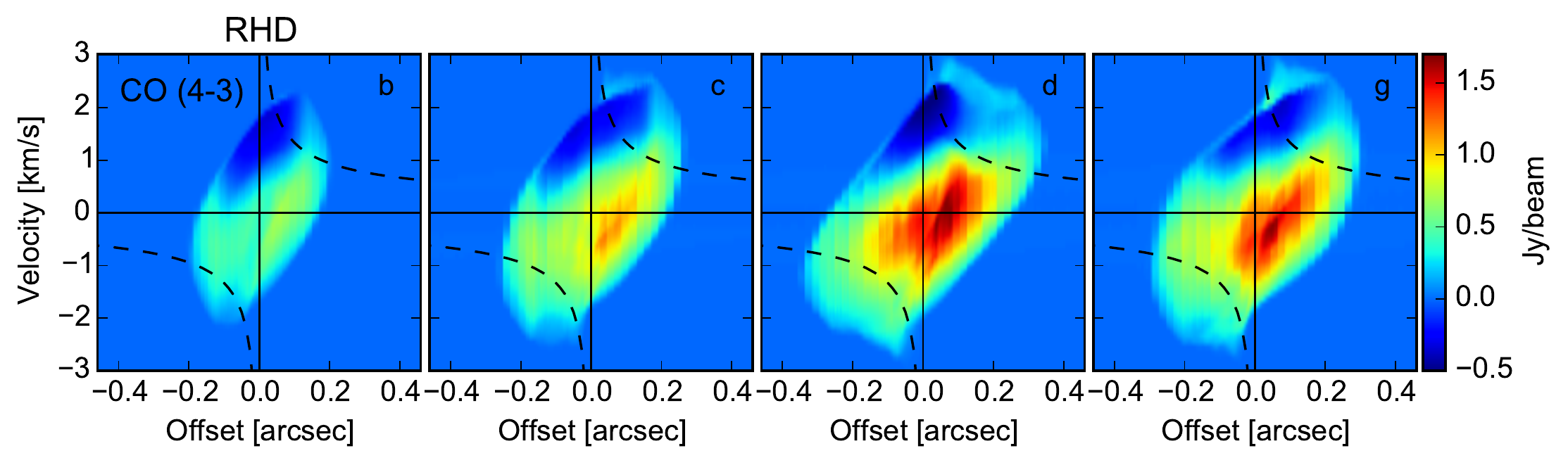}
\includegraphics[width=15cm]{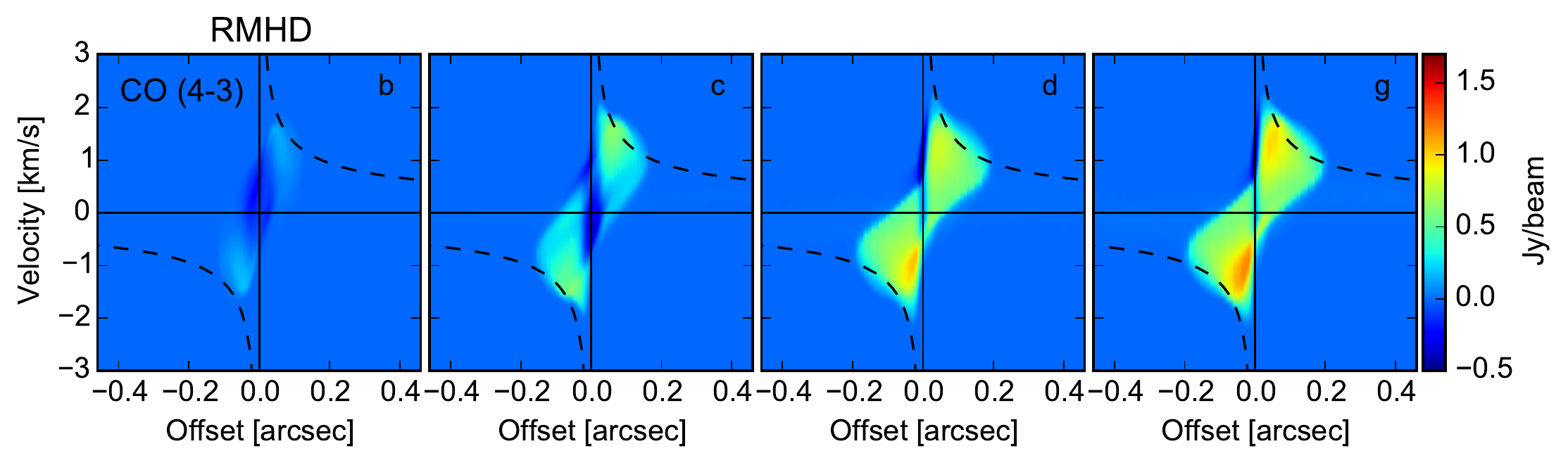}
\caption{Position--velocity diagrams for CO (4-3) for a series of snapshots of the FHSC (b, c, d) and newly formed stellar core (g) viewed  at $i=90^{\circ}$ inclination. The intensity was summed over an 0.3~arcsec slice centered on the plane of the disc. Dashed lines show the Keplerian rotation profile for a \SI{0.03}{\solarmass} central protostar.}
\label{fig:COpv}
\end{figure*}

\begin{figure}
\centering
\includegraphics[width=6cm]{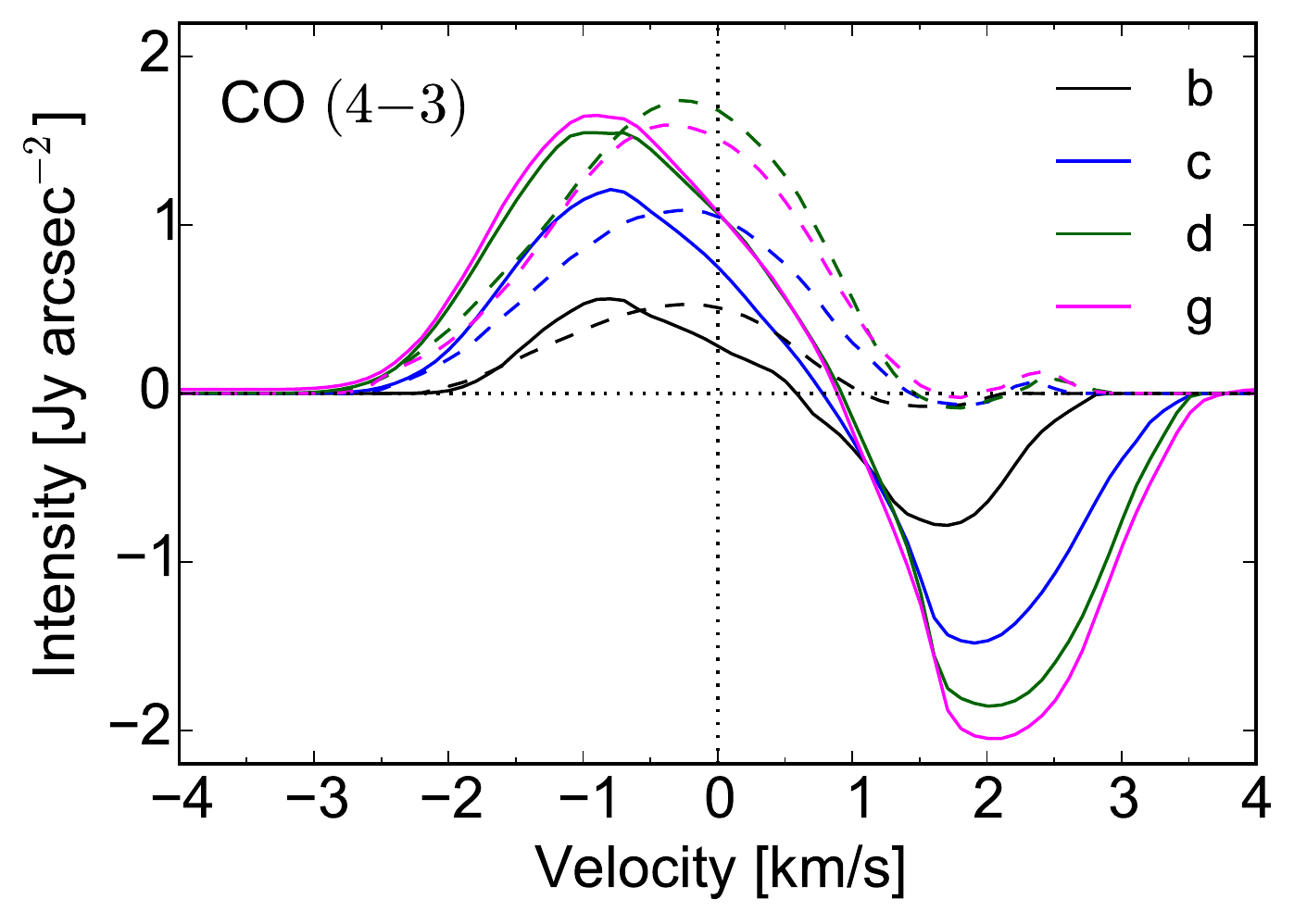}\\
\includegraphics[width=6cm]{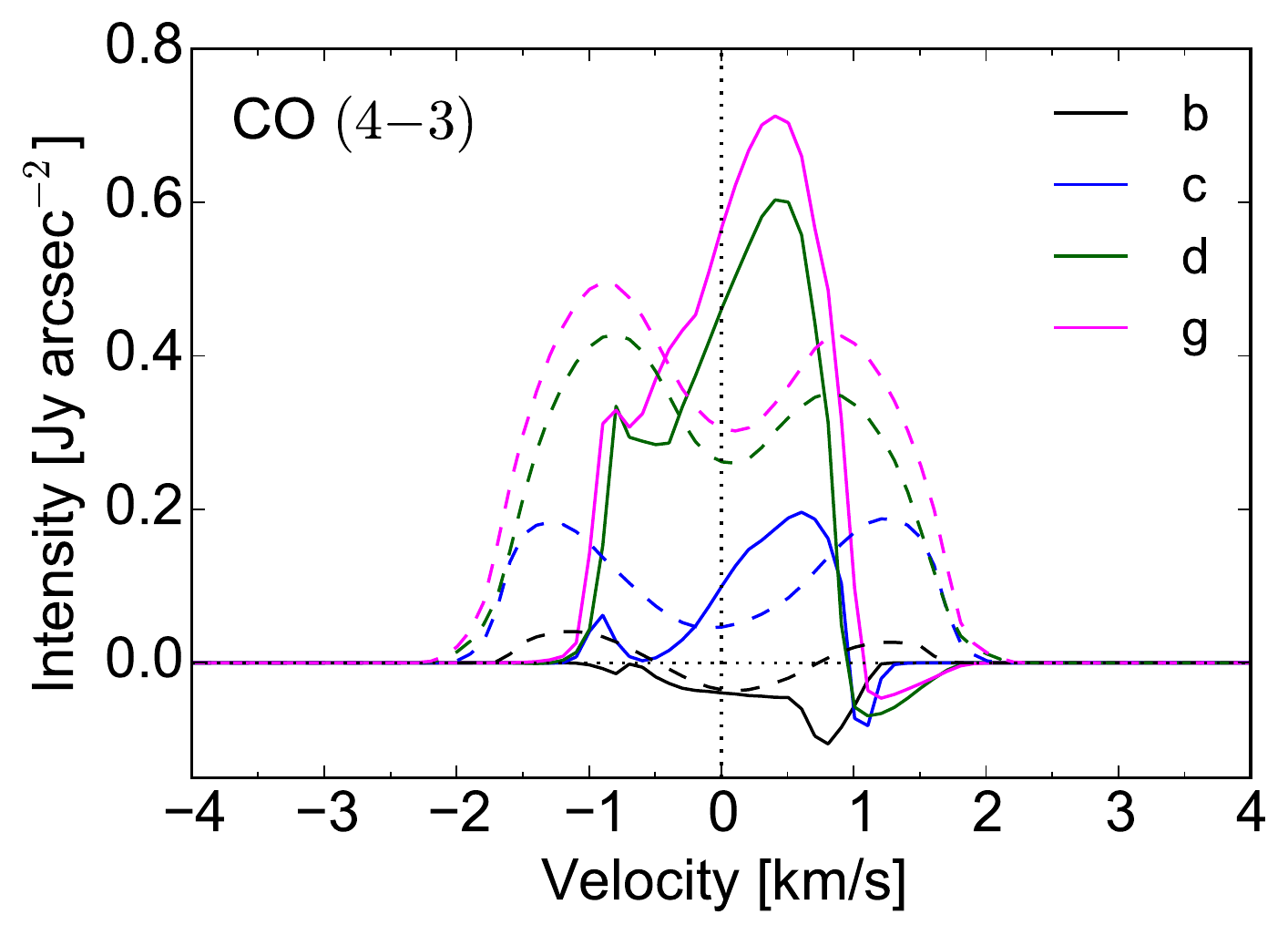}
\caption{Spectra averaged over a central 0.67 arcsec diameter aperture for RHD model (top) and MHD model (bottom). Solid lines: $i=0^{\circ}$, dashed lines: $i=90^{\circ}$. Letters refer to evolutionary stage as defined in Section~\ref{sec:morphology}.}
\label{fig:COspecs}
\end{figure}

From Fig.~\ref{fig:spiralabunds} it is clear that CO traces the region containing the spiral structures in the RHD model. The integrated intensity maps for a snapshot late in the FHSC stage for both the RHD and MHD models are shown in Fig.~\ref{fig:CO_mom0}. The CO~$(4-3)$ line appears to trace the rotational structures present in the RHD model in the face--on direction. The CO~$(4-3)$ emission reveals two spiral arms but the emission traces the lower density region rather than the spiral features seen in the column density and dust continuum emission (Fig.~\ref{fig:rhocont}).

There are significant absorption features in the integrated intensity maps where the FHSC continuum emission is absorbed by cooler gas in front of it. This effect is occasionally seen in observations of Class 0 protostars (e.g. \citealt{ohashi2014aa}). The FHSC is smaller in the MHD model so a smaller region is seen in absorption. The FHSC remains axisymmetric in the MHD model and the integrated intensity map shows a 50~AU diameter disc--like structure.

When viewed side--on (inclination $i=90^{\circ}$) the dense inner envelope is bright in emission for the RHD model. The outflow structure of the MHD model is apparent in the CO~$(4-3)$ emission. The CO abundance varies by less than an order of magnitude across the inner region which it traces, indicating that the outflow structure in Fig.~\ref{fig:CO_mom0}, bottom right, is not caused by an increased CO abundance in the outflow. The brighter structures are where the velocities are higher because emission at the higher velocities ($>1$~\si{\kilo\metre\per\second}) suffers much less from self--absorption.

The spectra of the late FHSC snapshot from the MHD model in Fig.~\ref{fig:CO_line0deg} reveal the kinematics that could be detected at $i=0^{\circ}$. The spectrum for each panel was calculated simply by summing the intensities for each pixel in the area of the image covered by the panel. The CO~$(4-3)$ spectra are brighter, more sharply peaked and have deeper absorption features than the CO~$(3-2)$ spectra. The line profiles for the snapshot just after stellar core formation are very similar, only slightly brighter.

At $i=0^{\circ}$ we are looking down the direction of the outflow. The double--peaked spectra in the four central panels are indicative of the blue-- and redshifted components with velocities of $\sim$~\SI{1}{\kilo\metre\per\second}. Self--absorption at low velocities also contributes to the central dip. There is a second absorption feature at $\sim$~\SI{1}{\kilo\metre\per\second} which is caused by infalling gas directly in front of the FHSC (see Fig.~\ref{fig:CO_mom0}, lower right). Outside of the central four panels, the spectra peak at $\sim$~\SI{+0.7}{\kilo\metre\per\second}. The redshifted emission from the far side outflow is brighter than the emission from the approaching part of the outflow, which causes this asymmetry in the spectra. The nearside emission is self--absorbed because the gas in the outflow, through which the emission propagates, is moving at similar velocities to the emitting gas and this is not the case for the far side emission. The effect of the near side infalling gas on the redshifted outflow CO emission is negligible because the CO abundance is very low in the envelope.

To examine the observable kinematics further we plot the position--velocity cuts for three snapshots during FHSC stages (b,c,d) and shortly after stellar core formation (g) in Fig.~\ref{fig:COpv} for both the RHD and MHD models. These were constructed from the $i=90^{\circ}$ velocity cubes as described in Section~\ref{sec:processing}.

The RHD PV diagrams show the characteristics of a rotating infalling structure. The broad absorption feature at $\sim 1 -2$~\si{\kilo\meter\per\second} indicates infalling gas and the velocity increases towards positive offsets because of the contribution of rotation. CO is frozen out just outwards of the spiral shocks at $r\gtrsim 40$~AU. The extent of the emission increases as the temperature increases and CO desorbs at increasing radii.
There is blueshifted emission to the right hand, receding, side of the object because the gas on the far side has a component in the direction of the observer as it spirals towards the disc. This emission is not self--absorbed due to the velocity difference between the emitting gas and the gas in the disc and near--side envelope through which it passes.

The PV diagram for the MHD model appears to be closer to a Keplerian rotation profile, although this is misleading because there is no rotationally supported disc and gas is spiralling inward through the pseudo--disc. CO is frozen out at $r\gtrsim 20$~AU so there is no emission beyond $\pm0.2$~arcsec.

Earlier in the FHSC stage (snapshots b and c) absorption occurs at negative velocities. For snapshot (c), this is likely to be due to absorption in the young outflow. By snapshot (d) absorption occurs only at $v\gtrsim$~\SI{0.4}{\kilo\metre\per\second} and is restricted to a very narrow region of $< 0.04$~arcsec. The nature of the CO~$(4-3)$ absorption feature could provide an indication of the evolutionary stage.

From the evolution of the spectra averaged within a 0.35~arcsec circle shown in Fig.~\ref{fig:COspecs}, the difference between the RHD and MHD models is even more apparent. The disc in the RHD model is seen almost entirely in absorption against the continuum. Early in the FHSC stage (b) the absorption and emission features are similar in magnitude in the RHD model. The brightness of the emission changes little through to when the stellar core forms but the depth of absorption increases by up to a factor of 3. There is no disc in the MHD model and much of the emission is from the outflow. At (b) the line is entirely in absorption at $i=0^{\circ}$. The development of the outflow leads to the emission brightening as gas--phase CO is present in high abundance in a more extended region.

At $i=90^{\circ}$ a double--peaked rotation signature is not seen in the RHD model due to the strong absorption at positive velocities by infalling material. The rotating outflow produces a double--peaked spectrum in the MHD model, however, which may be mistaken for a rotating disc (Fig.~\ref{fig:COspecs}, bottom panel, dashed lines).

{\referee Our models assume a turbulent velocity of \SI{0.1}{\kilo\metre\per\second} and we note that a higher value will result in less self--absorption at the systematic velocity of the FHSC as well as slightly broader line widths.}

\subsubsection{SO}

\begin{figure}
\centering
\includegraphics[width=\columnwidth,trim= 1cm 1.5cm 1cm 0cm,clip]{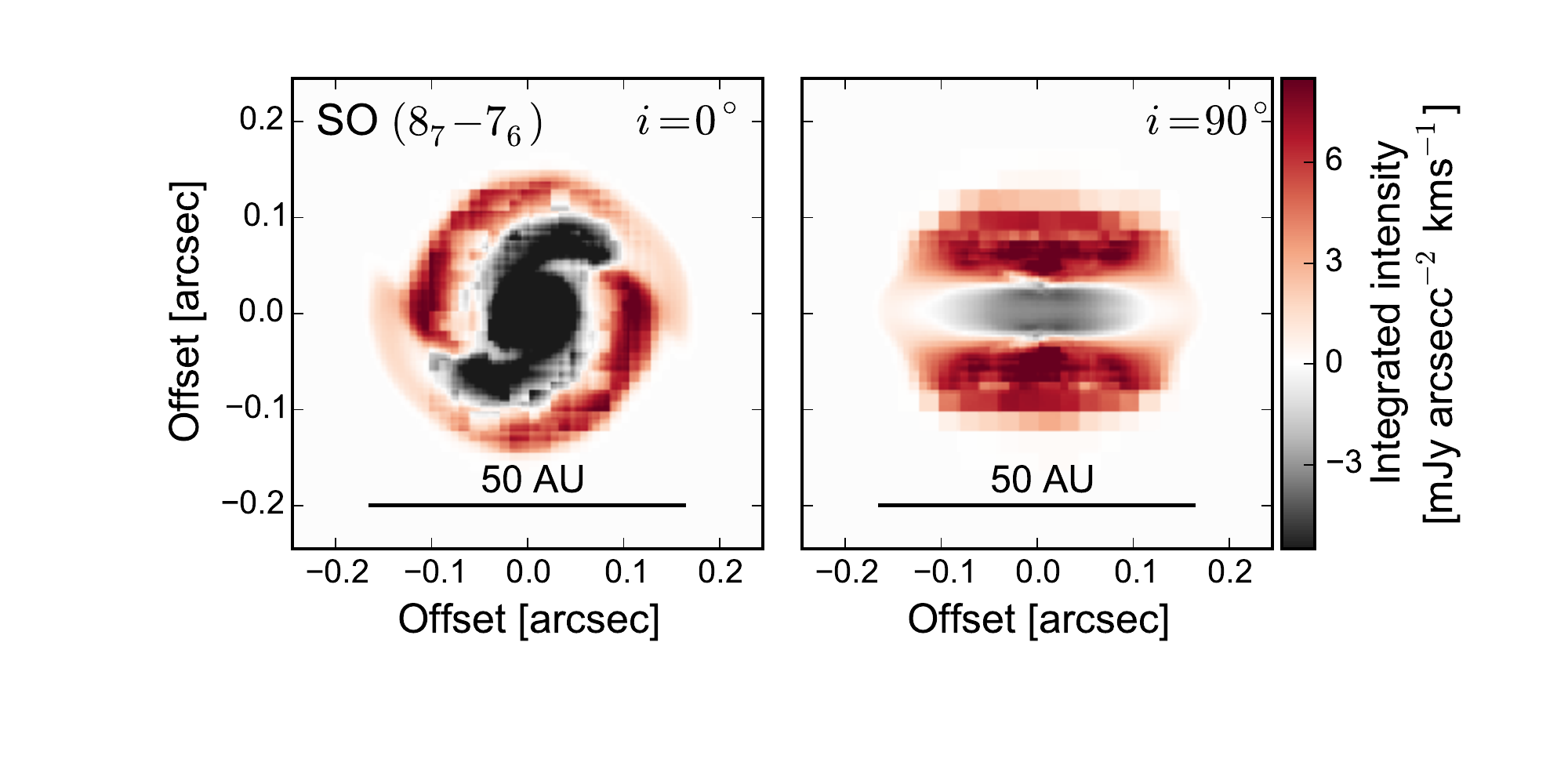}
\caption{SO integrated intensity maps for the RHD model. We do not show the MHD model because the scale of the emission would be too small to resolve.} 
\label{fig:SO_mom0}
\end{figure}

\begin{figure}
\centering
\includegraphics[width=6cm]{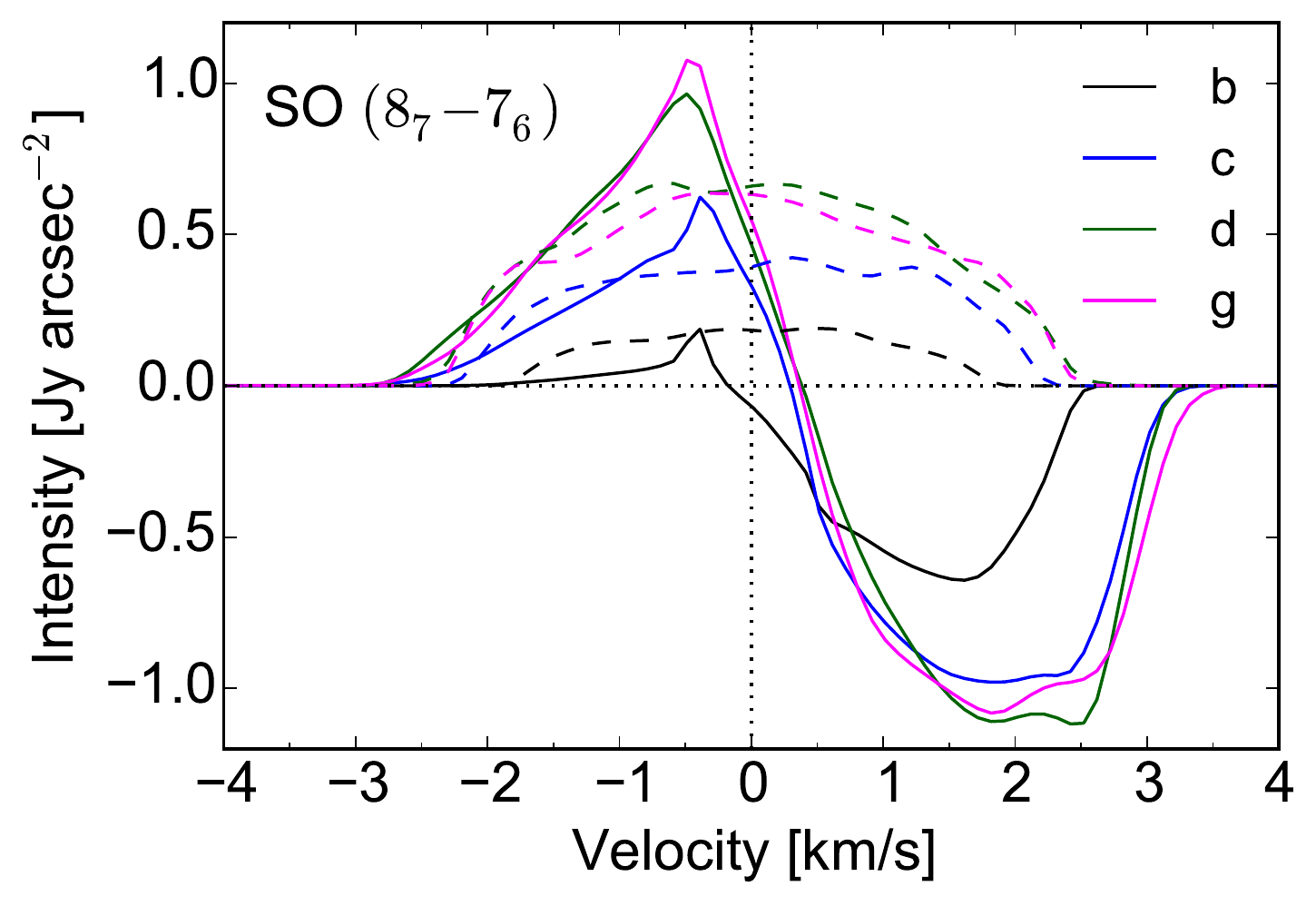}\\
\includegraphics[width=6cm]{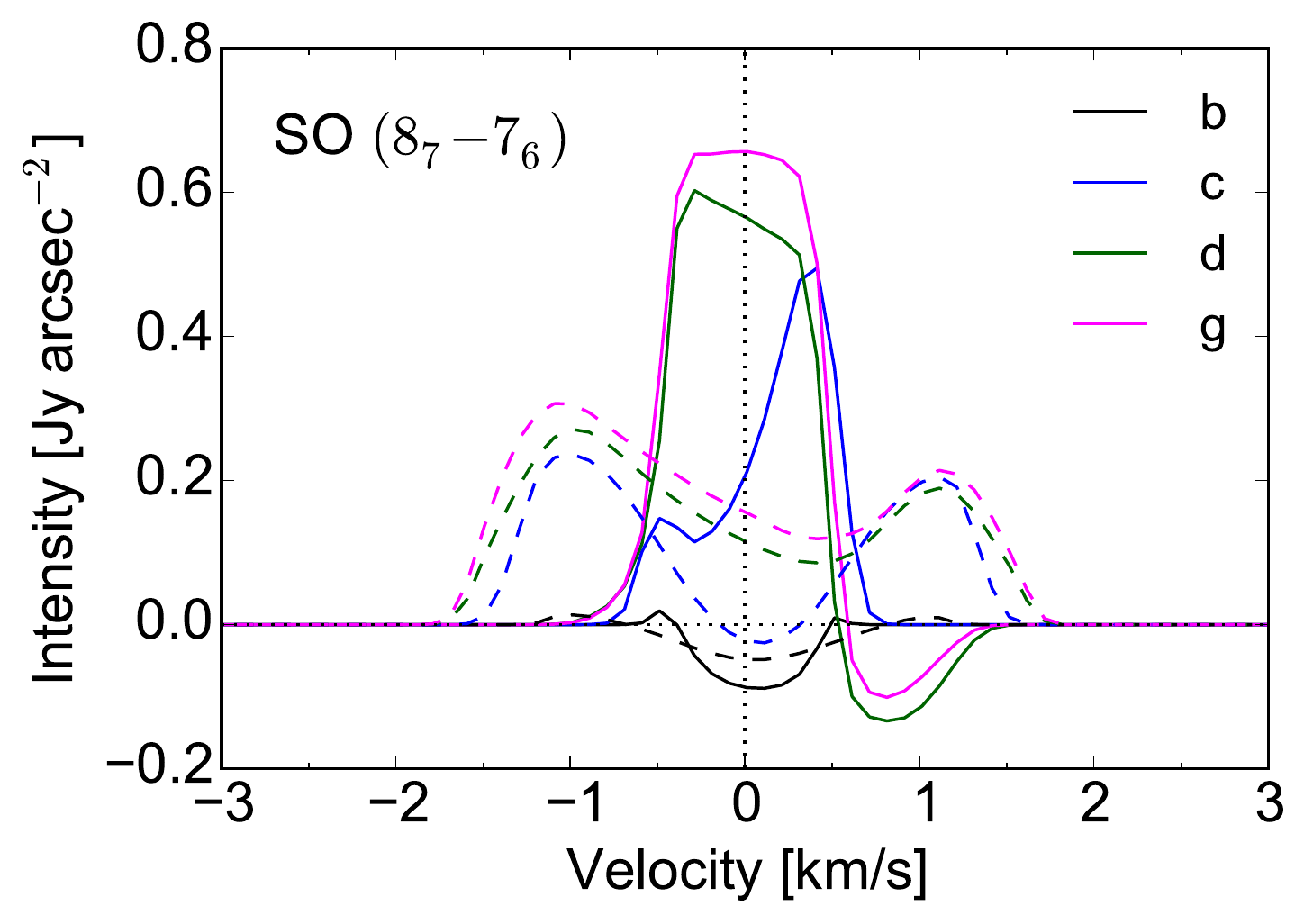}
\caption{Evolution of the SO $(8_7 - 7_6)$ line profile in the MHD model for RHD model (top) and MHD model (bottom). Each spectrum is averaged over pixels within a central 0.35~arcsec (53~AU) diameter circular beam. Solid lines: $i=0^{\circ}$, dashed lines: $i=90^{\circ}$.}
\label{fig:SOspec}
\end{figure} 

The abundance of SO is $\sim$~\num{1e-9} within $r<20$~AU in the RHD model and within $r<10$~AU in the MHD model and drops off rapidly outside this region, meaning it is a tracer of the FHSC only in the RHD model (Fig.~\ref{fig:spiralabunds}) and of the FHSC and very inner envelope in the MHD model (Fig.~\ref{fig:MuChem}). At a distance of 150~pc, the region traced in the MHD model is only 0.13~arcsec across so the object will not be well resolved. The abundance of SO is lower in the outer parts of the spiral arms of the RHD model, although SO traces the inner spiral well (Fig.~\ref{fig:spiralabunds}). In Fig.~\ref{fig:SO_mom0}, the integrated intensity plots reveal significant absorption as seen in CO, which actually traces out the central spiral structure. Emission is limited to a pair of spiral arms at $r\sim20$~AU. At $i=90^{\circ}$, we see a "hamburger" structure with emission from above and below the midplane.

SO~($8_7 - 7_6$) spectra for the MHD model are shown in Fig.~\ref{fig:SOspec}. The line brightens considerably during the early FHSC stage as the temperature increases quickly in the central few AU and SO traces an increasing radius. The region traced by SO incorporates only a small section of the outflow where the outflow velocities are $<$~\SI{0.5}{\kilo\metre\per\second}. In the same region, the infall and rotation velocities are $\sim$~\SI{1}{\kilo\metre\per\second}.

The spectra show a clear evolution during FHSC stage. At snapshot (b) the emission is undetectable and the line is symmetrical in absorption. By snapshot (c) the outflow has launched and the line is asymmetrical since it is affected by self--absorption by outflowing gas, giving rise to a redshifted peak. From snapshot (d) there is an absorption feature against the continuum. This becomes less pronounced after the stellar core formation because the density distribution becomes more sharply peaked. The peak becomes more symmetrical as emission from the outflow brightens.

At $i=90^{\circ}$, the emission peaks at $\pm$~\SI{1.2}{\kilo\metre\per\second} and the blueshifted peak is brighter, as expected for an infalling, rotating structure. The central absorption dip is skewed to positive velocities due to absorption by infalling material (with a positive recession velocity). The infall velocity is lower here than in the vertical direction due to the added effect of rotation.

Without an outflow, the spectra for the RHD model are quite different. There is a very strong absorption feature from 1.5 to 2.5~\si{\kilo\metre\per\second}. After stage (c) this does not deepen although the emission peak brightens between snapshots (c) and (d). The $i=90^{\circ}$ spectra do not display the characteristic double peak. Looking at the integrated intensity images in Fig.~\ref{fig:SO_mom0} we can see that the disc is seen nearly entirely in absorption. Rotation speeds are low above and below the disc where the emission originates and emission at higher velocities is less bright than the low velocity emission from nearer the centre.

\subsubsection{CS}
\begin{figure}
\centering	
\includegraphics[width=6cm]{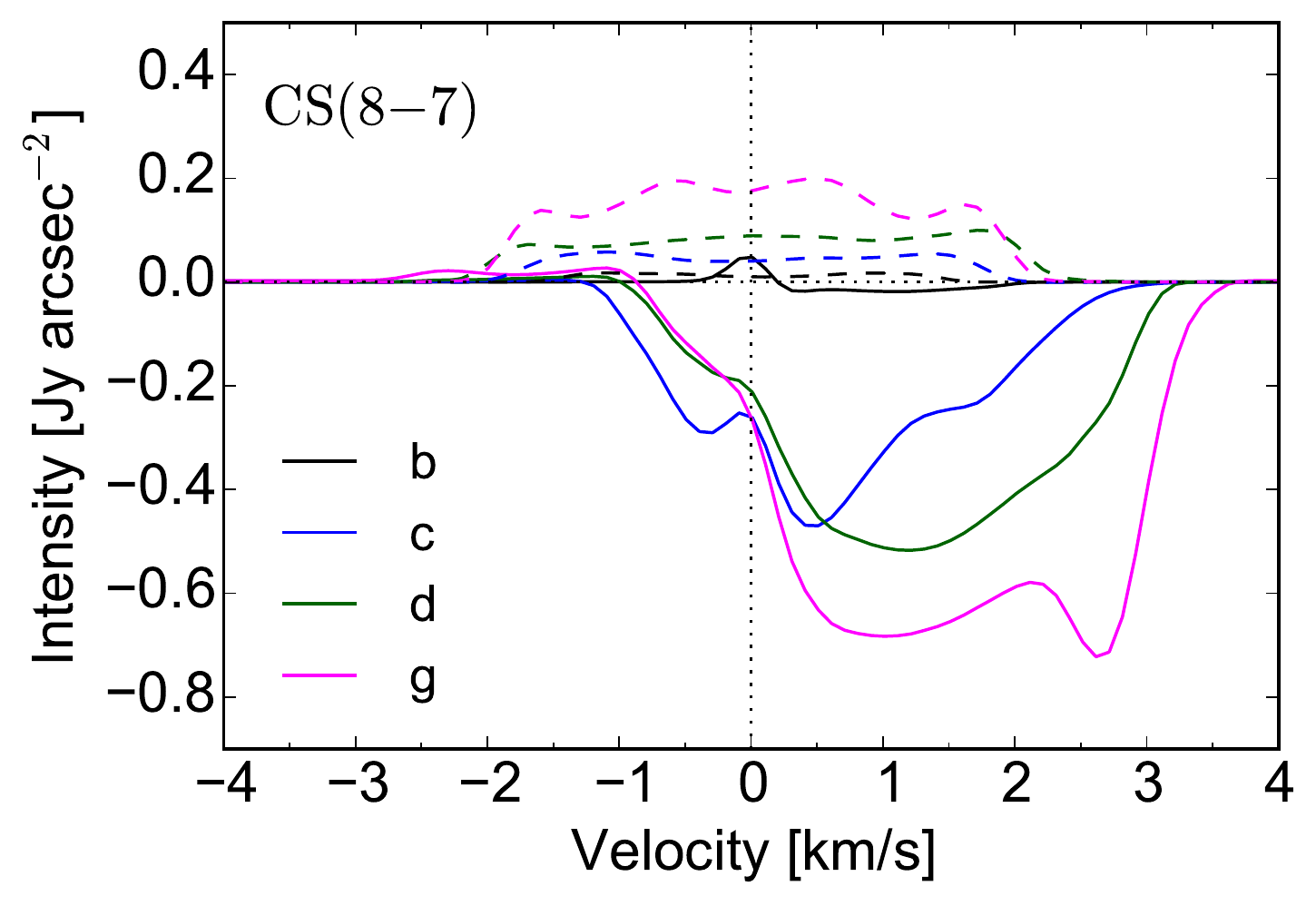}\\
\includegraphics[width=6cm]{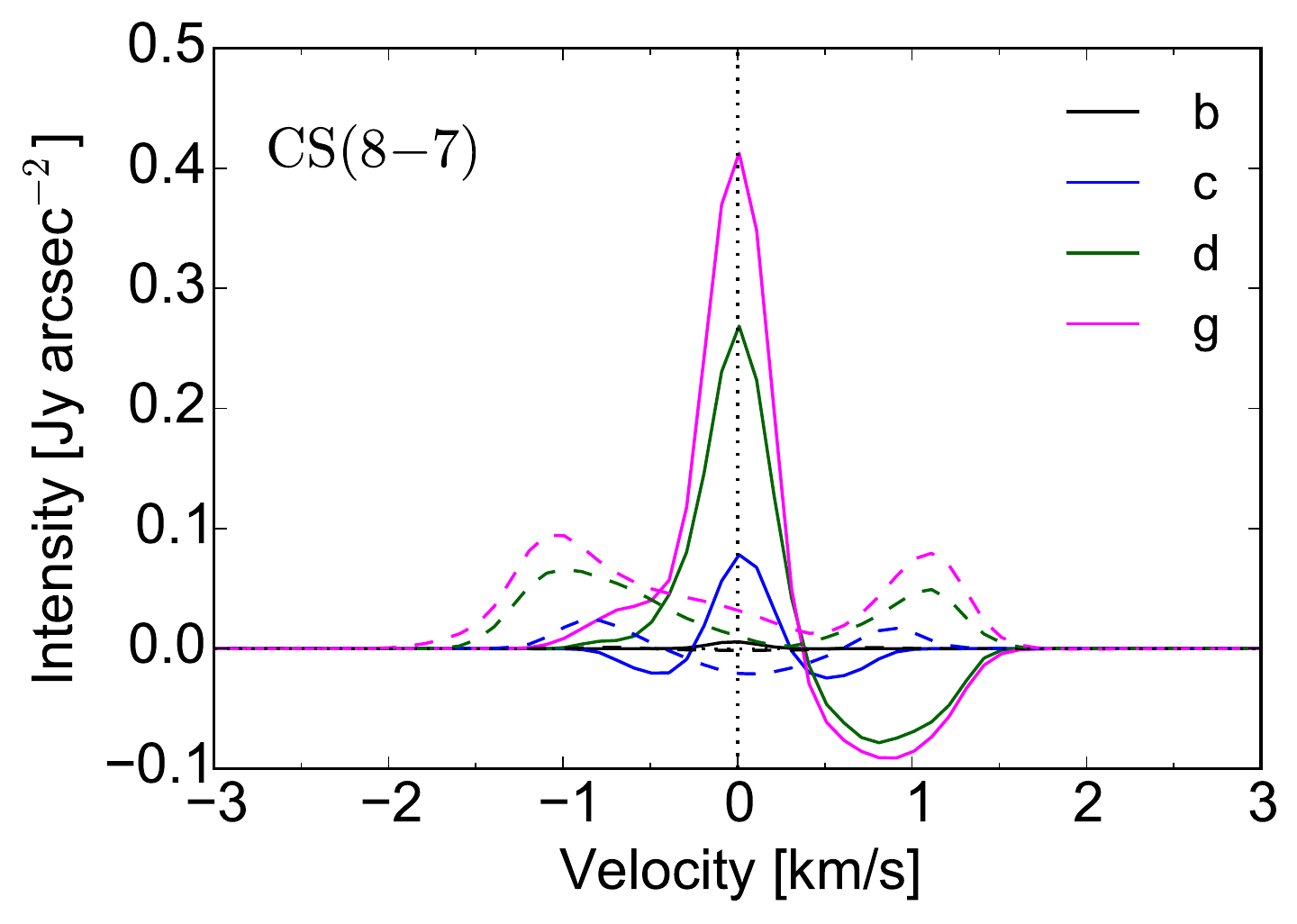}
\caption{Spectra for CS~$(8-7)$ obtained from a 0.35~arcsec aperture from the RHD model (top) and MHD model (bottom). Solid lines $i=0^{\circ}$; dashed lines $i=90^{\circ}$.}
\label{fig:CS_8_line}
\end{figure}

The average CS abundance for the late FHSC of the RHD model is presented in the final panel of Fig.~\ref{fig:spiralabunds}. The abundance peaks at a few \num{e-8} within $r<10$~AU in a symmetrical distribution and is several orders of magnitude lower elsewhere. We therefore do not expect CS to be useful for tracing rotational structures. The emission will not be well resolved so we consider only the spectra obtained from the total emission within a 0.35~arcsec aperture centred on the FHSC.

Fig.~\ref{fig:CS_8_line} shows the spectra for CS~$(8-7)$. CS is extremely depleted towards the centre of the cloud core prior to FHSC formation and the abundance soon after formation is still too low to be detectable. 

In the MHD model, at $i=0^{\circ}$ from snapshot (c) onwards, a sharp peak develops around the CS~$(8-7)$ line centre and this brightens considerably between the FHSC and stellar core stages. There is an absorption feature at $\sim$~\SI{+1}{\kilo\metre\per\second}. At the frequency of the CS~$(8-7)$ transition (\SI{391.8}{\giga\hertz}), the envelope is optically thin and most of the continuum emission originates within the FHSC or from the very inner envelope near to the FHSC. The optically thin gas directly in front of the FHSC is infalling with a velocity of $\sim$~\SI{+1}{\kilo\metre\per\second} and absorbs the continuum emission. The peaks of the spectra are symmetrical and show no evidence of the outflow. The CS~$(8-7)$ line brightens between during and beyond the FHSC stage.
%

At stage (c), mid--FHSC stage, we see symmetrical absorption features around the central peak. The outflow has launched by then with a velocity $<$~\SI{1}{\kilo\metre\per\second}. The gas is still relatively dense and therefore absorbs radiation from the FHSC, but does so predominantly at blueshifted frequencies since this gas is approaching the observer. Within $\lesssim$~\SI{5}{\au} from the centre, the gas is infalling at a similar speed to the outflow, which creates the redshifted absorption feature.

At $i=90^{\circ}$, the CS~$(8-7)$ line is double--peaked because of the rotation of the FHSC. The emission is fainter due to the increased optical depth caused by the disc structure such that the emission comes from a slightly larger radius where the temperatures are a little lower. The central dip is deepened by absorption and is skewed to positive velocities due to infall as discussed above.

Spectra from the RHD model show only absorption at $i=0^{\circ}$ and little asymmetry at $i=90^{\circ}$. Absorption at increasing velocities as the FHSC evolves indicates the increasing infall velocities. In the ideal MHD model, the additional support from the magnetic field reduces infall speeds so the spectra reveal a narrower velocity range.
\subsubsection{HCO$^+$}

\begin{figure}
\centering
\includegraphics[width=\columnwidth,trim= 1cm 1.5cm 0cm 0cm,clip]{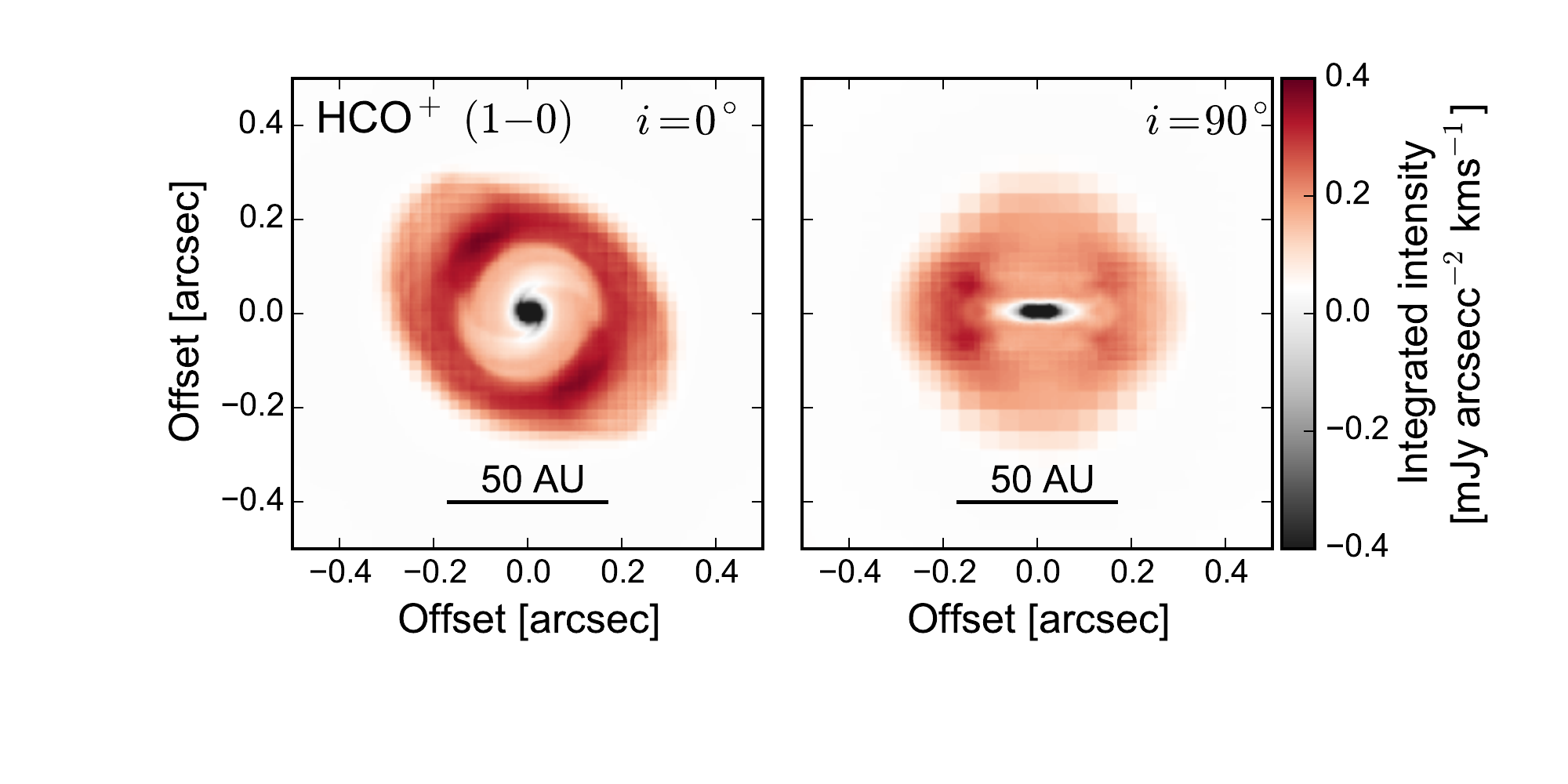}\\
\includegraphics[width=\columnwidth,trim= 1cm 1.5cm 0cm 0cm,clip]{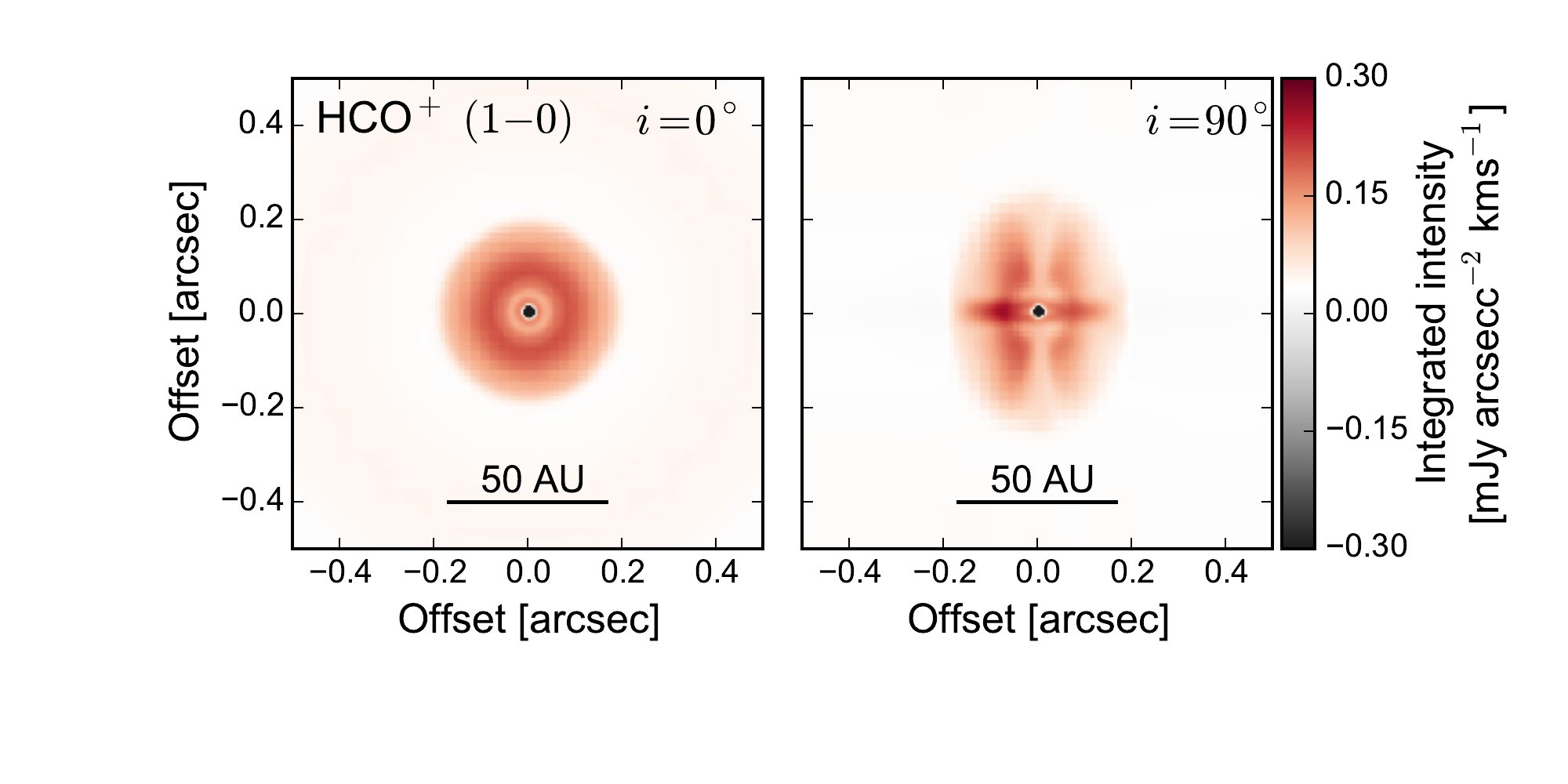}
\caption{HCO$^+$~($1-0$) Integrated intensity for snapshot (d). Middle: RHD model; Bottom: MHD model}
\label{fig:HCOplusmom0}
\end{figure}

\begin{figure}
\centering
\includegraphics[width=6cm]{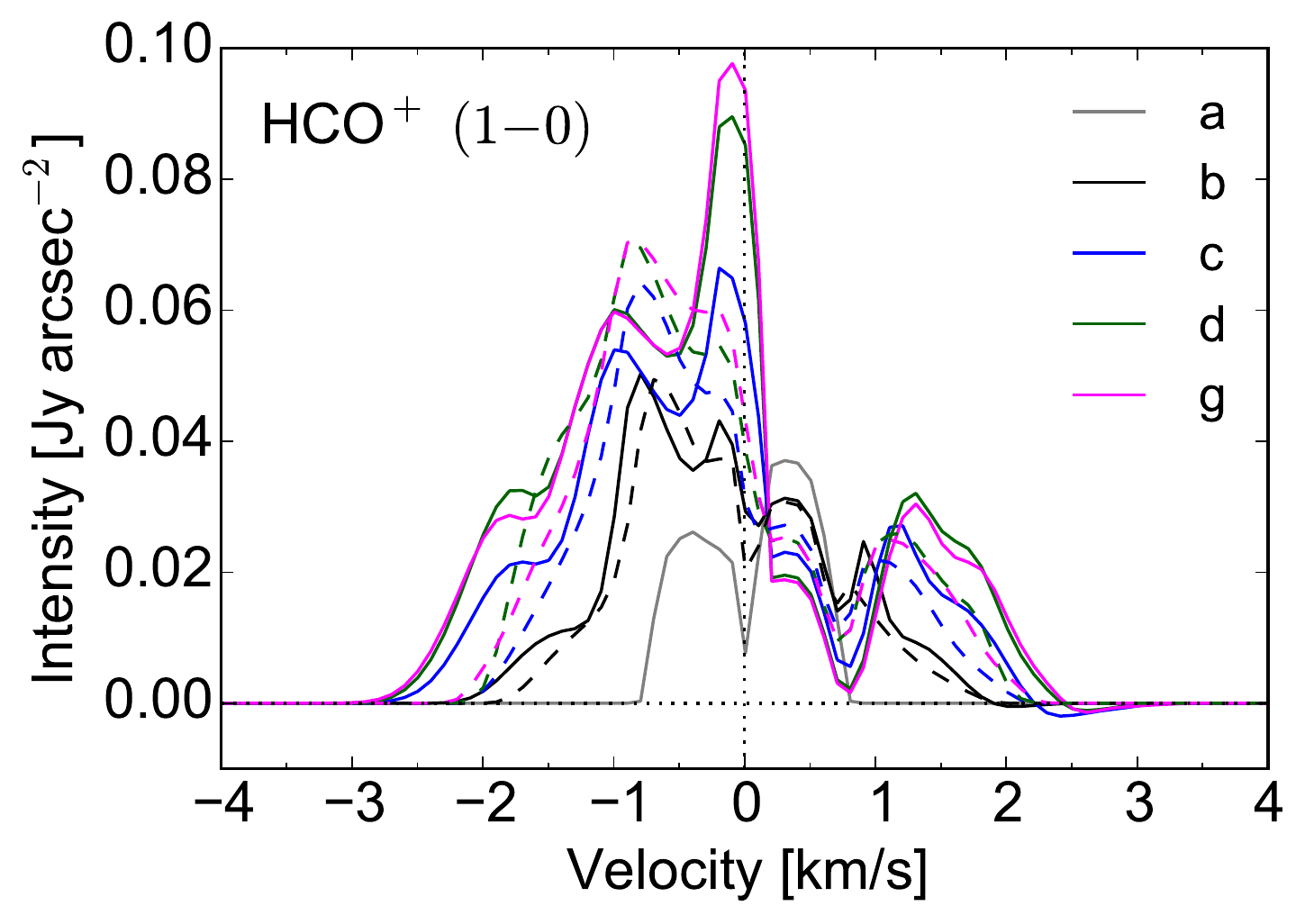}\\
\includegraphics[width=6cm]{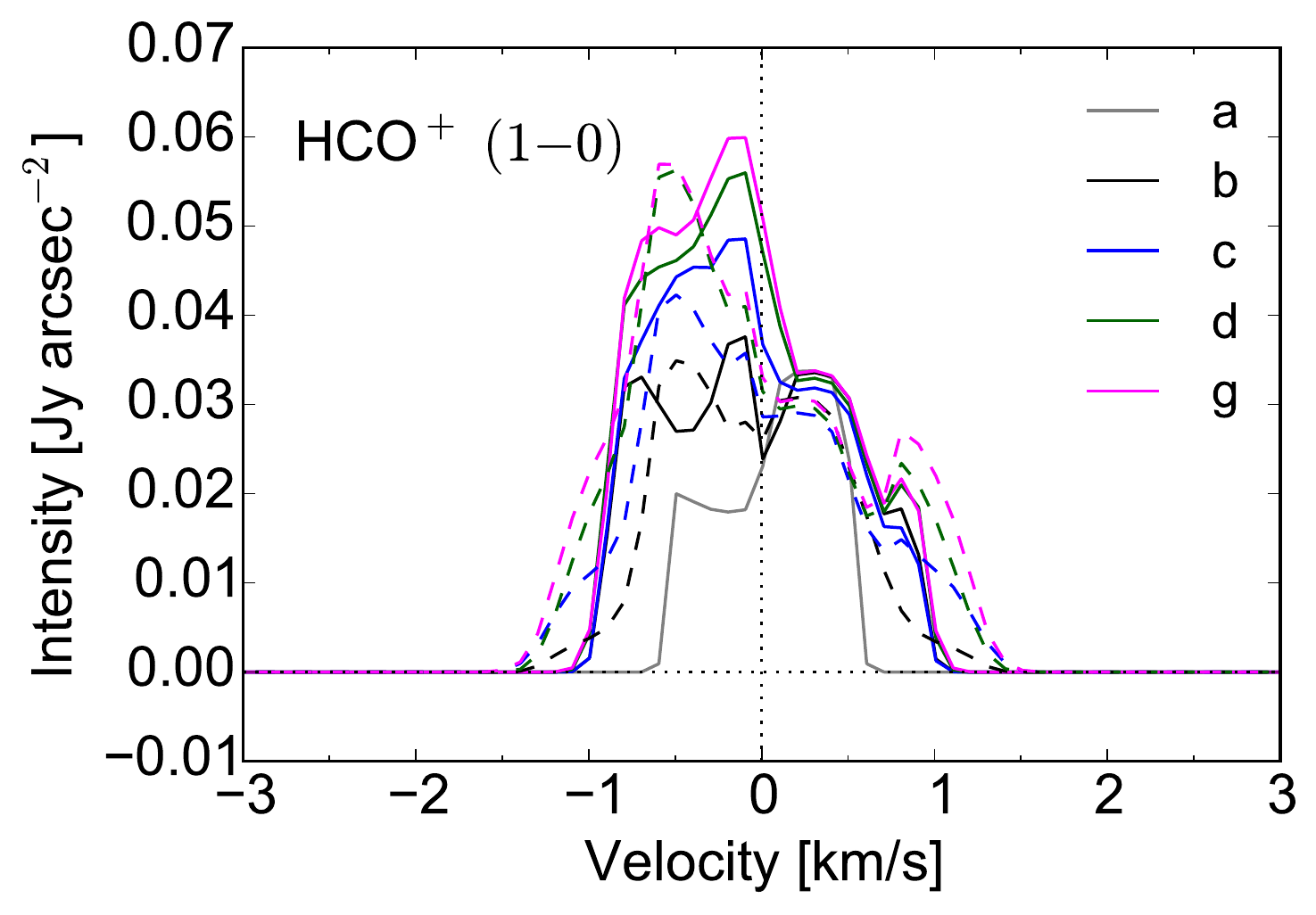}
\caption{HCO$^+$~($1-0$) spectra for RHD model (top) and MHD model (bottom) within a 0.67~arcsec aperture. N.B. These use a grid of 5350~AU rather than the 400~AU grid used for the other species.}
\label{fig:HCOplusSpectra}
\end{figure}

The abundance of HCO$^+$ is highest in the regions of the lowest density and it is depleted in the FHSC and disc (Fig.~\ref{fig:spiralabunds}). The integrated intensity map for the RHD model (Fig.~\ref{fig:HCOplusmom0}, top left) shows an asymmetrical ring with a central hole. This feature is commonly observed in HCO$^+$ near protostellar sources, albeit on scales of 100s AU rather than 10s AU (e.g. \citealt{jorgensen2013aa}). {\referee Only the inner envelope at $r\sim30$--$50$~AU is warm enough to produce significant HCO$^+$~$(1-0)$ emission and within the disc the HCO$^+$ emission is much fainter, primarily due to depletion at $r \lesssim 15$~AU (see Figs.~\ref{fig:B02abundh} - \ref{fig:MuChem})}. At $i=90^{\circ}$, the approaching (left) side is visibly brighter in both models. This is a well known phenomenon for rotating, infalling envelopes observed in an optically thin line.

The emission in the MHD model also has the appearance of a disc and a smaller region is traced. What we see is not just the pseudo--disc but also the conical outflow from above. At $i=90^{\circ}$, the emission is also less extended than in the RHD model, despite the presence of an outflow. Here the HCO$^+$ line traces the pseudo--disc and outflow and there are brighter structures in the outflow like those seen in the CO emission.

The spectra for both models are double--peaked at $i=0^{\circ}$. For the RHD model this is due to the combination of infall motions and the self--absorption dip at low velocities. The spectra for the MHD model have similar features but the infall speeds are lower. The emission from the outflow at $i=0^{\circ}$ in the MHD model at $\sim$~\SI{-0.5}{\kilo\meter\per\second} reduces the depth of the dip between the central and blue shifted peaks from snapshot (c) onwards.

Before FHSC formation there is no central emission peak at the line centre. This central peak brightens as the core evolves and, from late in the FHSC stage, it is brighter than the blueshifted peak. This is true for both RHD and MHD models and provides a possible indication of the presence of an FHSC. The relative strengths of the line centre and red-- and blueshifted peaks do depend on the aperture over which the spectra are averaged. The HCO$^+$ abundance is extremely low in the centre so this emission at the line centre is not simply coming from the FHSC. In the RHD model, the HCO$^+$ abundance increases within a shell between $\sim$~20-30~AU (c.f. Figs.~\ref{fig:B02abundh} and \ref{fig:B02abundv}). This coincides with the outer regions of the disc--shaped FHSC, which strongly suggests that we are seeing the build up of material in the midplane. This feature is weaker in the MHD model (c.f. Figs.~\ref{fig:Mu5abundh} and \ref{fig:Mu5abundv}), which makes sense because gas is infalling along the midplane, i.e. the radial velocity~=~0 for $i=0$, but no disc forms so the density is lower. 

\begin{figure*}
\centering
\includegraphics[width=5.8cm,trim= 0cm 0cm 0cm 0cm,clip]{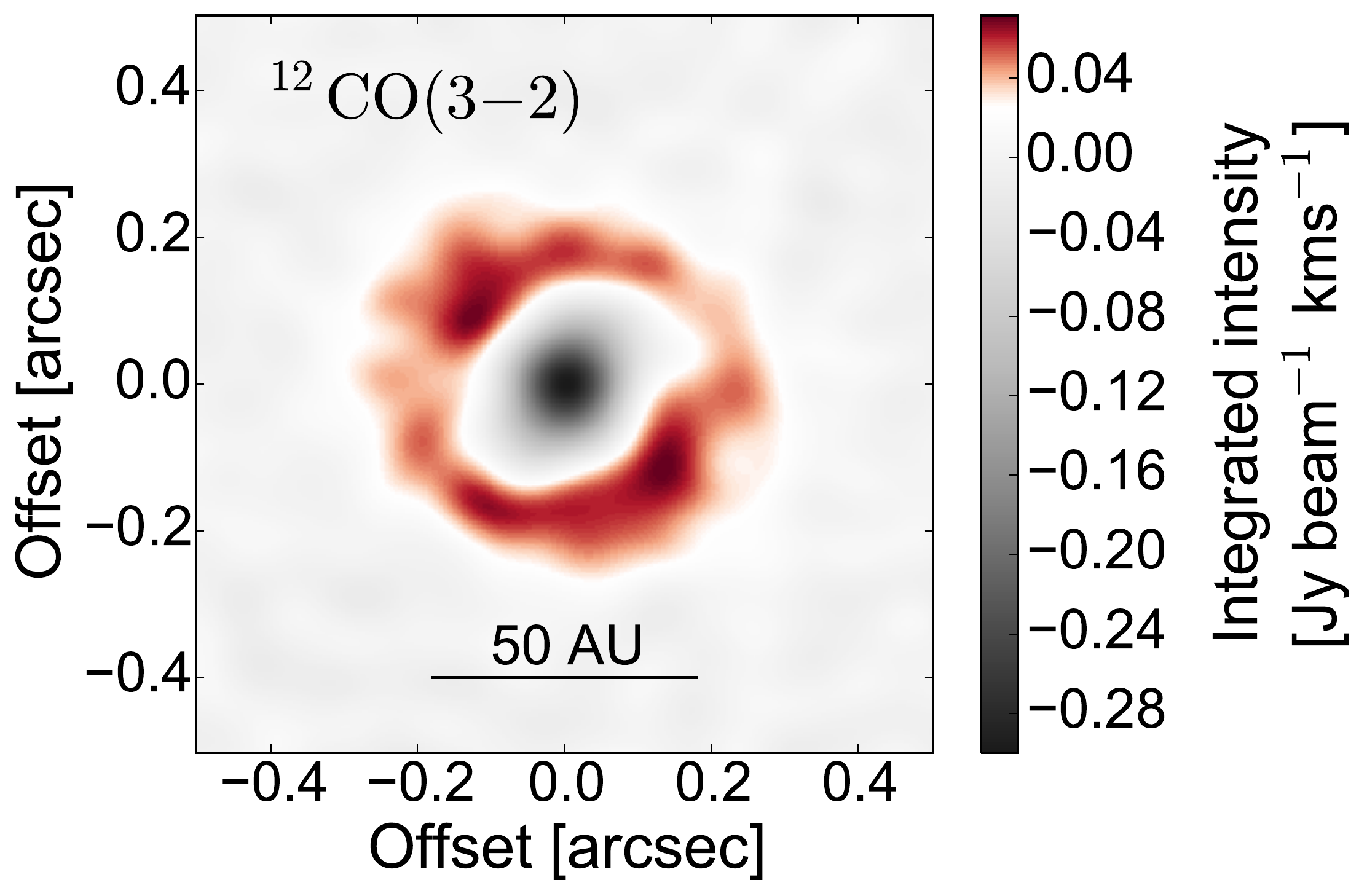}
\hfill
\includegraphics[width=5.8cm,trim= 0cm 0cm 0cm 0cm,clip]{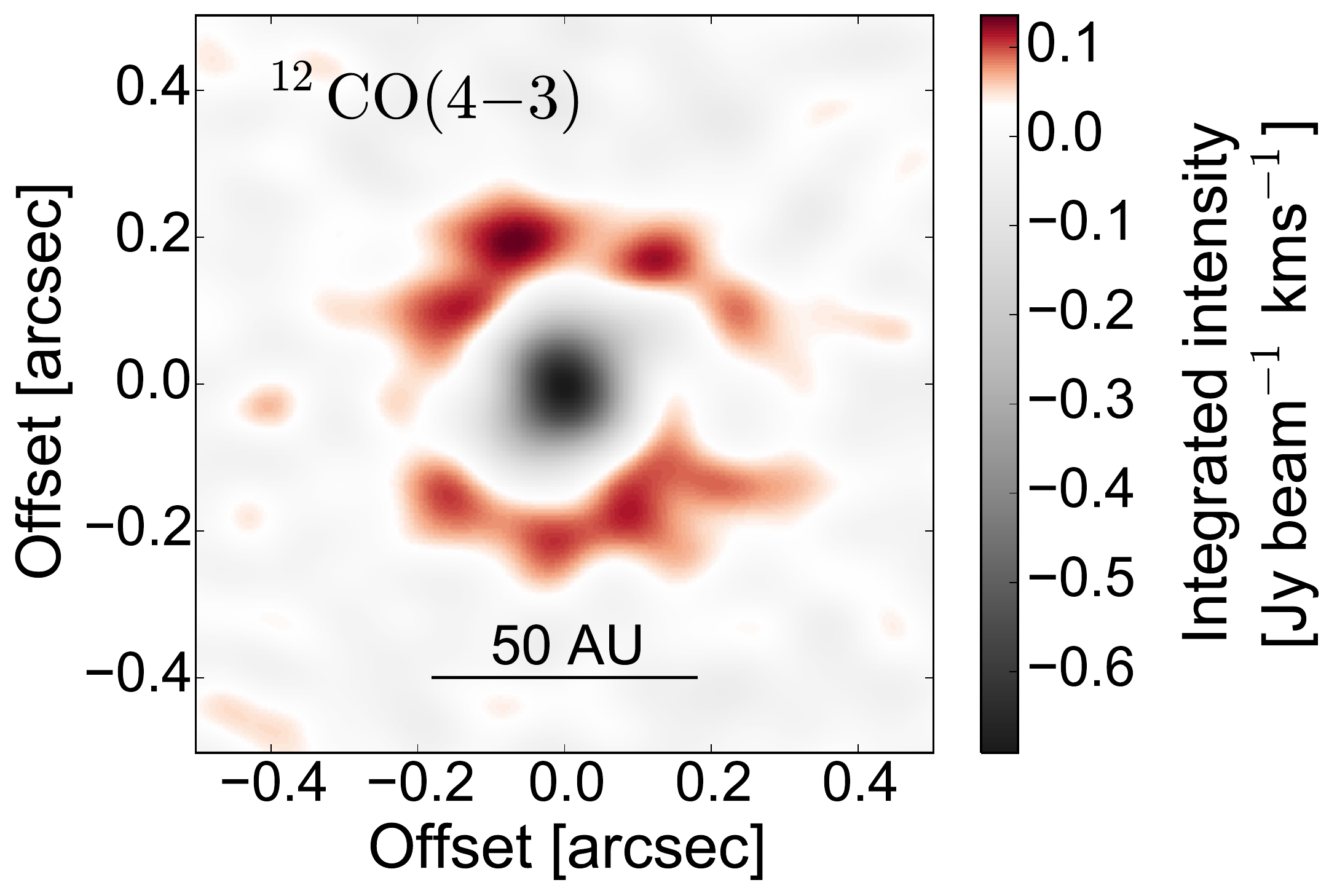}
\hfill
\includegraphics[width=5.8cm,trim= 0cm 0cm 0cm 0cm,clip]{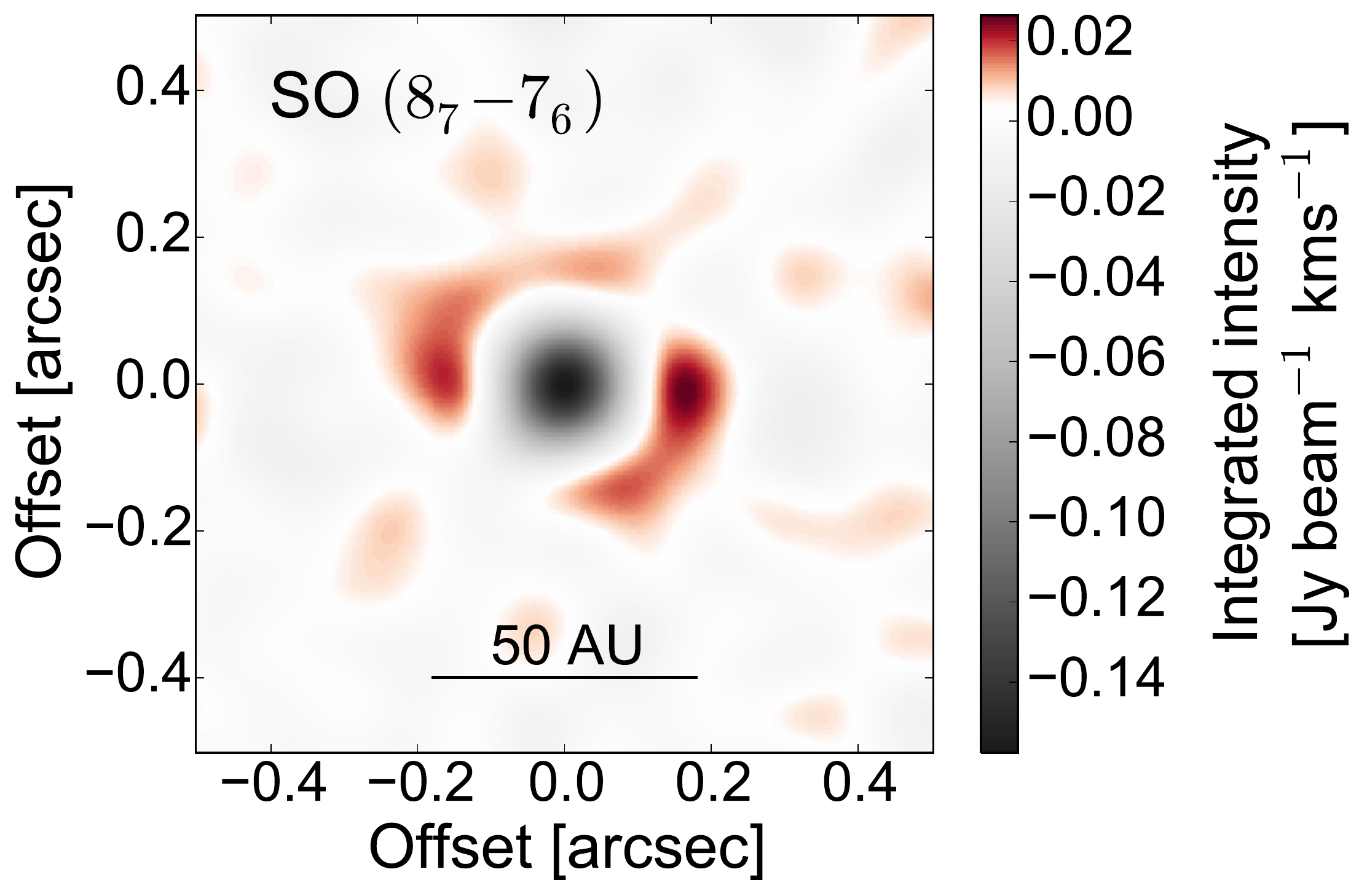}
\caption{Synthetic ALMA observations of snapshot (d) of the RHD model in the face--on direction all with pwv = 0.8, left to right: CO~$(3-2)$ 4h total time, ALMA out16 configuration giving $0.08 \times 0.1$~arcsec beam; CO~$(4-3)$, ALMA out14 configuration giving $0.11 \times 0.09$~arcsec beam, 8h total integration; SO~($8_7 - 7_6$), ALMA out14 configuration giving $0.15 \times 0.12$~arcsec beam, 8h total time.}
\label{fig:COalma}
\end{figure*}
\subsection{Detectability with ALMA}
\begin{figure}
\centering
\includegraphics[width=7cm,trim= -0cm 1.2cm -0.2cm 0cm,clip]{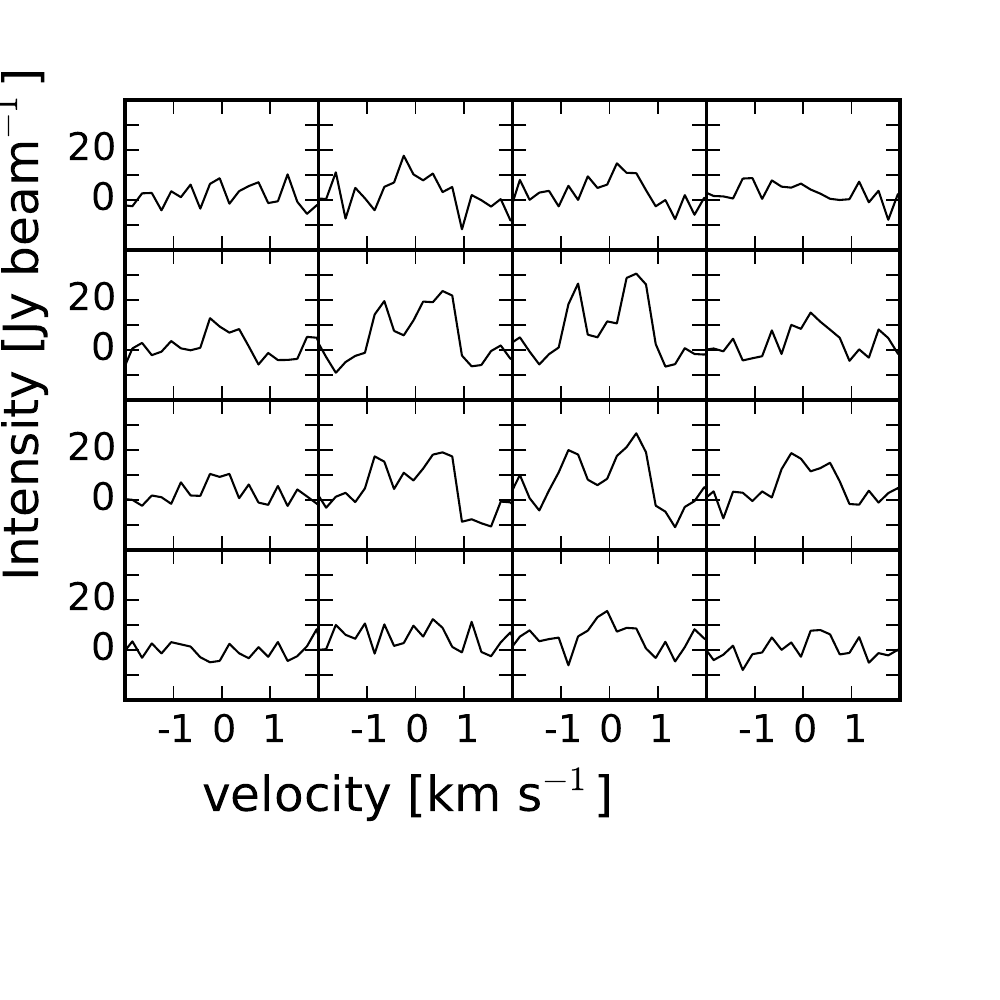}
\includegraphics[width=7cm,trim= 1.5cm 0cm 1cm 2cm,clip]{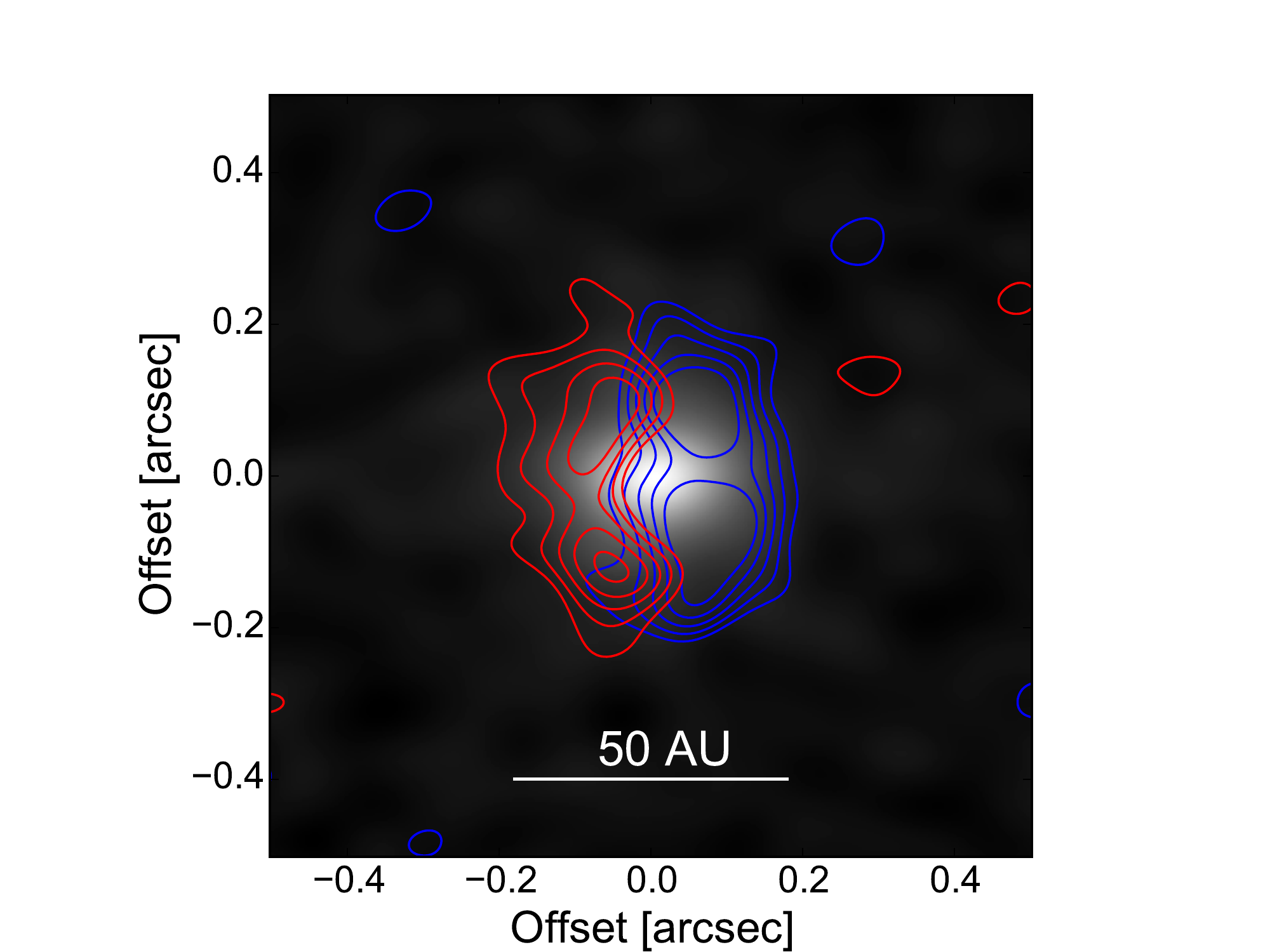}
\caption{Synthetic ALMA observations of the MHD model snapshot (d) in CO~($4-3$). Top: face--on direction, total integration time of 8 hours, pwv = 0.8~mm and $0.1 \times 0.09$~arcsec beam. Each panel is 0.12~arcsec across. Bottom: Red and blue contours show integrated intensity in the edge--on direction from \SIrange{0}{4}{\kilo\metre\per\second} and \SIrange{-4}{0}{\kilo\metre\per\second} respectively. Continuum is shown in greyscale. A total integration of 6 hours, beam size $0.1 \times 0.09$~arcsec and pwv = 0.5.}
\label{fig:CASAgridspec}
\end{figure}

\begin{figure}
\centering
\includegraphics[width=6cm]{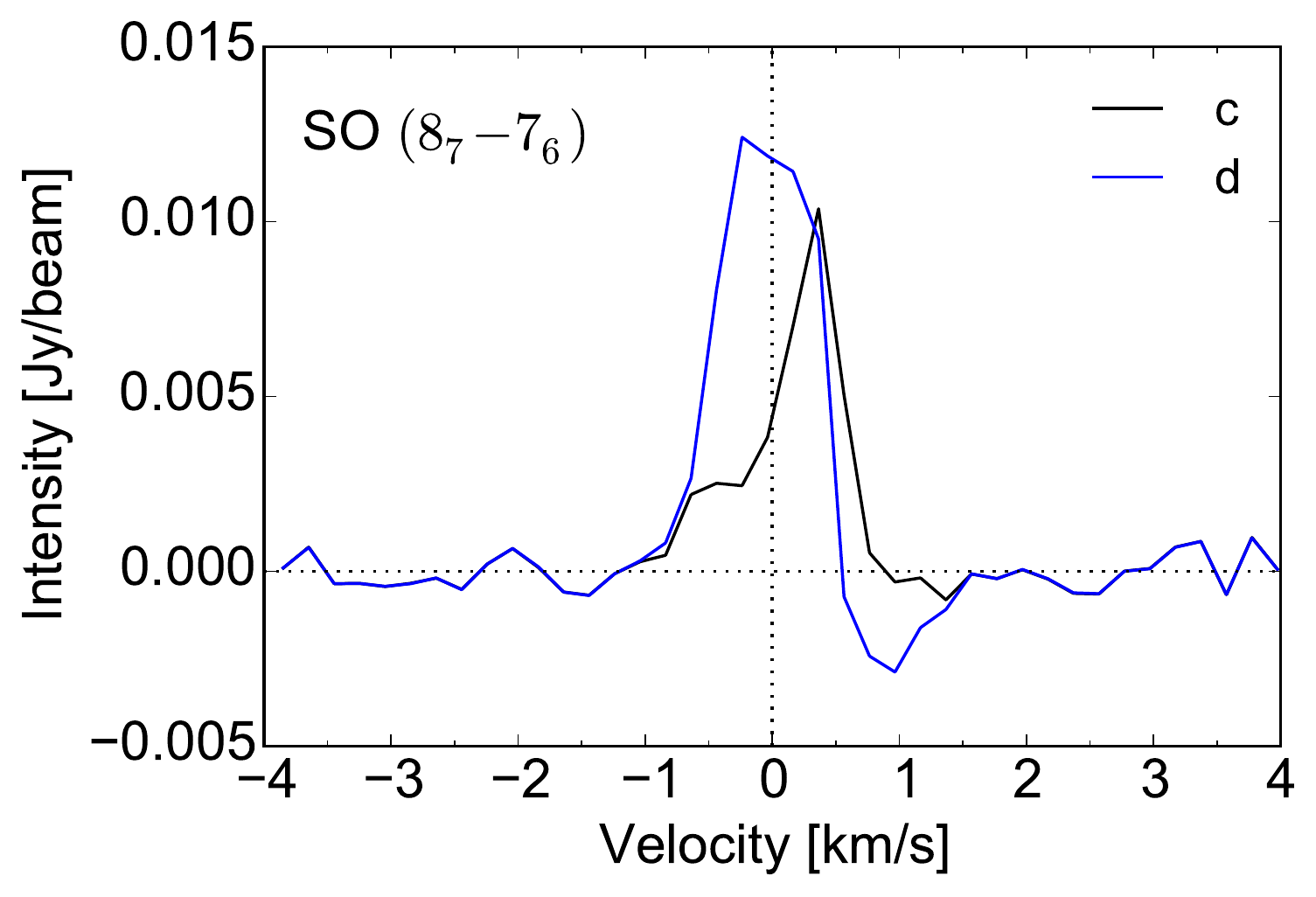}
\includegraphics[width=6cm]{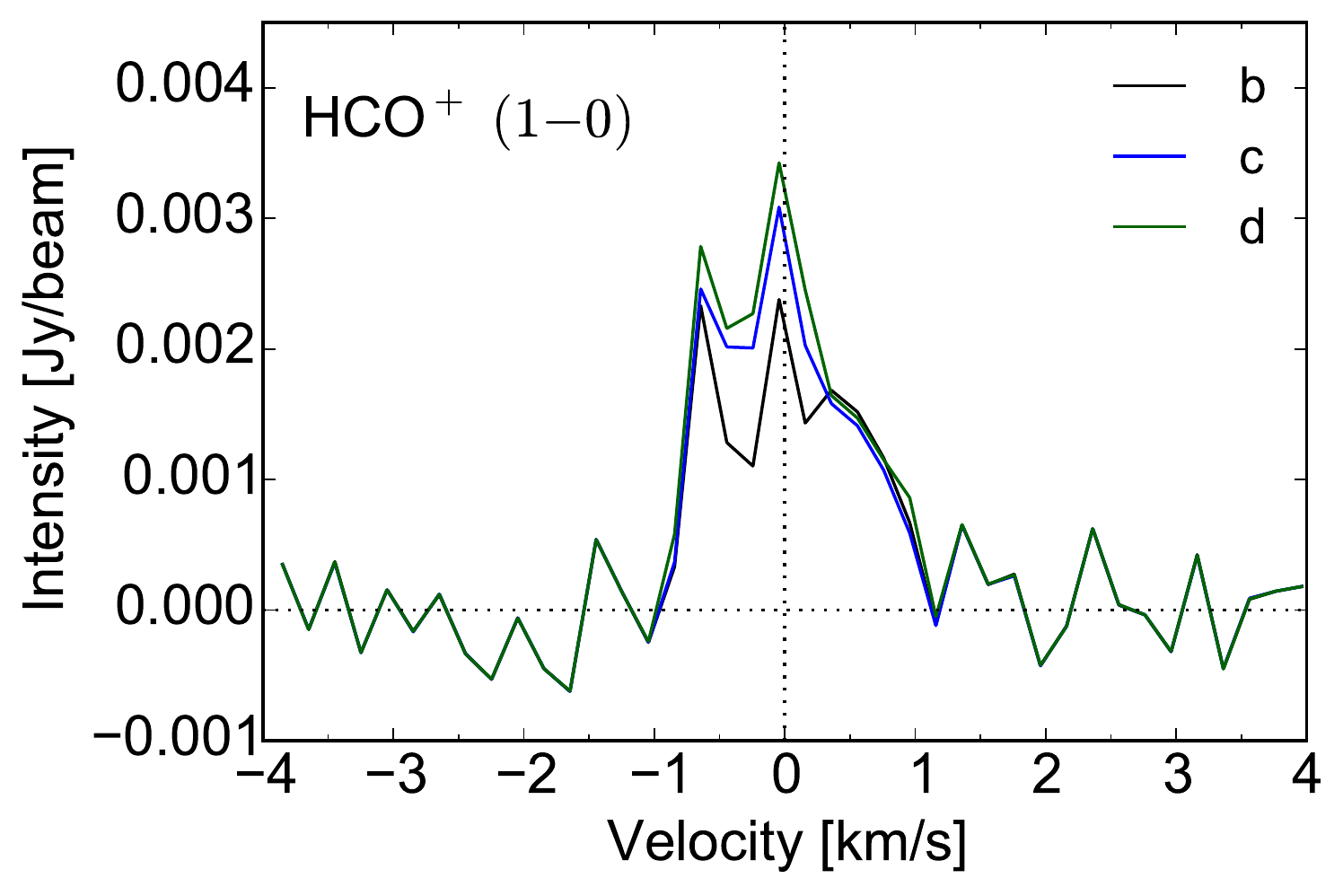}
\caption{Synthetic ALMA spectra of the MHD model in the face--on direction. Top: SO~($8_7 - 7_6$), 8 hours total integration, out14 configuration. It is possible to distinguish spectra early and late in FHSC stage at $i=0^{\circ}$ as the outflow develops. Bottom: MHD model HCO$^+$~(1-0), 8 hours total integration, $0.21 \times 0.25$~arcsec beam. As the FHSC grows emission increases at the line centre and becomes brighter than the blue peak and this is still discernible in the synthetic interferometric spectra.}
\label{fig:SOpostcasa}
\end{figure}

In this section we examine the observability of features identified in the previous section by simulating ALMA observations. For all synthetic ALMA observations we assume a realistic precipitable water vapour (pwv)~=~0.8~mm, unless otherwise stated, and use Briggs weighting. Simulated ALMA integrated intensity maps are presented in Fig.~\ref{fig:COalma} and spectra in Figs.~\ref{fig:CASAgridspec} and \ref{fig:SOpostcasa}.

The asymmetric CO~$(4-3)$ emission is discernible after a total on source time of 8 hours with the \texttt{out14} configuration (beam size $0.11 \times 0.09$~arcsec) (Fig.~\ref{fig:COalma}, centre panel). The absorption at the centre is very prominent and certainly detectable with a shorter integration. No additional detail is gained by using a smaller beam due to the increased noise. The CO~$(3-2)$ line (Fig.~\ref{fig:COalma}, left panel) is brighter and it is possible to obtain a similar level of detail with the \texttt{out16} configuration ($0.1 \times 0.08$~arcsec beam) with a far shorter 4 hour integration. Emission from the spiral arms clearly traces the asymmetric central region where the CO is seen in absorption. The CO~$(3-2)$ line is more likely to be contaminated by the foreground cloud so it is encouraging that the detection of nonaxisymmetric structure is achievable in the higher $(4-3)$ transition. SO~($8_7 - 7_6$) is also detectable (Fig.~\ref{fig:COalma}, right panel) and traces the spiral arms clearly with a total integration of 8 hours and the \texttt{out14} configuration ($0.09 \times 0.1$~arcsec beam).

The CO~$(4-3)$ spectrum in Fig.~\ref{fig:CO_line0deg} showed a double--peaked feature, characteristic of the outflow, and we find that it should be possible to detect this with ALMA. In Fig.~\ref{fig:CASAgridspec}, top, the panels each cover a larger area due to the lower resolution of the convolved image but the feature is present in the central panels after a total on source time of 8 hours. A high resolution of $\sim0.1$ arcsec is needed to detect the double--peak, otherwise it is hidden by low velocity emission from the pseudo--disc.

Many of the observed candidate FHSCs have accompanying detections of slow outflows. In Section~\ref{sec:COresults} we showed that CO emission should trace the inner regions of the outflow. However, in the simulated ALMA observation in Fig.~\ref{fig:CASAgridspec}, bottom panel, there is no clear outflow structure either in the continuum or CO~$(4-3)$ line emission. The line emission reveals red-- and blueshifted lobes, indicative of the rotation of the outflow. The rotation velocities are comparable to, or higher than, the outflow velocities. Even at low inclinations, the outflow motions were not detectable but rotation should be observable with ALMA with a 6 hour integration.

The SO~($8_7 - 7_6$) transition revealed a shift in the line centre and the development of an absorption feature between snapshots (c) and (d). This difference is still clear in the simulated ALMA observation (Fig.~\ref{fig:SOpostcasa}, upper panel) and may provide a method for distinguishing a source midway through FHSC stage from a more evolved object. In HCO$^+$ too (Fig.~\ref{fig:SOpostcasa}, lower panel), the difference between the spectra of snapshots (b) and (c) is clear although somewhat more subtle.

CS emission was found to be indistinguishable from noise even with an 8 hour integration with pwv~$=0.5$~mm for both the ($8-7$) and ($2-1$) transitions from the MHD model. The CS~($8-7$) line only became detectable for the last snapshot (g) with excellent observing conditions of pwv$=0.2$~mm and an 8 hour integration with the \texttt{out20} configuration.

\section{Discussion}
\subsection{Chemical abundances}
{\referee
Abundances of chemical species that form primarily on the surface of dust grains are highly underestimated and so we do not simulate observations for them. We now compare the abundances of several species presented in this paper with observed values, bearing in mind that the observed values were measured for dense starless cores. We will later compare our chemical abundances to the results of other simulations in \ref{sec:otherwork}.

The calculated initial abundance of CO is \num{1e-4} which is similar to the measured CO abundances in TMC-1 (\num{1.7e-4}) and L134N (\num{8.7e-5}) \citep{agundez2013aa,dickens2000aa}. As the core collapses and the density increases, the CO freezes out and the abundance is lower than TMC-1 and L134N, which is to be expected as the density exceeds that of L134N (n$_{\mathrm H2} \approx$~\SI{2e4}{\per\centi\metre\cubed} which corresponds to \SI{8e-20}{\gram\per\centi\meter\cubed}, \citealt{dickens2000aa}) and TMC-1 (n$_{\mathrm H2} \approx$~\SI{8e4}{\per\centi\metre\cubed} which corresponds to \SI{3e-20}{\gram\per\centi\meter\cubed}, \citealt{pratap1997aa}).

The initial abundance of HCO$^+$ is \num{7e-9} which is similar to the measured abundance of \num{7.9e-9} in L134N \citep{dickens2000aa} and a little lower than the value of \num{9.3e-9} found for TMC-1 \citep{pratap1997aa}.

The initial abundance of CS (\num{2e-8}) is an order of magnitude greater than that observed in TMC-1 \citep{pratap1997aa} and the KIDA network overestimates the CS abundance for t $<$\num{e6}~years. CS freezes out in the collapsing dense core and only exceeds \num{e-10} in the central few AU. Observations of the FHSC candidate Cha-MMS1 \citep{tsitali2013} find a CS abundance of \num{2.5e-9} at a radius of 8000~AU, which is consistent with our calculations if extrapolated to this radius.

The SO abundance is significantly lower by $\sim 4$ orders of magnitude initially than observed in TMC-1 and L134N. The initial abundance of NH$_3$ is initially ten times lower than observed in TMC-1 and L134N and reaches a few \num{e-9} in the core.
}

\subsection{Spatially resolved structures}
The spiral structure that develops in the RHD model should easily be spatially resolved in dust continuum observations. Indeed, there are now several examples of such structures observed in discs (e.g. \citealt{grady2013aa,perez2016aa,stolker2016aa}). We can examine whether the rotational or outflow structures are more prominent in line observations than the continuum after modelling the chemical evolution and the line emission.

The spiral structure remains more easily observed in continuum emission on scales of $\sim$~20~AU. This is, in part, because the lines of species that trace the FHSC are seen in absorption against the continuum across most of the FHSC. SO~$(8_7 - 7_6)$, CO~$(3-2)$ and CO~$(4-3)$ do however trace the spiral structure on scales of 50-60~AU where the continuum emission is much fainter.

While \mbox{HCO$^+$} and CO emission traces the outflow, the extent is very small at a representative distance of 150~pc. HCO$^+$ emission is also very faint. Any structure will be difficult to distinguish from noise even with an on source time of 8 hours.

Observations of CO~$(4-3)$ and SO~$(8_7 - 7_6)$ lines could be used to supplement continuum observations of rotating candidate FHSCs at low inclinations to provide evidence of rotational structures. For the MHD model here, it looks unlikely that the outflow will be detectable in HCO$^+$ line emission maps. The outflow is not well defined in the simulated ALMA integrated intensity map, but red and blue lobes are just detectable within 8 hours in CO($4-3$) (see Fig.~\ref{fig:CASAgridspec}, lower panel). 

\subsection{Kinematics}
Currently, the outflow velocity is one of the factors considered when classifying a faint, young source as a candidate FHSC. FHSC outflows are expected to be wide and to have velocities of a few \si{\kilo\meter\per\second} (e.g. \citealt{tomisaka2002aa,machida2008aa,bate2014,lewis2017,wurster2018aa}). {\referee Some sources identified as candidate FHSCs have later been reclassified when further observations revealed a collimated outflow, faster than $\sim$~\SI{10}{\kilo\metre\per\second} once inclination effects were considered (e.g. Per-Bolo 58, \citealt{dunham2006}, and B1-bS, \citealt{gerin2015})}. During the relatively short lifetime of the FHSC, the extent of the outflow is only a few 100~AU and infall velocities are comparable to outflow velocities so it is questionable whether the FHSC outflow would be detectable.

Of the species considered here, only CO spectra clearly show the signature of the outflow. This is because the CO abundance is much higher in the outflow out to $\sim$~30~AU above the midplane than in the envelope. However, the outflow velocities are so slow that once we consider a moderate inclination the outflow velocity component becomes negligible in comparison to the rotational and, in reality, turbulent motions. The contribution of the outflow to the HCO$^+(1-0)$ spectrum is less clear. This line is optically thin and still dominated by infall due to the higher relative abundance in the envelope than the denser central regions.

The different chemical species trace infall on different scales. Infall is apparent in the optically thick lines as redshifted absorption against the continuum. Species such as CS and SO trace regions within the FHSC and a few AU above the midplane and reveal infall onto the FHSC. As expected, HCO$^+$ traces larger scale infall onto the disc.

Most of these species display the double--peaked rotation signature in the spectra viewed from edge--on. This is not true of SO, CS or CO from the RHD model, however. In both CO and SO the disc is seen in absorption. SO traces a smaller radius than CO but most of the central part of the disc is optically thick. The SO emission comes from essentially a narrow ring offset from the midplane. The resulting spectrum is then missing the double peak. The same is true for CS but on a smaller scale. We emphasize that the rotation signatures appear to be stronger for the MHD model which does not have a rotationally supported disc but has a rotating outflow.

\subsection{Observing with ALMA}
Many species that may in theory distinguish features associated with the FHSC, including CS, are likely to be too faint to detect, even with ALMA. Temperatures are still very low in much of the object so abundances of ISM species are depleted with respect to the surrounding envelope. In addition, outflow and disc structures are still relatively compact (100s and 10s of AU respectively). Efforts to identify an FHSC should therefore concentrate on just the most abundant species.

Although these structures are still compact, the resolution is a less important consideration than minimising the noise. We find the optimum resolution is 0.08 to 0.1~arcsec for detecting rotational structures at 150~pc. With a synthesized beam of 0.1~arcsec, the largest recoverable scale is $\sim$~1.5~arcsec which means that emission from the inner envelope should be acceptably recovered. Observations were synthesized here with a realistic value for pwv = 0.8. For worse observing conditions, the FHSC structures are very likely to be indistinguishable from noise. In order to detect the outflow spatially in CO, pwv~$\leq0.5$ is necessary. An on source time of 8 hours is required for the lines modelled here. Only CO~($3-2$) is feasible with a shorter, 4 hour, integration.

{\referee
\subsection{Comparison to other work}
\label{sec:otherwork}
There have been only a few prior attempts to model chemistry of collapsing dense cores that we know of, some from analytical density and temperature structures and some from hydrodynamical models. We compare to the following work, where possible, since not every paper includes every chemical species. \citet{aikawa2008aa} calculated chemical abundances for a 1-D frequency--dependent radiation hydrodynamical model of a collapsing core. The chemical calculations included gas--phase and grain surface reactions and assumed low metal elemental abundances. \citet{van-weeren2009aa} performed similar calculations but from 2-D hydrodynamical models, allowing the effects of a disc structure to be explored. The chemical evolution was followed for 700 tracer particles by calculating gas--phase, gas--grain and grain surface reactions. \citet{furuya2012aa} used a 3-D RMHD model of a collapsing 1~M$_\odot$ cloud, calculating chemistry for \num{e5} tracer particles. Gas--phase and grain surface reactions were included. \citet{hincelin2016} specifically set out to look for chemical differences between different components of the core in 3-D. They performed several RMHD simulations of a collapsing 1~M$_\odot$ cloud and calculated the chemical evolution of \num{e6} tracer particles with gas--phase and grain surface reactions. \citet{dzyurkevich2016aa} used a reduced network of mainly gas--grain H-C-O chemistry to calculate abundances at the same time as calculating the dynamical evolution of the collapse of a dense core. \citet{maret2013aa} attempted to model the chemistry and two line profiles to compare to two observed pre--stellar cores. They took an analytical density profile which did not vary in time since they did not seek to model a collapsing core but these results are useful to compare with the abundances in the envelope here. The chemical model included gas--phase and gas--grain reactions. We note that the above papers quote abundances relative to the total density of hydrogen nuclei and we account for this in the comparison.

The highest abundances of CO that we obtained during the collapse of the pre--stellar core are approximately a factor of 2 higher than those reported by \citet{furuya2012aa}, which used a $\approx$ 50 per cent lower elemental C$^+$ abundance. The envelope abundance of CO was a few $\times 10^{-7}$ which is similar to the values of \citet{hincelin2016}, \citet{furuya2012aa} and to the abundance from the fiducial single sized dust grain model (S1) of \citet{dzyurkevich2016aa} at $\sim$~\num{2000}~AU.

In the envelope, we find an HCO$^+$ abundance of a few $\times 10^{-9}$, which is similar to that of \citet{van-weeren2009aa} and  \citet{dzyurkevich2016aa} at $\sim$~\SI{6000}{\au} in their models. This abundance is a little higher than that found for similar radii by \citet{maret2013aa} and \citet{aikawa2008aa}.

Abundances of CS in the FHSC and envelope have previously been calculated to be a few $\times 10^{-11}$ \citep{hincelin2016,van-weeren2009aa} or as high as \num{e-8} \citep{aikawa2008aa}. Between these regions CS is very highly depleted. We find the abundance in the centre to be a few $\times 10^{-9}$ and a few $\times 10^{-12}$ in the outer envelope.

The abundance of NH$_3$ is lower than the central values of a few \num{e-6} to a few \num{e-5} reported by others \citep{aikawa2008aa,van-weeren2009aa,hincelin2016}. The initial abundance of N$_2$H$^+$ agrees well with the envelope abundances of \citet{van-weeren2009aa} and \citet{hincelin2016}. The abundance in the centre is a few \num{e-20} as was found by \citet{hincelin2016} but we find the abundance in the disc and pseudo--disc to be far lower than the abundances they report for those regions.

Synthetic CS spectra were presented by \citet{tomisaka2011} and these are qualitatively very different to those here, but their models assumed a constant CS abundance of \num{4e-9} throughout whereas we expect CS to be depleted as the pre--stellar core collapses. It is not returned to the gas phase in the envelope until some time after the protostar has formed (see also \citealt{aikawa2008aa} for a study of the chemical changes that occur as the protostellar core warms up). We find that the CS sublimation radius extends to $\sim$~10--15~AU in the MHD model (Figs.~\ref{fig:B02abundh} and \ref{fig:B02abundv}) so does not trace the gas kinematics. In the RHD model, the CS sublimation radius extends to 20--30~AU (Figs.~\ref{fig:Mu5abundh} and \ref{fig:Mu5abundv}) because the FHSC is larger due to the greater rotational support. In addition, because of the freeze--out, the CS~$(8-7)$ emission is faint, and likely to be undetectable during much of the FHSC phase.

\citet{hincelin2016} find that species such as CS and HCO$^+$ may be useful for distinguishing the envelope from the outflow and pseudo--disc. We also find a difference in abundance between the envelope and these components but for CS the spectrum is still dominated by the central few AU where the abundance of CS and temperature are much higher. For HCO$^+$, the decrease in abundance in the disc and outflow is offset by the enhanced density and temperature such that the outflow and pseudo--disc are brighter than the envelope in emission maps. We agree with their conclusion that the chemistry of the envelope alone cannot distinguish between FHSC formation models. The temperature profile of the envelope changes very little and changes in the composition of the envelope are likely to occur only on longer timescales. Nevertheless, the spectra are very sensitive to the distribution of chemical species so chemical evolution should certainly be taken into account when simulating molecular line observations.
}

\subsection{Determining evolutionary stage}

The main objective of modelling the chemistry and synthetic observations of the FHSC is to determine how to distinguish it from "empty" pre--stellar cores and from very young protostars. The abundances of CS and SO are extremely low before FHSC formation and the SO~$(8_7 - 7_6)$ line becomes detectable midway through the FHSC stage. CS~$(8-7)$ remains undetectable with 8 hours integration with ALMA throughout FHSC stage and will still be close to the detection limit after stellar core formation, even in the best observing conditions. A detection of CS at subarcsecond scales would preclude a source from being an FHSC.

Changes in the shape of the SO~$(8_7 - 7_6)$ spectrum at low inclinations when the outflow is launched should be detectable with ALMA. The central peak of HCO$^+$~$(1-0)$ becomes brighter than the blueshifted peak late in the FHSC stage, offering another possible indicator of evolutionary stage.

Factors other than age cause considerable variation in the properties of the spectra. For this reason, it is probably not possible to provide generic criteria to distinguish an FHSC from a stellar core. The positive identification of the FHSC is going to require specific modelling of individual sources based on observationally derived morphology, magnetic field and dust properties to determine the spectra expected from those specific conditions.

\subsection{Comparison to observations of candidate FHSCs}
There are a few published observations of candidate FHSCs with better than 1000~AU resolution. We now consider whether these sources are genuinely in the FHSC stage in light of this work and also from the SED fitting of \citet{young2018aa}.

CS~$(2-1)$ and CS~$(5-4)$ transitions have been detected at the FHSC candidate Chamaeleon-MMS1 \citep{tsitali2013}. We do not expect there to be detectable CS emission from the FHSC and what is detected here is mostly emission from the surrounding envelope since the beam size (24.9~arcsec) is much larger than our models. It would be informative to obtain interferometric observations of CS emission with a smaller beam to determine its source.
The abundance of CS increases quickly within $r<50$~AU after stellar core formation, during which time the extent of the outflow would not have increased significantly. The stellar core outflow is expected to launch very quickly after formation 
unless the magnetic field is weak or misaligned. While it is unlikely that Cha-MMS1 is an FHSC, it is also unlikely to have evolved far beyond stellar core formation. Further observations suggest rotational motions and an absence of shocks, which also points to it being a very young object. The results of previous SED fitting indicated too that Cha-MMS1 is more likely to be a very young protostar than an FHSC \citep{young2018aa}.

Another candidate is L1451-mm where a slow, poorly collimated outflow was detected in CO~($2-1$) \citep{pineda2011aa,tobin2015aa}. We find that CO should indeed trace the outflow although in the synthetic observations the velocity channels trace the rotational motions of the outflow. The observed CO extends for several hundred AU which is a far larger region than we would expect, unless the lifetime of the FHSC is longer than expected.

Two candidate FHSCs in Ophiuchus presented by \citet{friesen2018aa} are also consistent with the results here. Blue-- and redshifted CO~$(2-1)$ emission was detected at SM1N without a clear outflow morphology. This is what we find from simulated ALMA observations: the FHSC outflow has a small extent and once the image was convolved with the ALMA beam and noise was added we did not see a clear outflow structure. N6-mm also appears to have a compact and broad outflow. No absorption against the continuum was observed at either source but this is not expected in CO from the MHD model.

The SED of CB17-MMS1 is consistent with a young FHSC \citep{young2018aa}. There is also a rather confusing structure seen in CO~($2-1$) emission thought to be associated with the source \citep{chen2012}. If this is tracing an outflow then the extent is probably too great for an FHSC since we expect only the central part of the outflow to be detectable where temperatures are high enough for CO to desorb from dust grains. 

B1-bN is another FHSC candidate that has been observed several times now. Previous SED modelling indicated that the source is consistent with an FHSC but only if it has a substantial disc and is viewed at a near side--on inclination \citep{young2018aa}. An outflow has been detected in CO~($2-1$) \citep{hirano2014}, {\referee H$_2$CO$~(2_{02}-1_{01})$ and several CH$_3$OH $(3-2)$ lines \citep{gerin2015}. This outflow is thought to extend to $\approx$~1000~AU (see the Table 3 of \citealt{gerin2015}). We find that CO only traces the inner few 10s of AU of the outflow and H$_2$CO and CH$_3$OH are unlikely to trace a much more extended region than CO.} Therefore it is difficult to explain this observation as an FHSC outflow with such a large extent. \citet{gerin2015} also point out that the outflow is not aligned with the magnetic field which means it is possible that the source is a slightly more evolved object but with a shorter and less well collimated outflow than might be expected for its age.

\subsection{Limitations}
The chemical network used here omitted grain--surface reactions and we therefore excluded species known to be governed by surface reactions. Species such as formaldehyde and methanol are very likely to be good tracers of the outflow so it is worth modelling observations of these in future. The chemical network we have used is valid between 10-800~K and is extrapolated in the inner 1~AU after late FHSC stage. While this is unlikely to affect the results here which concern larger scales, the calculated abundances in the centre of the FHSC may be different when additional reactions and correct rates are considered.

In this paper we used two extreme models: one which forms a disc and another which forms an outflow and pseudo--disc. Other initial conditions will result in a combination of these structures. Specific modelling is therefore required to compare with particular sources in detail.

\section{Conclusions}
We have presented synthetic CO, CS, SO and HCO$^+$ observations produced from hydrodynamical and chemical modelling of the formation and evolution of the first hydrostatic core. We showed that rotational structures should be detectable by ALMA in CO~($4-3$) and SO~$(8_7 - 7_6)$ transitions in realistic observing conditions. CO emission did not reveal the outflow morphology in simulated ALMA maps but red and blue lobes caused by rotation, rather than outflowing motions, were detectable.

Regarding determining the evolutionary stage, we find changes in SO and HCO$^+$ spectra as the FHSC develops when viewed at a low inclination. We find that CS is difficult to detect during the FHSC stage with a reasonable integration time and observing conditions with ALMA. This means that candidate FHSCs with CS detections on subarcsecond scales (i.e. without significant contribution from the envelope) are unlikely to be FHSCs; they are likely to be more evolved.

FHSC structures are still very compact and most chemical species are frozen out in much of the inner envelope. Although chemical evolution alone is unlikely to provide a diagnostic of the FHSC, synthetic molecular line observations should account for chemical evolution. Using constant spatial abundances is likely to lead to incorrect synthetic observations.

Finally, we compare these results with selected candidate FHSCs. We rule out B1-bN due to the detection of an outflow which extends $\sim$1000~AU. Cha-MMS1 is more likely to be a very young protostar but we cannot rule it out completely because there is no reported detection of an outflow and existing observations resolve regions no smaller than 1000~AU. L1451-mm is unlikely to be in FHSC stage because the extent of the observed outflow is greater than we would expect. Similarly, if the observed CO emission near CB17-MMS1 is tracing an outflow driven by the source then it is too extended to be driven by an FHSC. This leaves Oph A SM1N and N6-mm as the most promising sources to follow up.

Identifying the FHSC in nature will require interferometric observations of lines such as CO~($4-3$) and SO~$(8_7 - 7_6)$ with a spatial resolution $\sim0.05$~arcsec and a velocity resolution of \SI{0.2}{\kilo\metre\per\second} to detect structures on the scale of a few 10s AU. Detection of these lines should be achievable within an 8 hour integration with ALMA.

\section*{Acknowledgements}
We thank Jennifer Hatchell and Stefan Kraus for helpful discussions regarding the interpretation of observations and interferometry. We are also grateful to Tom Douglas for his advice on processing data files and to Catherine Walsh for a useful conversation about chemical networks. We also thank the anonymous referee for suggesting that we compare both the KIDA 2011 and 2014 chemical networks.

This research made use of Astropy,\footnote{http://www.astropy.org} a community--developed core Python package for Astronomy \citep{astropy:2013, astropy:2018}. Some figures were produced using the publicly available SPLASH visualization software \citep{price2007}.

This work was supported by the European Research Council under the European Commission's Seventh Framework Programme (FP7/2007-2013 Grant Agreement No. 339248).
Calculations discussed in this paper were performed on the University of Exeter Supercomputer, a DiRAC Facility jointly funded by STFC, the Large Facilities Capital Fund of BIS, and the University of Exeter. This work made use of the Cambridge Service for Data Driven Discovery (CSD3), part of which is operated by the University of Cambridge Research Computing on behalf of the STFC DiRAC HPC Facility (\url{www.dirac.ac.uk}). The DiRAC component of CSD3 was funded by BEIS capital funding via STFC capital grants ST/P002307/1 and ST/R002452/1 and STFC operations grant ST/R00689X/1. This work also used the DiRAC Complexity system and the DiRAC Data Intensive service, operated by the University of Leicester IT Services, which forms part of the STFC DiRAC HPC Facility. The Complexity system was funded by BIS National E-Infrastructure capital grant ST/K000373/1 and STFC DiRAC Operations grant ST/K0003259/1. The Data Intensive service is funded by BEIS capital funding via STFC capital grants ST/K000373/1 and ST/R002363/1 and STFC DiRAC Operations grant ST/R001014/1. DiRAC is part of the National e-Infrastructure.




\bibliographystyle{mnras}
\bibliography{paper}




\appendix
{\referee
\section{Comparison to the KIDA 2014 network}
We performed the chemical calculations again for the MHD model with the gas phase reactions from KIDA 2014 network \citep{wakelam2015aa} to examine how the results might differ. The gas--grain reactions which were added to the KIDA network separately remain unchanged. Figs.~\ref{fig:K2014_h} and \ref{fig:K2014_v} show the resulting abundance profiles averaged in the plane of the disc and the rotation axis respectively.

The abundance profiles are mostly very similar with the 2011 and 2014 KIDA networks because it is the gas--grain reactions which drive the abundances outside of the FHSC. The snapshot taken during the first collapse (a) shows higher central abundances for CO and HCO$^+$. This is simply because the desorption of these species happens very quickly at this stage so the abundances increase rapidly and the exact abundance is sensitive to the exact moment it is measured. For CO, the peak abundance in the FHSC is slightly lower early in the FHSC stage but reaches $\sim$~\num{1e-4} by late in the FHSC stage, as it did with the KIDA 2011 network. The sublimation radius extends to 20~AU in the plane of the disc, as before. The SO abundance is slightly higher deep within the FHSC but the profiles are very similar. The abundances profiles of HCO$^+$ are very similar for both networks.
With the 2014 network, the CS abundance may be nearly an order of magnitude higher deep within the FHSC ($r \lesssim3$~AU). Comparing all of the molecular profiles shown in Figs.~\ref{fig:Mu5abundh} and \ref{fig:Mu5abundv} with those in Figs.~\ref{fig:K2014_h} and \ref{fig:K2014_v}, only OCS has substantially different abundances over a wide range of radii.

\begin{figure*}
\centering
\includegraphics[width=17.2cm]{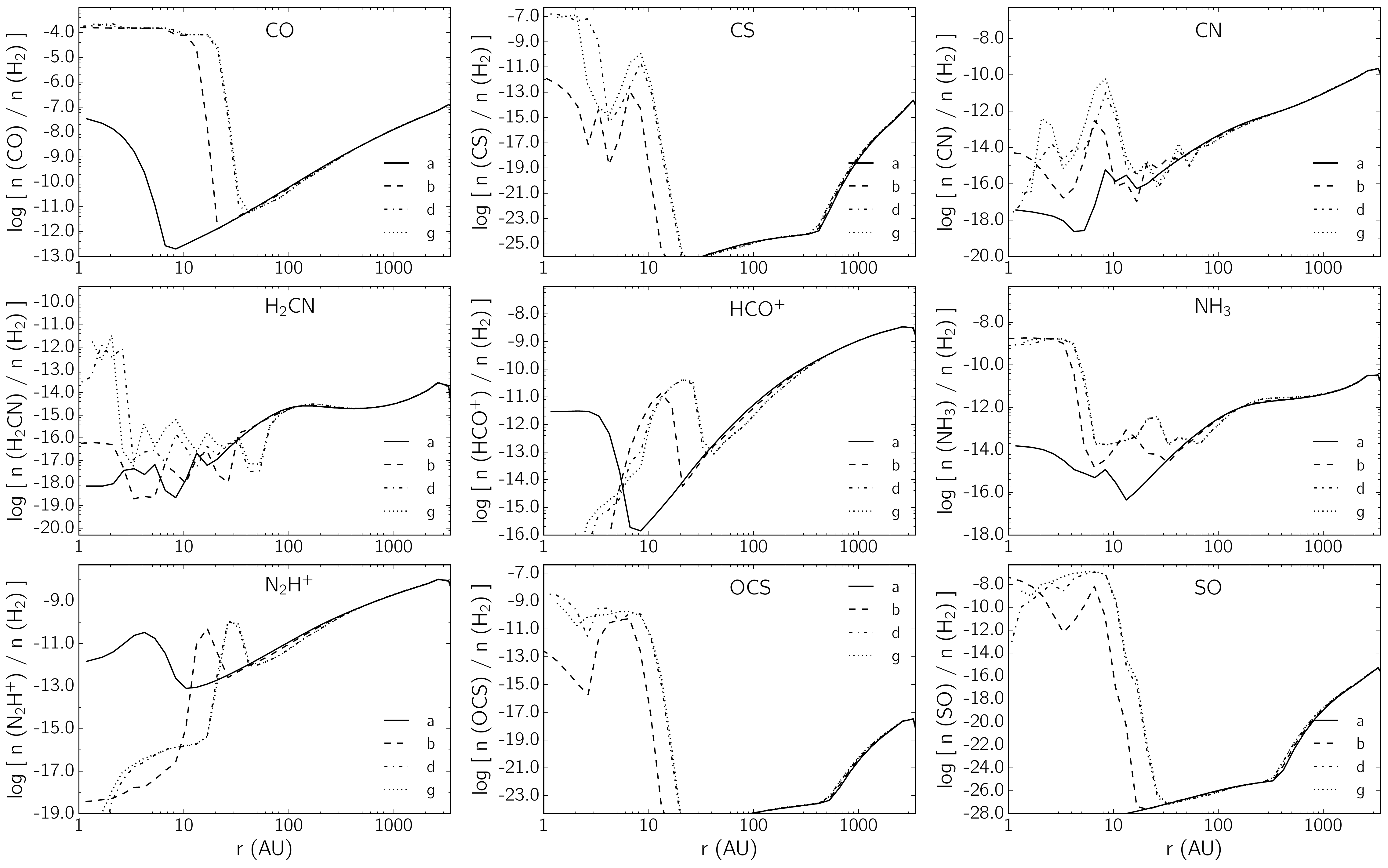}
\caption{Average abundances of selected species from the MHD model perpendicular to the rotation axis, calculated using the KIDA 2014 chemical network.}
\label{fig:K2014_h}
\end{figure*}

\begin{figure*}
\centering
\includegraphics[width=17.2cm]{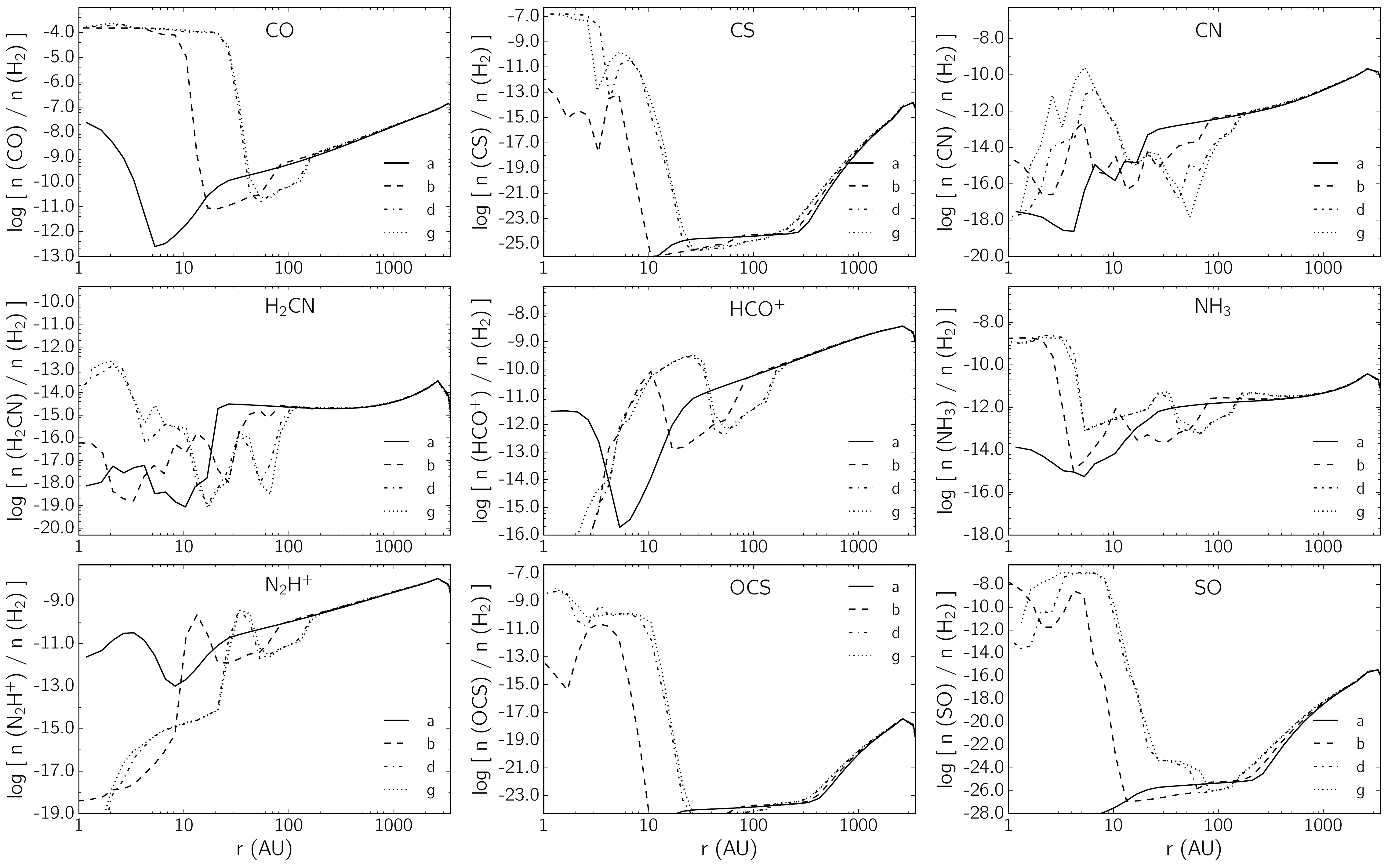}
\caption{Average abundances of selected species from the MHD model parallel to the rotation axis, calculated using the KIDA 2014 chemical network.}
\label{fig:K2014_v}
\end{figure*}
}

\bsp	
\label{lastpage}
\end{document}